\documentclass[10pt]{article}

\usepackage{amsmath,amsthm,amssymb}
\usepackage[top=2cm, bottom=2cm, left=2cm, right=2cm]{geometry}
\usepackage{url}
\usepackage[colorlinks=true,bookmarks=false,citecolor=blue,urlcolor=blue]{hyperref} 
\usepackage{comment}
\usepackage{orcidlink}
\usepackage{caption}
\usepackage{float}
\usepackage{microtype}
\usepackage{subcaption}
\usepackage{graphicx}
\usepackage[backend=biber, maxbibnames=99, giveninits=true, sorting=nyt]{biblatex}
\DeclareNameAlias{author}{family-given}
\DeclareDelimFormat{nameyeardelim}{\addcomma\space} 

\newcommand{\textRoman}[1]{\textup{\uppercase\expandafter{\romannumeral#1\relax}}}

\usepackage{mathtools}
\usepackage{authblk}
\usepackage{bbm}
\usepackage{etoolbox}
\usepackage{hyperref}
\usepackage{multirow}
\usepackage{fancyhdr}
\usepackage{ulem}
\usepackage{multicol}

\title{\textbf{Optimal Execution under Incomplete Information}}

\author{
Etienne CHEVALIER\footnote{Laboratoire de Math\'ematiques et Mod\'elisation d'Evry, Universit\'e Paris-Saclay, UEVE,
 UMR 8071 CNRS, France; email: etienne.chevalier@univ-evry.fr} $\,$
 Yadh HAFSI \footnote{Laboratoire de Math\'ematiques et Mod\'elisation d'Evry, Universit\'e Paris-Saclay, UEVE,
 UMR 8071 CNRS,  France; email: yadh.hafsi@universite-paris-saclay.fr} 
 $\,$ 
 Vathana LY VATH \footnote{Laboratoire de Math\'ematiques et Mod\'elisation d'Evry, Universit\'e Paris-Saclay, ENSIIE, UEVE,
 UMR 8071 CNRS, France; email: vathana.lyvath@ensiie.fr}
}
\newcommand{\blfootnote}[1]{\begingroup\renewcommand\thefootnote{}\footnote{#1}\addtocounter{footnote}{-1}\endgroup}

\newcommand{\nunder}[2][5]{\mathrlap{\mkern\the\numexpr#1/2mu\relax\underline{\phantom{\mathrm{#2}\mkern-#1mu}}}#2}

\bibliography{main.bib}

\begin{document}
\maketitle
\blfootnote{The authors are grateful to Claudia Ceci and Johannes Muhle-Karbe for their insightful remarks, which contributed to substantially improving the paper.}
\pagenumbering{arabic}
\newcounter{axiom}
\newtheorem{axiome}[axiom]{Axiome}
\newtheorem{defi}{Definition}[section]
\newtheorem{theo}{Theorem}[section]
\newtheorem{prop}{Proposition}[section]
\newtheorem{corro}{Corollary}[section]
\newtheorem{rque}{Remark}[section]
\newtheorem{example}{Example}[section]
\newtheorem{nota}{Notation}[section]
\newtheorem{demo}{Demonstration}
\newtheorem{propt}{Propriété}
\newtheorem{lemma}{Lemma}
\newtheorem{assump}{Assumptions}[section]
\newcounter{cases}
\newcounter{subcases}[cases]
\newenvironment{mycase}
{
    \setcounter{cases}{0}
    \setcounter{subcases}{0}
    \newcommand{\case}
    {
        \par\indent\stepcounter{cases}\textbf{Case \thecases.}
    }
    \newcommand{\subcase}
    {
        \par\indent\stepcounter{subcases}\textit{Subcase (\thesubcases):}
    }
}
{
    \par
}
\renewcommand*\thecases{\arabic{cases}}
\renewcommand*\thesubcases{\roman{subcases}}


\begin{abstract}
We study optimal liquidation strategies under partial information for a single asset within a finite time horizon. We propose a model tailored for high-frequency trading, capturing price formation driven solely by order flow through marked Hawkes processes. The model assumes a limit order book framework, accounting for both permanent price impact and transient market impact. Importantly, we incorporate liquidity as a hidden Markov process, influencing the intensities of the point processes governing bid and ask prices. Within this setting, we formulate the optimal liquidation problem as an impulse control problem. We elucidate the dynamics of the hidden Markov chain's filter and determine the related normalized filtering equations. We then express the value function as the limit of a sequence of auxiliary continuous functions, defined recursively. This characterization enables the use of a dynamic programming principle for optimal stopping problems and the determination of an optimal strategy. It also facilitates the development of an implementable algorithm to approximate the original liquidation problem. We enrich our analysis with numerical results and visualizations of candidate optimal strategies.
\end{abstract}\hspace{10pt}

\section{Introduction}
In modern financial markets, the execution of large orders within short timeframes presents unique challenges. Traders must develop strategies to maximize profits while minimizing risks, navigating a complex landscape influenced by immediate market depth and liquidity constraints. The studies by \textcite{BOUCHAUD200957}, \textcite{Zhou_impact}, and \textcite{taranto_impact}, for instance, highlight how large trades impact asset prices by depleting available market liquidity. This depletion occurs because the market's immediate depth is limited, meaning that a single large order can exhaust all current buyers or sellers. Consequently, splitting large orders into smaller blocks often proves advantageous, reducing the price impact and allowing for more efficient execution.

Research in market microstructure has extensively explored the optimal execution problem across various models of market impact and cost functions. For instance, \textcite{BERTSIMAS19981} introduced one of the earliest frameworks for optimal trade execution, focusing on minimizing the total trading cost given a linear price impact. \textcite{Almgren2000OptimalEO} extended this line of research by considering a trade-off between volatility risk and liquidation costs. They incorporated linear permanent and temporary market impact models, providing a more dynamic approach to the execution problem. Their work laid the groundwork for subsequent studies, including \textcite{LyVath2007}, \textcite{gatheral} and \textcite{OBIZHAEVA20131}, who examined different aspects of market impact and execution costs, proposing models that account for both temporary and permanent price impacts. Different extensions of these models have been introduced since then to include nonlinear impacts, reflecting more realistic trading conditions; see e.g. \textcite{bouchard_lehalle}, \textcite{gueant2012}, \textcite{GuilbaudPham}, \textcite{ChevalierLVRoch}, \textcite{Kalsi2020}, \textcite{Zhou}, \textcite{Becherer2018}, \textcite{Guanxingfu}, \textcite{Cartea2023}, and \textcite{carmona2022}.

Since high-frequency traders manage small inventories, their transaction costs are driven mainly by short-term order-flow imbalances rather than by volume effects. As emphasized by \textcite{lehalle_rosenbaum2021}, the order-book imbalance provides an effective representation of short-term liquidity in this regime. Our study builds on this foundation by proposing a model tailored for high-frequency trading, capturing price formation driven solely by order flow. To accommodate a general specification of the limit order book and hence a general form of market impact, we adopt a framework inspired by \textcite{predoiu2011}. We complement the transient impact component described in the latter with a permanent impact term. Previous works have examined nonlinear price impact models but considered order arrival processes as Poisson processes. In contrast, our model represents order arrivals through Hawkes processes. The use of mutually stimulating point processes to model order flow, as introduced by \textcite{Bacry2015}, allows for an accurate description of market dynamics (see \textcite{abergel2016, Lu01022018}). This approach recognizes the feedback loops present in high-frequency trading, where the occurrence of trades influences the likelihood of subsequent trades. \textcite{alfonsi2016dynamic} were the first to introduce Hawkes dynamics in an optimal execution setting. In contrast, our approach adopts a general limit order book shape which is not flat, broadening the representation of market impact. We also take into account fixed transaction costs to model the operational costs related to trading infrastructure.

As a subsequent step, our modeling approach incorporates the stochastic behavior of liquidity. Our study acknowledges that liquidity regimes are unobservable. These hidden aspects of market microstructure have been shown to substantially influence trading outcomes. For instance, the recent work by \textcite{chevalier2023uncovering}, based on real market data, exhibits that the distribution of market liquidity is not directly observable and, furthermore, exhibits substantial intraday variation. Studies like \textcite{BayraktarLudoExecution}, \textcite{COLANERI20201913} and \textcite{Dammann2023} have begun addressing this gap by modeling hidden liquidity as a stochastic process. These works emphasize that ignoring hidden liquidity can lead to suboptimal execution strategies, as traders may underestimate the true market depth and the associated risks. We propose an approach that enables the integration of the dynamics of hidden liquidity, which we model as a hidden Markov process. This hidden Markov process influences the intensities of the point processes that govern the bid and ask prices, resulting in prices driven by Markov-modulated Hawkes processes.

Given the non-observable nature of liquidity within the limit order book, market participants must operate under a state of incomplete information. Consequently, they are required to estimate the prevailing liquidity conditions based solely on the observable order flow data. To address this challenge, we use stochastic filtering to derive our state variables under complete information. More precisely, we utilize the innovation approach for point processes described in the work of \textcite{Bremaud} and \textcite{Ceci2000}. This allows us to derive the Kushner-Stratonovich equations, which characterize the dynamics of our estimates for these hidden liquidity states as new information becomes available. This enables us to make informed trading decisions, ensuring that our strategies adapt to the evolving market conditions. 

Once we have a method for estimating liquidity states, we move on to tackle an optimal liquidation problem formulated as an impulse control problem. The impulse control problem involves determining the optimal times and sizes of trades. We apply the separation principle, which simplifies the problem by treating the estimation and control tasks separately. This principle, rooted in control theory, allows us to manage the complexity of partial observation and control, as described by \textcite{menaldi81}, \textcite{fleming1982}, \textcite{separation_partial}, and \textcite{Mazliak1993}. This approach has been particularly effective in systems where the underlying state variables cannot be directly observed, requiring robust estimation techniques to guide optimal control actions. Next, we derive an approximating sequence of auxiliary functions defined recursively. This approach has also been utilized in the works of \textcite{BAYRAKTAR20091792} and \textcite{sezer_ludo}, among others. We prove that this sequence of functions is continuous and converges locally uniformly to the original value function. 
This characterization enables the use of a dynamic programming principle for closed loop optimal stopping problems, providing us with a critical argument for the explicit determination of the optimal liquidation strategy. Our model introduces complexities and specific features that make it challenging to adapt the previously established mathematical tools. Overall, our approach to optimal execution incorporates a diverse range of market conditions and dynamics, taking into account both hidden liquidity states and generalized limit order book shapes with fixed transaction costs.\\

\noindent\textbf{Our contributions.} our main contribution is to solve an impulse control problem whose observation process is both self-exciting and modulated by the hidden state. Because the order flow excites itself, the observed intensities depend on its whole past and are unbounded. The posterior probabilities no longer form a Markov process on their own, and boundedness no longer hold. Our analysis rests on a specification in which every candidate intensity is observable, so that the filter, once augmented by these intensities, becomes a finite-dimensional PDMP by construction (Remark \ref{terminology_mmhp}), and moment, nonexplosion and continuity estimates (Proposition \ref{model_construction}, Lemmas \ref{cont_expectation_bankruptcy} and \ref{continue_u_0}) that take the place of boundedness up to the bankruptcy boundary, which makes the horizon state-dependent. Explicit methods offer no alternative. The strategies of \textcite{alfonsi2016dynamic}, for example, depend on the unobservable intensities, while the general book shape and the fixed transaction cost remove the quadratic structure on which their computations are based. The optimal strategy is therefore constructed through the recursive characterization and the verification theorem of Section \ref{Characterization of the Value Function}.\\

\noindent\textbf{Outline.} The article is structured in the following manner. In Section \ref{model_setup}, the primary aim is to develop a model that characterizes the dynamics of a Limit Order Book (LOB) within a framework that accounts for the stochastic nature of market liquidity. In Section \ref{The filtering equations}, we elucidate the dynamics of the filter using the Kushner-Stratonovich equations, which allows to update the estimates of the hidden liquidity states based on new information (see Theorem \ref{ks_equation_final}). Section \ref{The Separated Impulse Control Problem} formulates the impulse control problem and transitions to a full information setup using the separation principle. It also introduces an approximating sequence of functions to characterize the value function and states the dynamic programming principle in various forms. In Section \ref{Characterization of the Value Function}, we investigate the regularity of the value function, which will pave the way for constructing an optimal strategy under the given market conditions through the Verification Theorem \ref{verfication}. Finally, in Section \ref{Numerical Results}, we provide numerical illustrations of the shape of the optimal exercise region while assessing the influence of the state variables and the price impact on the candidate optimal liquidation policy.

\section{Model Setup}
\label{model_setup}
In this section, our objective is to characterize the dynamics of a limit order book subject to recurrent buy and sell market orders within a framework of stochastic liquidity.
 Expanding upon the model proposed by \textcite{alfonsi2016dynamic} for a single asset, we generalize the impact function and move beyond the constraints of a block-shaped LOB. This model is selected due to its suitability for high-frequency trading, as it captures price formation driven by order flow through mutually exciting point processes. We will further extend this framework across various liquidity regimes by employing Markov-modulated processes.
\subsection{Liquidity Dynamics}
Consider a complete filtered probability space $(\Omega ,\mathbb{F}, \mathcal{F} = \{\mathcal{F}_t\}_{t \geq 0},\mathbb{P})$, where $\{\mathcal{F}_t\}_{t \geq 0}$ is a right-continuous filtration. Throughout this paper, any equality between random variables is understood to hold 
$\mathbb{P}$-almost surely, though this is not always explicitly stated.

The space supports a continuous-time Markov chain $I$ with values in a finite state space $E = \{1, \ldots, d\} \subset \mathbb{N}$ and with càdlàg sample paths, and a pair $(M^{+},M^{-})$ of independent Poisson random measures on $\mathbb{R}_+^{3}$ with intensity measure $\mathrm{d}t\,\mathrm{d}v\,\mathrm{d}z$, independent of $I$, from which the observed order flow $(n^{+},n^{-})$ is constructed by thinning in \eqref{thinning_def} below. The filtration $\mathcal{F}$ is the usual augmentation (right-continuous and $\mathbb{P}$-complete) of the filtration generated by these primitives,
$$t\;\mapsto\;\sigma\big(I_s;\ s\leq t\big)\ \vee\ \sigma\big(M^{\pm}\big((0,s]\times B\big);\ s\leq t,\ B\in\mathcal{B}(\mathbb{R}_+^{2})\big).$$
In particular, $M^{+}$ and $M^{-}$ remain Poisson random measures with respect to $\mathcal{F}$, the chain $I$ retains the Markov property with respect to $\mathcal{F}$, and all the processes appearing in the paper, being constructed from the primitives $(I,M^{+},M^{-})$, are $\mathcal{F}$-adapted.
\paragraph{Markov Chain.} Let $\mathcal{F}^I = \{\mathcal{F}_t^I\}_{t \geq 0}$ denote the natural filtration generated by $I$, augmented by the $\mathbb{P}$-null sets. The process $I$ represents the liquidity state within the limit order book. This process is driven by a possibly time-dependent transition rate matrix $\psi(t) = (\psi_{jk}(t))_{1 \leq j,k \leq d}$, such that, for all $t \geq 0$,
$$\psi_{ij}(t) = \underset{h\rightarrow 0}{\lim}\frac{\mathbb{P}(I_{t+h} = j \mid I_t = i)}{h}, \quad \forall i \neq j.$$
The transition rate matrix $\psi(t)$ captures the probabilities of transitioning between different liquidity states over infinitesimally small time intervals. The off-diagonal elements $\psi_{ij}(t)$ for $i \neq j$ represent the instantaneous rate of transitioning from state $i$ to state $j$. We assume that $I$ is stable and conservative, i.e., for all $(j,k)\in E^2$ and $t\geq 0$,
$$\psi_{jj}(t) = -\sum_{k \neq j} \psi_{jk}(t),~~\text{and}~~ \psi_{jj}(t) < +\infty.$$
\paragraph{Order Arrivals.} Consider the sequences $(\tau^{+}_k)_{k \geq 1}$ and $(\tau^{-}_k)_{k \geq 1}$ of strictly increasing positive $\mathcal{F}$-measurable stopping times, representing the arrival times of buy and sell market orders, respectively, and $(v^{+}_k)_{k \geq 1}$ and $(v^{-}_k)_{k \geq 1}$ the corresponding $\mathbb{R}_+$-valued $\mathcal{F}$-measurable order volumes. We define $n^{+}$ and $n^{-}$ as the associated counting measures on the measurable space $\left(\mathbb{R}_+\times\mathbb{R}_+,\mathcal{B}(\mathbb{R}_+)\otimes\mathcal{B}(\mathbb{R}_+)\right)$, where $\mathcal{B}(\mathbb{R}_+)$ denotes the Borel $\sigma$-field on $\mathbb{R}_+$. These measures correspond to the sets of points $\left\{(\tau_k^{+},v^{+}_k); k \geq 1\right\}$ and $\left\{(\tau_k^{-},v^{-}_k); k \geq 1\right\}$, respectively. Therefore, for all $C\in\mathcal{B}(\mathbb{R}_+)\otimes\mathcal{B}(\mathbb{R}_+)$,
\begin{align*}
  n^{+}(C) := \sum_{k\geq 1}\mathbbm{1}_C(\tau_k^{+},v^{+}_k),~~\text{and}~~ n^{-}(C) := \sum_{k\geq 1}\mathbbm{1}_C(\tau_k^{-},v^{-}_k). 
\end{align*}
    We denote by $\mathcal{F}^{N} = \left\{\mathcal{F}^{N}_t\right\}_{t\geq 0}$ the natural filtration of the observed order flow $(n^+,n^-)$, that is,
    $$
    \mathcal{F}_t^{N} := \underset{\substack{k\in\{+,-\}}}{\bigvee}\sigma\left(n^{k}((0,s]\times A); s \leq t, A \in \mathcal{B}(\mathbb{R}_+)\right),
    $$ 
    and we introduce the full-information filtration $\mathcal{F}^{N,I} = \left\{\mathcal{F}^{N,I}_t\right\}_{t\geq 0}$, enlarged by the $\mathbb{P}$-null sets $\mathcal{N}$, such that, for all $t \geq 0$, 
    $$
    \mathcal{F}^{N,I}_t := \left(\mathcal{F}^{N}_t \vee \mathcal{F}^{I}_t\right) \vee \mathcal{N}.
    $$
    We also set $N^{\pm}_t := n^{\pm}\left((0,t]\times \mathbb{R}_+\right)$, the processes counting the number of incoming buy and sell orders, respectively, up to time $t$.
\paragraph{Intensity Dynamics.} On each regime $i\in E$ we are given a volume distribution $\nu_i$ on $\mathbb{R}_+$. Throughout the paper we work under the following assumption, which is needed for the change-of-measure and filtering arguments of Section \ref{The filtering equations} and Appendix \ref{filtering_proofs}.
\begin{assump}
\label{ass_density}
For every $i\in E$, the mark distribution $\nu_i$ is absolutely continuous with respect to the Lebesgue measure, $\nu_i(\mathrm{d}v) = \nu_i(v)\,\mathrm{d}v$, with a density $v\mapsto\nu_i(v)>0$ on $\mathbb{R}_+$ (we use the same symbol for the measure and for its density).
\end{assump}
We next define, for every regime $i\in E$, the candidate intensity processes $\lambda^{i,+}$ and $\lambda^{i,-}$: they return to their baseline levels $\lambda^i_{\infty}$ at rate $\beta_i$ and are excited by the observed order arrivals through the kernels $\varphi_s^i$ (self-excitation) and $\varphi_c^i$ (cross-excitation). In other words,
\begin{equation}
\label{intensity_dynamics}
    \left\{\begin{array}{ll}        
& \mathrm{d} \lambda_t^{i,+}=-\beta_i\left(\lambda_t^{i,+}-\lambda^{i}_{\infty}\right)\, \mathrm{d} t+\int_{\mathbb{R}_+}\varphi^{i}_{\mathrm{s}}\left(\frac{v}{m_{1,i}}\right)\,n^{+}(\mathrm{d}t,\mathrm{d}v)+\int_{\mathbb{R}_+}\varphi^{i}_{\mathrm{c}}\left(\frac{v}{m_{1,i}}\right)\,n^{-}(\mathrm{d}t,\mathrm{d}v), \\
& \mathrm{d} \lambda_t^{i,-}=-\beta_i\left(\lambda_t^{i,-}-\lambda^{i}_{\infty}\right) \,\mathrm{d} t+\int_{\mathbb{R}_+} \varphi^{i}_{\mathrm{c}}\left(\frac{v}{m_{1,i}}\right)\, n^{+}(\mathrm{d}t,\mathrm{d}v)+\int_{\mathbb{R}_+}\varphi^{i}_{\mathrm{s}}\left(\frac{v}{m_{1,i}}\right)\,n^{-}(\mathrm{d}t,\mathrm{d}v),\\&\lambda_0^{i,+} = \kappa^{i,+},~~ \lambda_0^{i,-} = \kappa^{i,-},
\end{array}\right.
\end{equation}
where $\varphi^i_c,\varphi^i_s:\mathbb{R}_+\rightarrow\mathbb{R}_+$ are measurable positive functions, $(\beta_i)_{1\leq i\leq d}$ and $\lambda^i_{\infty}$ are positive reals, and $m_{1,i} := \int_{\mathbb{R}_+}v\,\nu_i(v)\,\mathrm{d} v<+\infty$, for $t \geq 0$. 

Since all the candidate intensities are driven by the same observed flow $(n^+,n^-)$, each $\lambda^{i,\pm}_t$ is an explicit functional of the order flow on $[0,t]$ and is therefore $\mathcal{F}^{N}$-adapted. This feature is crucial for the finite-dimensional filter of Section \ref{The filtering equations} (see Remark \ref{terminology_mmhp} below).

Next, we introduce the aggregated intensities
\begin{align}
\label{intensities}
  \lambda^+_t := \sum_{i =1}^d \mathbbm{1}_{\left\{I_{t^-}=i\right\}}\, \lambda^{i,+}_{t^-} ,~~\text{and}~~  \lambda^-_t := \sum_{i =1}^d \mathbbm{1}_{\left\{I_{t^-}=i\right\}}\, \lambda^{i,-}_{t^-},\quad\forall t\geq 0.
\end{align}
These intensities reflect the aggregated effect of the individual candidate intensities $\lambda^{i,+}$ and $\lambda^{i,-}$, weighted by the current liquidity state. This aligns with the conclusions drawn by \textcite{chevalier2023uncovering}, where it is shown that the order arrival intensities fluctuate with the order book's liquidity. The pair $(n^+,n^-)$ is then defined as a two-dimensional marked point process admitting the $(\mathbb{P},\mathcal{F}^{N,I})$-predictable intensity kernels
\begin{equation}
\label{kernel_def}
\lambda^{k}_t(v) := \sum_{i =1}^d \mathbbm{1}_{\left\{I_{t^-}=i\right\}}\, \lambda^{i,k}_{t^-}\, \nu_i(v),\quad \forall k\in\{+,-\},
\end{equation}
in the sense that, for every nonnegative $\mathcal{F}^{N,I}$-predictable process $H$ and $k\in\{+,-\}$,
\begin{equation*}
\mathbb{E}\bigg[\int_0^{+\infty}\int_{\mathbb{R}_+} H_t(v)\, n^k(\mathrm{d}t,\mathrm{d}v)\bigg] = \mathbb{E}\bigg[\int_0^{+\infty}\int_{\mathbb{R}_+} H_t(v)\, \lambda^{k}_t(v)\,\mathrm{d}v\,\mathrm{d}t\bigg].
\end{equation*}
Equivalently, the $\mathcal{F}^{N,I}$-dual predictable projection of $n^k(\mathrm{d}t,\mathrm{d}v)$ is $\lambda^k_t(v)\,\mathrm{d}v\,\mathrm{d}t$. In particular, $\lambda^{\pm}_t = \int_{\mathbb{R}_+}\lambda^{\pm}_t(v)\,\mathrm{d}v$ is the total arrival rate of buy (resp.\ sell) orders.
\paragraph{Construction of the model.} The system consisting of \eqref{intensity_dynamics} and \eqref{kernel_def} is constructed explicitly by thinning the Poisson random measures $M^{+}$ and $M^{-}$ carried by the stochastic basis introduced at the beginning of this section (see \textcite{massioule_bremaud}). One sets
\begin{equation}
\label{thinning_def}
n^{k}(\mathrm{d}t,\mathrm{d}v) := M^{k}\Big(\mathrm{d}t,\mathrm{d}v, \Big[0, \textstyle\sum_{i=1}^d \mathbbm{1}_{\{I_{t^-}=i\}}\lambda^{i,k}_{t^-}\nu_i(v)\Big]\Big),\quad\forall k\in\{+,-\},
\end{equation}
where the candidate intensities $\lambda^{i,\pm}$ solve \eqref{intensity_dynamics} driven by the measures $n^{\pm}$ so constructed. The modulation by the hidden chain thus acts only through the current value $I_{t^-}$, both on the selected candidate intensity and on the mark distribution $\nu_{I_{t^-}}$. The well-posedness of the scheme \eqref{thinning_def} and the structural properties of the construction used throughout the paper are collected in Proposition \ref{model_construction} below.
\begin{rque}
\label{terminology_mmhp}
In our model, the coefficients of \eqref{intensity_dynamics} are constant within each regime, and the hidden chain enters the dynamics of the observed flow only through the kernel \eqref{kernel_def}, i.e.\ through the current value $I_{t^-}$ at time $t$. This is precisely what makes a finite-dimensional filter possible. Since each candidate intensity $\lambda^{i,k}$ is $\mathcal{F}^N$-adapted, the $\mathcal{F}^N$-projection of \eqref{kernel_def} is $\sum_{i=1}^d \pi_{t^-}(i)\,\lambda^{i,k}_{t^-}\,\nu_i(v)$, a function of the $d$-dimensional filter and of observable quantities only (see Theorem \ref{ks_equation_final}). A modulation acting on the whole past of the intensity 
would not enjoy this property.
\end{rque}
Definition \eqref{kernel_def} states that, conditionally on the current regime $\{I_{t^-}=i\}$, the marks are identically and independently distributed according to the law $\nu_i$. In the terminology of Definition 6.4.III and Section 7.3 in \textcite{daley2007introduction}, for any $A \in \mathcal{B}(\mathbb{R}_+)$, $k\in\{+,-\}$ and $t \geq 0$,
$$
\lambda_t^{k}\, \nu_{I_t}(A)=\lim _{h \rightarrow 0^{+}} \frac{1}{h} \mathbb{E}\bigg[\int_t^{t+h}\int_A n^{k}(\mathrm{d}u, \mathrm{d}v) \mid \mathcal{F}^{N,I}_t\bigg],\quad \mathbb{P}\text{-a.s.},
$$
the limit being justified, under Assumptions \ref{existence_uniqueness} below, by the martingale property of the compensated measure and dominated convergence. To ensure the well-definedness of the subsequent processes and to guarantee the non-explosiveness of $n^+$ and $n^-$, we introduce the following assumptions:

\begin{assump}
\label{existence_uniqueness}
The following conditions must hold:

\begin{enumerate}
    \item[(A1)] For every $i\in E$, the first and second moments are finite:
    $$
    m_{1,i} = \int_{\mathbb{R}_+} v \, \nu_i(v)\,\mathrm{d}v < +\infty, \quad m_{2,i} := \int_{\mathbb{R}_+} v^2 \, \nu_i(v)\,\mathrm{d}v < +\infty.
    $$
    
    \item[(A2)] For all $(i,j)\in E^2$ and $k\in\{s,c\}$, the following integrals are finite:
    $$
    \int_{\mathbb{R}_+} \varphi^i_k\left(\frac{v}{m_{1,i}}\right) \, \nu_j(v)\,\mathrm{d}v < +\infty, \quad \int_{\mathbb{R}_+} (\varphi^i_k)^2\left(\frac{v}{m_{1,i}}\right) \, \nu_j(v)\,\mathrm{d}v < +\infty.
    $$
    The marks exciting the candidate intensity of regime $i$ are drawn from the distribution $\nu_j$ of the active regime $j$, whence the cross integrals.
    
    \item[(A3)] For every $i\in E$, the branching matrix
    $$\Gamma_i := \frac{1}{\beta_i}\begin{pmatrix} \displaystyle\int_{\mathbb{R}_+}\varphi^i_s\left(\tfrac{v}{m_{1,i}}\right)\nu_i(v)\,\mathrm{d}v & \displaystyle\int_{\mathbb{R}_+}\varphi^i_c\left(\tfrac{v}{m_{1,i}}\right)\nu_i(v)\,\mathrm{d}v \\[6pt] \displaystyle\int_{\mathbb{R}_+}\varphi^i_c\left(\tfrac{v}{m_{1,i}}\right)\nu_i(v)\,\mathrm{d}v & \displaystyle\int_{\mathbb{R}_+}\varphi^i_s\left(\tfrac{v}{m_{1,i}}\right)\nu_i(v)\,\mathrm{d}v \end{pmatrix}$$
    has spectral radius strictly smaller than $1$. By symmetry of $\Gamma_i$, this is equivalent to
    $$\int_{\mathbb{R}_+}\left(\varphi^i_s+\varphi^i_c\right)\left(\frac{v}{m_{1,i}}\right)\nu_i(v)\,\mathrm{d}v < \beta_i,\quad \forall i\in E.$$
    
    \item[(A4)] The decay functions $\gamma^{ik}(t) := e^{-\beta_i t} \int_{\mathbb{R}_+}\varphi^i_k\left(\frac{v}{m_{1,i}}\right) \nu_i(v)\, \mathrm{d}v$ satisfy:
    $$
    \int_0^{+\infty} t\, \gamma^{ik}(t) \, \mathrm{d}t < +\infty, \quad \forall (i,k) \in \{1,\dots,d\} \times \{s,c\},
    $$
    which holds automatically here since $\beta_i>0$ and the integrals in (A2) are finite.
\end{enumerate}
\end{assump}
We now verify that the preceding construction is well posed and collect the moment, nonexplosion and structural properties that will be used throughout the paper.
\begin{prop}
\label{model_construction}
\label{existence_intensities}
Suppose that Assumptions \ref{ass_density} and \ref{existence_uniqueness} are satisfied. Then:
\begin{enumerate}
\item The system consisting of \eqref{intensity_dynamics} and \eqref{thinning_def} admits a pathwise unique (up to indistinguishability) $\mathcal{F}$-adapted solution $\big(n^{+},n^{-},(\lambda^{i,\pm})_{i\in E}\big)$. The counting processes $N^{\pm}$ are nonexplosive and, for every $t\geq 0$,
$$\mathbb{E}\bigg[\int_0^t \lambda^{\pm}_s\,\mathrm{d}s\bigg]<+\infty,\quad \max_{i\in E}\,\mathbb{E}\bigg[\sup_{s\in[0,t]}\big(\lambda^{i,\pm}_s\big)^2\bigg]<+\infty.$$
\item The chain $I$ and the counting processes $(N^{+},N^{-})$ have no common jump times almost surely. In particular, $[\mathbbm{1}_{\{I=i\}},N^{k}]\equiv 0$ for every $i\in E$ and $k\in\{+,-\}$.
\item Setting $n^{i,k}(\mathrm{d}t,\mathrm{d}v) := M^{k}\big(\mathrm{d}t,\mathrm{d}v,\big[0,\lambda^{i,k}_{t^-}\nu_i(v)\big]\big)$ for $i\in E$ and $k\in\{+,-\}$, one has, almost surely,
$$n^{k}(\mathrm{d}t,\mathrm{d}v) \;=\; \sum_{i=1}^d \mathbbm{1}_{\{I_{t^-}=i\}}\; n^{i,k}(\mathrm{d}t,\mathrm{d}v),\quad k\in\{+,-\}.$$
\end{enumerate}
\end{prop}
\begin{proof}
    See Appendix \ref{prop: exuni}
\end{proof}
\begin{prop}
\label{compensator_immersion}
Suppose that Assumptions \ref{ass_density} and \ref{existence_uniqueness} are satisfied. Then:
\begin{enumerate}
    \item For $k\in\{+,-\}$, the $\mathcal{F}^{N,I}$-dual predictable projection of $n^k(\mathrm{d}t,\mathrm{d}v)$ is $\displaystyle\sum_{i=1}^d \mathbbm{1}_{\{I_{t^-}=i\}}\lambda^{i,k}_{t^-}\nu_i(v)\,\mathrm{d}v\,\mathrm{d}t$.
    \item Every $(\mathbb{P},\mathcal{F}^I)$-martingale remains a $(\mathbb{P},\mathcal{F}^{N,I})$-martingale. In particular the martingale part $M^I$ of $\varphi(I)$ is a $(\mathbb{P},\mathcal{F}^{N,I})$-martingale.
\end{enumerate}
\end{prop}
\begin{proof}
(1) For $t\geq 0$, set $\widetilde{\mathcal{F}}_t := \sigma\left(M^{\pm}((0,u]\times B);\, u\leq t,\ B\in\mathcal{B}(\mathbb{R}_+^2)\right)\vee \mathcal{F}^{I}_{\infty}\vee\mathcal{N}$. The acceptance indicator $\mathbbm{1}_{\{z\leq \sum_i\mathbbm{1}_{\{I_{t^-}=i\}}\lambda^{i,k}_{t^-}\nu_i(v)\}}$ is $\widetilde{\mathcal{F}}$-predictable, so, by the thinning theorem for Poisson random measures (see \ \textcite{massioule_bremaud}), the $\widetilde{\mathcal{F}}$-dual predictable projection of $n^k(\mathrm{d}t,\mathrm{d}v)$ is $\sum_i\mathbbm{1}_{\{I_{t^-}=i\}}\lambda^{i,k}_{t^-}\nu_i(v)\,\mathrm{d}v\,\mathrm{d}t$. This kernel is $\mathcal{F}^{N,I}$-predictable and $\mathcal{F}^{N,I}_t\subset\widetilde{\mathcal{F}}_t$. Hence, by the tower property of conditional expectations, the compensation identity $\mathbb{E}\big[\int\!\!\int H\, \mathrm{d}n^k\big]=\mathbb{E}\big[\int\!\!\int H\,\lambda^k_t(v)\,\mathrm{d}v\,\mathrm{d}t\big]$ transfers from $\widetilde{\mathcal{F}}$-predictable to $\mathcal{F}^{N,I}$-predictable integrands $H$, which is statement (1).

\smallskip

\noindent (2) Let $\varphi:E\to\mathbb{R}$ and $M^I_t := \varphi(I_t)-\varphi(I_0)-\int_0^t\mathcal{L}^I_s\varphi(I_s)\,\mathrm{d}s$. For $s\leq t$,
$$\mathbb{E}\big[M^I_t\mid\mathcal{F}^{N,I}_s\big] = \mathbb{E}\Big[\mathbb{E}\big[M^I_t\mid \mathcal{F}^I_s\vee\widetilde{\mathcal{F}}^M_s\big]\ \Big|\ \mathcal{F}^{N,I}_s\Big],\quad \widetilde{\mathcal{F}}^M_s := \sigma\left(M^{\pm}((0,u]\times B);\, u\leq s,\ B\in\mathcal{B}(\mathbb{R}_+^2)\right),$$
since $(M^+,M^-)$ is independent of $I$, the inner conditional expectation equals $\mathbb{E}[M^I_t\mid\mathcal{F}^I_s] = M^I_s$.
\end{proof}

\begin{rque}
    By construction, the Markov chain $I$ and the jump processes $N^+$ and $N^-$ have no common jump times, hence no mutual covariation. Allowing simultaneous jumps would require specifying their compensators
explicitly, as in \cite{Ceci_Colaneri_2012, Ceci2014}. The proofs in Appendix \ref{filtering_proofs} nevertheless cover mutual covariations for completeness.
\end{rque}
\begin{figure}[H]
    \centering
        \includegraphics[width=0.6\textwidth]{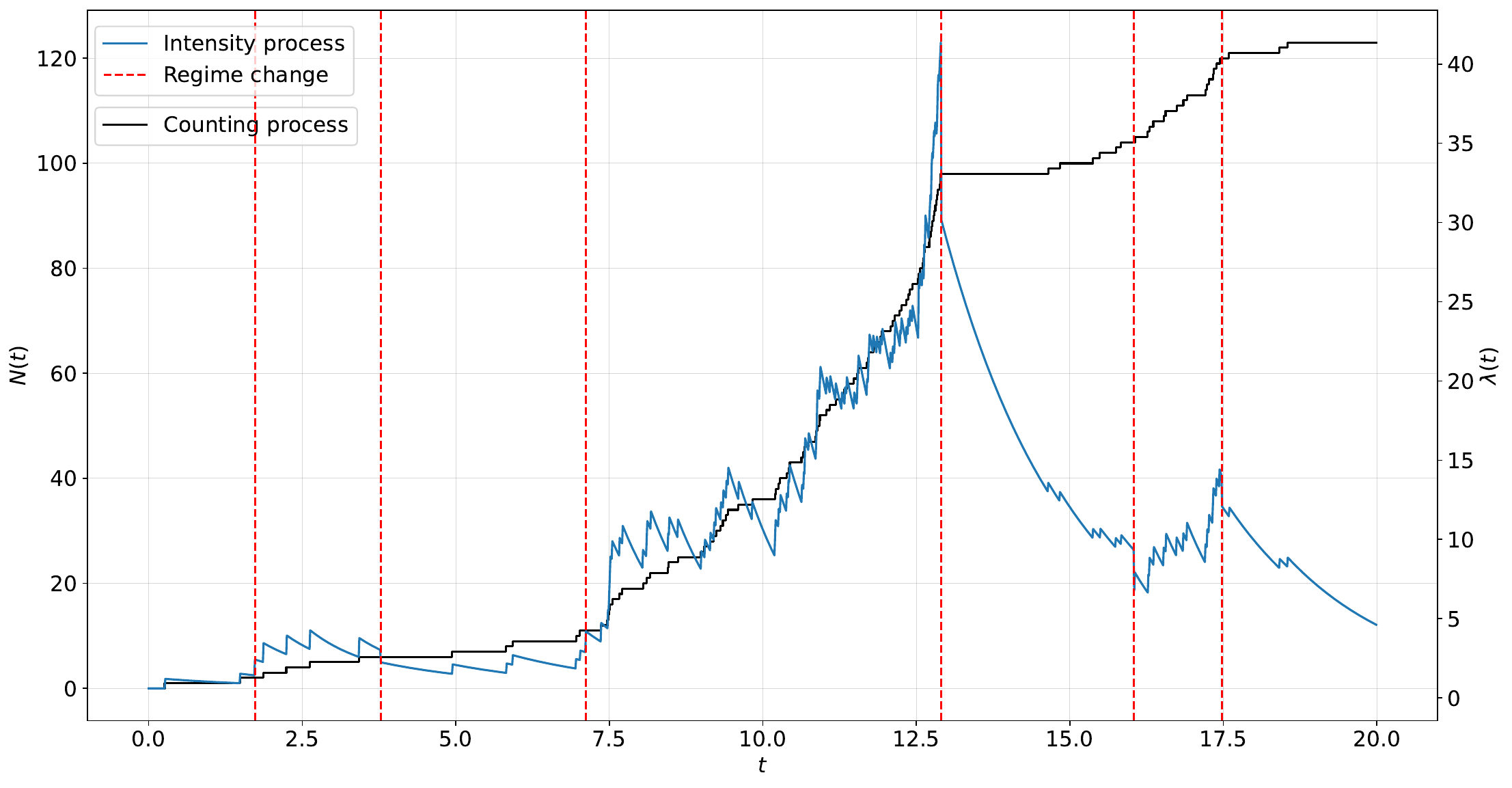}
       \caption{Representation of a Markov-Modulated Hawkes Process (MMHP) featuring an exponential Hawkes process intertwined with a two-state Markov chain. }
\end{figure} 
\subsection{Price Modelling}
\label{price_modelling}
 Empirical evidence suggests that an order's price impact is non-linear (see \textcite{power_impact}). Given our interest in short-term liquidity, we derive the market impact function by directly integrating the density of the limit order book. We choose a general "shape function" of the LOB (see \textcite{predoiu2011}, \textcite{OBIZHAEVA20131} and \textcite{general_shape} for example) that is represented by a depth density function $f:\mathbb{R}^+\rightarrow\mathbb{R}^+$. The density function measures the frequency of orders per unit price in the LOB. Throughout the paper, we assume that $f$ is strictly positive on $(0,+\infty)$ with $\int_0^{+\infty}f(x)\mathrm{d}x=+\infty$, so that $F$ is a strictly increasing bijection of $\mathbb{R}_+$ onto itself and $Q:=F^{-1}$ is globally well defined, continuous and non-decreasing with $Q(0)=0$. In this order book configuration, the quantity available at a distance $\Delta p\geq 0$ from the mid-price $p$ is expressed as $F(\Delta p) := \int_0^{\Delta p} f(x)\mathrm{d}x$. The price impact $Q:\mathbb{R}_+\rightarrow\mathbb{R}_+$ of an order executed with a size $v$ is continuous on $\mathbb{R}^*_+$ and is determined by the inverse function of $F$, denoted as $Q(v) := F^{-1}(v)$. Here, $p - Q(v)$ represents the post-trade price resulting from a large investor trading a position of $v$ shares of stock, where $p$ represents the pre-trade mid-price. Generally, $Q(0)$ equals zero, so that no trading activity leads to no impact on the price.
\begin{example}[Block-Shape]
In the work by \textcite{OBIZHAEVA20131}, the authors introduce an example of a block-shaped limit order book, where the density functions are represented as $x\mapsto f(x) = q$, with $q$ being a positive constant.
\end{example}
\begin{example}[Power-Shape]
An illustrative instance that aligns with the concave price impact and similar findings by \textcite{power_impact} and \textcite{Avellaneda_stoikoiv} is the power density function $x\mapsto f(x) = c x^{-1+e}$, where $c$ and $e$ are positive constants.
\end{example}
\begin{assump}
\label{Q_moments}
The impact function $Q$ satisfies $\int_{\mathbb{R}_+}\big(1+Q(v)\big)^2\,\nu_i(v)\,\mathrm{d}v<+\infty$ for every $i\in E$.
\end{assump}
We define the price process $P$ based on the order flow $(n^+,n^-)$ and suppose that it is $\mathcal{F}^{N}$-measurable. This means we rely exclusively on the filtration $\mathcal{F}^N$, without observing the filtration $\mathcal{F}^I$ generated by the liquidity process $I$. Consequently, the liquidity status of the traded asset might not be directly observable. As per convention, a buy order contributes positively to $P$, whereas a sell order results in a decrease in $P$. It is assumed that the price $P$ is a càdlàg process. We depict it as a composition of a fundamental price component $S$ and a price deviation process $D$, i.e.,
\begin{equation*}
    P_t := S_t + D_t, \quad \forall t\geq 0.
\end{equation*}
We propose to maintain the framework of the impact model introduced by \textcite{OBIZHAEVA20131} and to assess it under different liquidity regimes in the order book in order to account for these variations when evaluating transaction costs. We consider the following price dynamics:
\begin{equation}
\label{uncontrolled_sde}
\left\{\begin{array}{ll}
\mathrm{d} S_t  = \nu\int_{\mathbb{R}_+} Q(v) n(\mathrm{d}t,\mathrm{d}v),\\
\mathrm{d} D_t =-\rho D_{t^-} \mathrm{d}t+ (1-\nu)\int_{\mathbb{R}_+}Q(v) n(\mathrm{d}t,\mathrm{d}v), \\
S_0=s,~~ D_0 = d,
\end{array}\quad\quad\forall t\geq 0, \right.
\end{equation}
where $n := n^+ - n^-$. This means that a fraction $\nu$\footnote{The scalar impact fraction $\nu\in[0,1]$ should not be confused with the mark distributions $\nu_i(\cdot)$, $i\in E$: the latter always carry a regime index.} of the price impact persists permanently, while the complementary fraction $1-\nu$ exhibits transient behavior, decaying exponentially with a rate parameter $\rho>0$. Finally, the asset \textit{bankruptcy time} is defined as the $\mathcal{F}^N$-stopping time \begin{equation}
    \label{bankruptcy_time}\tau_S := \inf\left\{u \in [t, T] : P_u < 0\right\}\wedge T,\end{equation}
    with $\inf\{\emptyset\} = +\infty$.
 \begin{rque}
     Adapting Corollary $1$ from \textcite{delattre_hoffman}, we expect similar results to \textcite{alfonsi2016dynamic} regarding the convergence in law of $\frac{S_{mt}}{\sqrt{m}}$ to a non-standard Brownian motion with zero drift when $m$ goes to infinity. This is significant because it implies that the price dynamics are governed by a diffusion process in low-frequency asymptotic, aligning with the classical literature on the topic.
 \end{rque}
\section{The Filtering Equations}
\label{The filtering equations}
In the subsequent discussion, we define $\mathcal{F}^I$ as a sub-$\sigma$-field generated by the process $I$, where $\mathcal{F}^I_t$ is defined as $\sigma\left(I_s; s \in[0, t]\right)$ for all $t\in[0,T]$. We also consider the enlarged filtration $\mathbb{G} = \{\mathcal{G}_t\}_{t \ge 0}$ of $\mathcal{F}^N$, with $\mathcal{G}_t := \mathcal{F}_t^N\vee \mathcal{F}^I_t$ for all $t \ge 0$, so that $\mathbb{G}$ coincides with $\mathcal{F}^{N,I}$ up to $\mathbb{P}$-null sets. The filtration $\mathbb{G}$ encapsulates the information available to a hypothetical observer who, in addition to the order flow, observes the trajectory of $I$ up to the present time. Given that the liquidity status of the limit order book is not directly observable, there exists inherent information incompleteness. Consequently, market participants must estimate it using the data they have gathered. Here, we present a scenario where they must rely solely on observed order flow $\mathcal{F}^N$ to make such estimations. 
 
In essence, there are two principal approaches to address the filtering problem in this setting. The first involves solving the Zakai filtering equations. It requires a change of measure: we fix the reference probability density $\widetilde{\nu} := \tfrac{1}{d}\sum_{i\in E}\nu_i$ on $\mathbb{R}_+$, which is strictly positive under Assumptions \ref{ass_density} and satisfies $\nu_i\leq d\,\widetilde{\nu}$ for every $i\in E$, so that all the mark-likelihood ratios $\nu_i/\widetilde{\nu}$ below are bounded by $d$. 
\begin{rque}
    Any fixed positive density could serve as reference, none of the quantities of interest depending on this choice.
\end{rque}
One constructs, from $\mathbb{P}$, an equivalent measure $\mathbb{Q}$ under which $n^+$ and $n^-$ become independent Poisson random measures with intensity $\widetilde{\nu}(v)\,\mathrm{d}v\,\mathrm{d}t$, independent of the latent process $I$ (see \textcite{Brachetta2023}). The corresponding Dol\'eans-Dade exponential process $Z = (Z_t)_{t \in [0,T]}$ is
\begin{align*}
Z_t
:=
\prod_{k\in\{+,-\}}
\exp\bigg\{& 
-\int_0^t\int_{\mathbb{R}_+}
\log\bigg(\frac{\lambda^{k}_s(v)}{\widetilde{\nu}(v)}\bigg)\,
n^k(\mathrm{d}s,\mathrm{d}v)
+
\int_0^t
\big(\lambda^{k}_s- 1 \big)\,
\mathrm{d}s
\bigg\},
\quad \mathbb{P}\text{-a.s.},
\end{align*}
with $\lambda^k_s(v)$ the intensity kernel \eqref{kernel_def} and $\lambda^k_s = \int_{\mathbb{R}_+}\lambda^k_s(v)\,\mathrm{d}v$. Assumptions \ref{ass_density} is needed for $Z$ to be well defined. The intensities being unbounded, the fact that $Z$ is a true $(\mathbb{P},\mathbb{G})$-martingale (equivalently, that $\mathbb{E}^{\mathbb{P}}[Z_T]=1$) is not automatic (see Proposition \ref{prop_Z_martingale} in Appendix \ref{filtering_proofs}). The second approach relies on deriving the normalized Kushner-Stratonovich equations, which we present next.
\begin{nota}
    To improve the clarity of our findings, we define $\widehat{Y}$ as the optional projection onto the filtration $\mathcal{F}^{N}$ of an $\mathcal{F}^{N,I}$-progressively measurable process $Y$ in what follows. 
\end{nota}
As the process $I$ is partially observable, deriving an estimate based on the available information $\mathcal{F}^N$ becomes essential to address the control problem \eqref{prob_formulation}. This estimation is defined as the projection of $I$ within the domain of stochastic processes, denoted as $L= \{Y_t \in L^2(\mathbb{P}), \forall t\geq 0 :~Y~is~\mathcal{F}^{N}-measurable\}$. We define the optional projection of $\left(\varphi\left(I_t\right)\right)_{t\geq 0}$ as
\begin{equation}
\label{filter}
\begin{split}
\pi_{\tau}(\varphi)&:=\widehat{\varphi(I_{\tau})}\, \mathbbm{1}_{\{\tau<+\infty\}}\\&=\mathbb{E}\bigg[\varphi(I_{\tau}) \,\mathbbm{1}_{\{\tau<+\infty\}} \mid \mathcal{F}^{N}_{\tau}\bigg], \quad \mathbb{P}-a.s,
\end{split}
\end{equation}
where $\tau$ is an $\mathcal{F}^{N}$-stopping time and $\varphi$ are bounded measurable functions from $\mathbb{R}$ to $\mathbb{R}$.  
\begin{rque}
    As specified in Section {\MakeUppercase{\romannumeral 2}} of \textcite{bain2008fundamentals}, it is crucial to correctly construct an $\mathcal{F}$-adapted process when dealing with a process that is not adapted to the filtration $\mathcal{F}$. Unlike discrete time scenarios, directly using the process defined by the conditional expectation $\mathbb{E}[Y_t\mid \mathcal{F}_t]$, for all $t\in[0,T]$, is not viable due to its potential lack of uniqueness in null sets dependent on $t$. This mean that the process $t\mapsto\pi_t(\varphi)$ would only be defined up to modification. Such a definition could lead to an unspecified process over a set of strictly positive measure, which is undesirable. As specified in \textcite{Rao1972OnMT}, a proper way to uniquely characterize the optional projection up to indistinguishability for a bounded measurable process is through \eqref{filter}.
\end{rque}
Now let $\left(\pi(1),\dots,\pi(d)\right)$ describe the conditional probabilities varying on the unit simplex $\mathcal{S}:=\{\left(\pi(1),\dots,\pi(d)\right)\in\mathbb{R}_+^d;\,\pi(1)\geq 0,\dots\pi(d)\geq 0,\sum_{i=1}^d\pi(i)=1\}$, such that
\begin{equation}
    \label{cond_proba}
    \pi_{\tau}(i) = \mathbb{P}\left(\{I_{\tau} = i\}\mid\mathcal{F}^{N}_{\tau}\right),
\end{equation}
where $\tau$ is an $\mathcal{F}^{N}$-stopping time with values in $[0,T]$. Since the filter \eqref{filter} can be fully described by the conditional probability processes $(\pi_t(i))_{t\in[0,T]}$, we intend to state the filtering equations for $(\pi_t(i))_{t\in[0,T]}$ only in the following, with $i\in E$. Complete proofs of these equations are provided in the appendix for the case of a filtering problem where the observed processes and unobserved variables are marked point processes.
\begin{theo}[Kushner–Stratonovich equations]
\label{ks_equation_final}
    Let $t \in [0, T]$ and $i \in E$. The filter defined in \eqref{cond_proba} is the unique $\mathcal{F}^{N}$-adapted, càdlàg, $\mathcal{S}$-valued solution of the following system of stochastic differential equations, indexed by $i\in E$,
 {\small\begin{equation}
\begin{aligned}
\mathrm{d}\pi_t(i)
&=
\sum_{l=1}^d
\psi_{li}(t)\pi_{t^-}(l)\,\mathrm{d}t\\
&\quad+
\sum_{k\in\{+,-\}}
\int_{\mathbb{R}_+}
\pi_{t^-}(i)
\left(
\frac{
\lambda^{i,k}_{t^-}\nu_i(v)
}{
\sum_{j=1}^d
\pi_{t^-}(j)\lambda^{j,k}_{t^-}\nu_j(v)
}
-1
\right)\times
\left[
n^k(\mathrm{d}t,\mathrm{d}v)
-
\sum_{j=1}^d
\pi_{t^-}(j)\lambda^{j,k}_{t^-}\nu_j(v)
\,\mathrm{d}t\,\mathrm{d}v
\right].
\end{aligned}
\end{equation}}
    with $\pi_0 = \mu\in \mathcal{S}$. Additionally, the joint process $\left(\pi,\Bar{\lambda}^{+},\Bar{\lambda}^{-}\right)$, where $\Bar{\lambda}^{\pm} := (\lambda^{i,\pm})_{i\in E}$, is a piecewise-deterministic Markov process (PDMP).
\end{theo}
\begin{proof}
Let $i\in E$, and $(\tau^0_k)_{k \geq 0}$ be the jump times of the counting process $N^+ + N^-$. We apply Proposition \ref{ks_equation} with $(n^+,n^-)$ and $\varphi = \mathbbm{1}_{\{\cdot\, =\, i\}}$. By Proposition \ref{compensator_immersion}, the measure $n^k$, $k\in\{+,-\}$, admits the $(\mathbb{P},\mathbb{G})$-intensity kernel $\lambda^k_t(v) = \sum_{j=1}^d\mathbbm{1}_{\{I_{t^-}=j\}}\lambda^{j,k}_{t^-}\nu_j(v)$ of \eqref{kernel_def}, whose coefficients $\lambda^{j,k}_{t^-}$ are $\mathcal{F}^N$-predictable. Thus, the optional projections in Proposition \ref{ks_equation} are computed explicitly as
$$\pi_{t^-}\big(\mathbbm{1}_{\{\cdot\,=\,i\}}\,\lambda^k(v)\big) = \pi_{t^-}(i)\,\lambda^{i,k}_{t^-}\,\nu_i(v),\quad \pi_{t^-}\big(\lambda^k(v)\big) = \sum_{j=1}^d\pi_{t^-}(j)\,\lambda^{j,k}_{t^-}\,\nu_j(v).$$
Moreover, by Proposition \ref{model_construction}(2), $I$ and $(N^+,N^-)$ have almost surely no common jumps, so the covariation kernels of Proposition \ref{ks_equation} vanish. Substituting these expressions into Proposition \ref{ks_equation} yields the stated system. At each jump time $t$ of $N^k$ with mark $v$, the equation prescribes the Bayes update $\pi_t(i) = \pi_{t^-}(i)\lambda^{i,k}_{t^-}\nu_i(v)\big/\sum_{j=1}^d\pi_{t^-}(j)\lambda^{j,k}_{t^-}\nu_j(v)$, and between the jumps it reduces to the ordinary differential equation \eqref{pdmp_ode} below. Pathwise uniqueness therefore holds recursively from jump time to jump time. In general, the uniqueness of the solution is established by the results presented in \textcite{Ceci2000} and \textcite{Brachetta2023}. Moreover, this implies that the trajectories of $\left(\pi(1),\dots,\pi(d)\right)$ are guided by a system of ordinary differential equations (ODEs) during the intervals between the jump times of $N^+$ and $N^-$. In fact, based on the definition of the dynamics of the intensities \eqref{intensity_dynamics}, we have that, for all $\tau^0_k\leq t\leq u<\tau^0_{k+1}$,
    \begin{equation*}
    \begin{split}        
& \mathrm{d} \lambda_u^{i,+}=-\beta_i\left(\lambda_u^{i,+}-\lambda^{i}_{\infty}\right) \mathrm{d} u ,~~\text{and}~~\mathrm{d} \lambda_u^{i,-}=-\beta_i\left(\lambda_u^{i,-}-\lambda^{i}_{\infty}\right) \mathrm{d} u.
\end{split}
    \end{equation*}
    Hence, for all $\tau^0_k\leq t\leq u<\tau^0_{k+1}$,
    \begin{equation*}
    \begin{split}        
& \lambda_u^{i,+}=(\lambda^{i,+}_t-\lambda^{i}_{\infty})e^{-\beta_i(u-t)} + \lambda^{i}_{\infty},~~\text{and}~~\lambda_u^{i,-}=(\lambda^{i,-}_t-\lambda^{i}_{\infty}) e^{-\beta_i(u-t)} + \lambda^{i}_{\infty}. 
\end{split}
    \end{equation*}
    Additionally, we know that $\pi(i)$ is governed by the following differential equation between two jumps and that, for all $\tau^0_k\leq t\leq u<\tau^0_{k+1}$,\begin{equation} 
\label{pdmp_ode}
    \begin{split} \mathrm{d}\pi_u(i)&=\sum\limits_{\substack{l=1 }}^d\psi_{li}(u)\pi_{u^-}(l)\mathrm{d} u-\sum_{k \in\{+,-\}}\int_{\mathbb{R}_+} \pi_{u^-}(i)\bigg(\lambda^{i,k}_{u^-}\nu_i(v)-\sum_{j=1}^d\pi_{u^-}(j)\lambda^{j,k}_{u^-}\nu_j(v) \bigg)\mathrm{d}v\, \mathrm{d} u 
,
         \end{split}
    \end{equation}
with $\pi_t = \mu\in \mathcal{S}$. Consequently, between the jump times of $N^++N^-$ the triple $\left(\pi,\Bar{\lambda}^{+},\Bar{\lambda}^{-}\right)$ follows a deterministic flow, and at each jump time it is updated by a deterministic map of its left limit and of the observed mark; the joint process $\left(\pi,\Bar{\lambda}^{+},\Bar{\lambda}^{-}\right)$ is therefore characterized as a piecewise-deterministic Markov process (PDMP).
\end{proof}
\section{Optimal Liquidation Problem}
\label{The Separated Impulse Control Problem}
\subsection{Control Problem Formulation}
Our objective is to address an optimal liquidation problem within a finite time horizon $T$. Given the structure of the limit order book under consideration (see Subsection \ref{price_modelling}), the transaction price for a trade of volume $v$ is expressed as
\begin{equation}
\label{reward_per_trade}
C(p,v):=
\begin{cases}
\displaystyle
pv-\int_0^{Q(v)}yf(y)\,\mathrm{d}y-c_0,
& v>0,\\[1.2ex]
0,
& v=0,
\end{cases}
\end{equation}
with $(p,v)\in \mathbb{R}_+\times \mathbb{R}_+$, and $c_0>0$ a fixed constant. Since each transaction incurs the fixed cost $c_0$, continuous trading would generate infinite costs and the agent trades through a discrete, albeit continuously timed, sequence of orders. Impulse controls are therefore the natural formulation. The fixed cost also deters small-sized orders by enforcing a minimum profitable order size.
\begin{defi}[Admissible strategies]
    Define $\mathcal{A}_t(x)$ as the set of admissible controls for an agent with a position $x\in\mathbb{R}_+$ at time $t\in[0,T]$. An admissible liquidation strategy $\alpha\in\mathcal{A}_t(x)$ consists of a sequence $\alpha =\left(\tau_k, \xi_k\right)_{k \geq 1}$, where $\left(\tau_k\right)_{k \geq 1}$ is an increasing sequence of $\mathcal{F}^N-$stopping times and $(\xi_k)_{k\geq1}$ a sequence of nonnegative
$\mathcal{F}^N_{\tau_k}$-measurable random variables such that $\underset{t\leq \tau_k\leq T}{\sum}\xi_k = x$. The remaining quantity $x-\sum_{t\leq \tau_k<\tau_S}\xi_k$ being liquidated at the terminal time $\tau_S$, in accordance with \eqref{prob_formulation}. Interventions occurring after $\tau_S$ generate no revenue and are therefore irrelevant.
\label{admissible_strat}
\end{defi}
The process $X^{\alpha}$ denotes the remaining quantity to be
executed and is $\mathcal{F}^N$-progressively measurable. We define the controlled dynamics of the processes $S^{\alpha}$, $D^{\alpha}$, and $X^{\alpha}$ as 
\begin{equation}\label{price_dynamics}
\left\{
\begin{array}{ll}
\mathrm{d} S^{\alpha}_u  =\nu\int_{\mathbb{R}_+} Q(v) n(\mathrm{d}u,\mathrm{d}v) , \text { for } \tau_k < u\leq\tau_{k+1} \\
S^{\alpha}_{\tau^+_k} = S^{\alpha}_{\tau_k}- \nu Q(\xi_k),\\
\mathrm{d} D^{\alpha}_u =-\rho D^{\alpha}_{u^-} \mathrm{d}u+ (1-\nu)\int_{\mathbb{R}_+}Q(v) n(\mathrm{d}u,\mathrm{d}v), \text { for } \tau_k < u\leq\tau_{k+1} \\
D^{\alpha}_{\tau^+_k} = D^{\alpha}_{\tau_k}- (1-\nu)Q(\xi_k),\\
X^{\alpha}_{\tau_k^+} = X^{\alpha}_{\tau_k} - \xi_k,\\
S^{\alpha}_t=s,~~ D^{\alpha}_t = d,~~ X^{\alpha}_t = x,
\end{array}
\right.
\end{equation}
with $u\in[t,\tau_S]$, $k\in\mathbb{N}^*$, $\nu\in [0, 1]$, and $\alpha = \left(\tau_k, \xi_k\right)_{k \geq 1}$ an admissible strategy (see Definition \ref{admissible_strat}). \\
By applying classical theory (see Theorem $2$ in \textcite{Bremaud}), we can readily conclude that the aforementioned system of stochastic differential equations has a unique strong solution. Note that the dynamics of $P$, $S$ and $D$ are partially observable while $P$, $S$ and $D$ themselves are observable. The fact that $P$, $S$ and $D$ are $\mathcal{F}^N$-adapted implies that there are no additional terms driving the filter \eqref{filter}. 
\begin{nota}
     We denote the domain $\mathbb{R}_+\times\mathbb{R}^2\times(\mathbb{R}_+^d)^2 \times \mathcal{S}$ by $\mathcal{D}$. The conditional expectation given $(X^{\alpha}_t,S^{\alpha}_t,D^{\alpha}_t, \Bar{\lambda}^+_t, \Bar{\lambda}^-_t, \pi_t) = (x,s,d,\kappa^{+},\kappa^{-},\mu)$ under the probability measure $\mathbb{P}$ is denoted by $$\mathbb{E}_{t,y}[.] = \mathbb{E}^{\mathbb{P}}\bigg[.\mid X^{\alpha}_t,S^{\alpha}_t,D^{\alpha}_t, \Bar{\lambda}^+_t, \Bar{\lambda}^-_t, \pi_t= x,s,d,\kappa^{+},\kappa^{-},\mu\bigg],$$
     where $y = \left(x,s,d,\kappa^{+},\kappa^{-},\mu\right)\in \mathcal{D}$, $\alpha\in\mathcal{A}(x)$, $\Bar{\lambda}^+ = \left(\lambda^{i,+}\right)_{i\in E}$, $\Bar{\lambda}^- = \left(\lambda^{i,-}\right)_{i\in E}$, and $\pi = \left(\pi(i)\right)_{i\in E}$.
     
\end{nota}
We aim to solve an optimal execution problem in feedback form from the perspective of a market participant seeking to liquidate a position $X_t = x$ within a finite time horizon on a single stock, ensuring a zero terminal inventory at time $\tau_S$. The optimization involves minimizing costs while adhering to execution constraints. We will examine $\mathcal{F}^{N}$-adapted execution strategies involving a series of discrete trades. We aim to maximize the revenue $J$ over the set of admissible strategies, such that
\begin{equation}
\label{prob_formulation}
\begin{split}
    J\left(t,y,\alpha\right)& :=\mathbb{E}_{t,y}\bigg[\sum_k \mathbbm{1}_{\{t\leq \tau_k < \tau_S\}}C\left({S}^{\alpha}_{\tau_k} + {D}^{\alpha}_{\tau_k},|\Delta X^{\alpha}_{\tau_k}|\right)+C\left({S}^{\alpha}_{\tau_S} + {D}^{\alpha}_{\tau_S},X^{\alpha}_{\tau_S}\right)\bigg].
\end{split}
\end{equation}
with $t\in[0,T]$, $y \in \mathcal{D}$, $\alpha\in \mathcal{A}(x)$, and $\tau_S$ the \textit{bankruptcy time} defined in \eqref{bankruptcy_time}. The remaining inventory $X^\alpha_{\tau_S^-}$ is required to be liquidated at time $\tau_S$, so that the position is zero immediately
after the terminal transaction. The corresponding value function $V:[0,T]\times\mathcal{D}\rightarrow\mathbb{R}$ is equal to
\begin{equation}
    \label{val_fun}
    V\left(t,y\right) := \sup_{\alpha\in\mathcal{A}_t(x)}J\left(t,y,\alpha\right).
\end{equation}
The problem at hand is equivalent to one with partial observation, as the separation principle (see \textcite{fleming1982} and \textcite{separation_partial}) enables control and filtering to be disentangled, allowing them to be handled as separate tasks.

We conclude this section with a standard result regarding the finiteness of the value function.
 \begin{prop}
 \label{finite_val_fun}
     The value function $V$ described in \eqref{val_fun} is finite.
 \end{prop}
 \begin{proof}
     Let $t\in[0,T]$, $y = \left(x,s,d,\kappa^{+},\kappa^{-},\mu\right)\in\mathcal{D}$ and $\alpha\in\mathcal{A}_t(x)$. Based on \eqref{reward_per_trade},
     \begin{equation*}
\begin{split}
    C\left(P_t,x\right) &\leq P_t \int_0^{Q(x)}  f(y)\mathrm{d}y - c_0\leq P_t x. 
    \end{split}
\end{equation*}
 Using Assumptions \ref{existence_uniqueness} and \ref{Q_moments}, we get that $\sup _{i \in E} \mathbb{E}_{t,y}\bigg[\sup_{s \in[0, t]}\left(\int_0^s Q(v) n^{i,\pm}(\mathrm{d}v,\mathrm{d}u)\right)^2\bigg]<+\infty$. This leads to $ \mathbb{E}_{t,y}\bigg[\sup_{s \in[0, t]}S^2_s\bigg]<+\infty$ and $\mathbb{E}_{t,y}\bigg[\sup_{s \in[0, t]}D^2_s\bigg]<+\infty$, for all $t\in[0,T]$. Conversely, given that $p\mapsto C(p,x)$ is non-decreasing on $\mathbb{R}$, we have that
 \begin{equation*}
     \begin{split}
         J\left(t,y,\alpha\right)& \leq \mathbb{E}_{t,y}\bigg[\sum_{\tau_k \in [t, \tau_S)}C\bigg(\sup _{s \in[t, T]} {S}_{s} + {D}_{s},\xi_{\tau_k}\bigg)+{S}_{\tau_S} + {D}_{\tau_S}\bigg] \\&\leq \mathbb{E}_{t,y}\bigg[\sup _{s \in[t, T]} \left({S}_{s} + {D}_{s}\right)\bigg(1+\sum_{\tau_k \in [t, \tau_S)}\xi_{\tau_k}\bigg)\bigg] \\&\leq  \left(1+x\right)\mathbb{E}_{t,y}\bigg[\sup _{s \in[t, T]} \left({S}_{s} + {D}_{s}\right)\bigg]<+\infty.
     \end{split}
 \end{equation*}
 Since $\alpha$ is arbitrary, we conclude that $V$ is finite. 
 \end{proof}
\subsection{Dynamic Programming Principle}
When addressing the impulse control problem, it is standard practice to start by defining the value function as the solution to its corresponding Hamilton-Jacobi-Bellman Quasi-Variational Inequality (HJBQVI) using the dynamic programming principle. The expected approach is to solve the equation
\begin{equation}
\begin{split}
\min\left\{-\mathcal{L}V,V-\mathcal{M}V\right\}&=0,~\text{on}~[0,T)\times\mathbb{R}_+\times\mathbb{R}_+^2\times(\mathbb{R}_+^d)^2 \times \mathcal{S}.
\end{split}
\end{equation}
The boundary/terminal conditions here are
\begin{equation}
\begin{split}
V\left(t,0,s,d,\kappa^{+},\kappa^{-},\mu\right) &=  0,\\V\left(T,x,s,d,\kappa^{+},\kappa^{-},\mu\right) &=  C(s+d,x) ,\\V\left(t,x,s,d,\kappa^{+},\kappa^{-},\mu\right) &=  C(s+d,x), \quad \text{if } s+d<0. 
\end{split}
\end{equation}
The partial integro-differential operator $\mathcal{L}$ is defined as
\begin{equation*}
\begin{split}
\mathcal{L}\varphi(t,x,s,d,\kappa^{+},\kappa^{-},\mu)&:= \partial_t \varphi -\rho  d\partial_{d} \varphi - \sum_{i=1}^d \beta_i\bigg[(\kappa^{i,+}- \lambda^i_{\infty})\partial_{\kappa^{i,+}}\varphi + (\kappa^{i,-}- \lambda^i_{\infty})\partial_{\kappa^{i,-}} \varphi\bigg]\\ &\quad\quad+ \sum_{i=1}^d\bigg(\sum_{j=1}^d\psi_{ji}\mu_j-\mu_i\sum_{k \in\{+,-\}} (\kappa^{i,k}-\sum_{j=1}^d\mu_j\kappa^{j,k}) \bigg) \partial_{\mu_i}\varphi+ \mathcal{I}\varphi,
\end{split}
\end{equation*}
with,
\begin{equation*}\resizebox{\textwidth}{!}{$
    \begin{aligned}   
        \mathcal{I}\varphi := &\int_{\mathbb{R}_+} \bigg[ \varphi \left( t, x, s + \nu Q(z), d + (1-\nu)Q(z),  \Big(\kappa^{i,+} + \varphi^i_s(\frac{z}{m_{1,i}})\Big)_{i\in E}, \Big(\kappa^{i,-} + \varphi^i_c(\frac{z}{m_{1,i}})\Big)_{i\in E}, (\mu_i + \Delta \mu_i)_i \right) \\
        &\quad\quad\quad  - \varphi \left( t, x,  s, d, \kappa^{+}, \kappa^{-}, \mu \right) \bigg] \sum_{j=1}^d \mu_j\,\kappa^{j,+}\, \nu_j(z)\,\mathrm{d}z\\
        +&\int_{\mathbb{R}_+} \bigg[ \varphi \left( t, x, s - \nu Q(z), d - (1-\nu)Q(z), \Big(\kappa^{i,+} + \varphi^i_c(\frac{z}{m_{1,i}})\Big)_{i\in E}, \Big(\kappa^{i,-} + \varphi^i_s(\frac{z}{m_{1,i}})\Big)_{i\in E}, (\mu_i + \Delta \mu_i)_i \right) \\
        &\quad\quad\quad - \varphi \left( t, x,  s, d, \kappa^{+}, \kappa^{-}, \mu \right) \bigg] \sum_{j=1}^d \mu_j\,\kappa^{j,-}\, \nu_j(z)\,\mathrm{d}z,
\end{aligned}$}
\end{equation*}
where the update of the filter after an order of sign $k\in\{+,-\}$ and volume $z$ is
$$\Delta \mu_i := \frac{\mu_i\,\kappa^{i,k}\,\nu_i(z)}{\sum_{j=1}^d\mu_j\,\kappa^{j,k}\,\nu_j(z)} - \mu_i,\qquad i\in E,$$ in accordance with Theorem \ref{ks_equation_final}, and the intervention operator $\mathcal{M}$ is given by
\begin{equation}
 \label{intervention}
 \begin{aligned}
    &\mathcal{M}\varphi\left(t, x, s, d, \kappa^{+}, \kappa^{-}, \mu\right) :=  
    \begin{cases}
        \sup\limits_{\xi \in a(x)}  C(s+d, \xi) + \varphi\left(\Gamma\left(t, x, s, d, \kappa^{+}, \kappa^{-}, \mu, \xi\right)\right), & \text{if } a(x) \neq \emptyset, \\
        -\infty, & \text{otherwise},
    \end{cases} \\ &~\text { with } ~ a(x)=
\begin{cases}
(0,x], & x>0,\\
\varnothing, & x=0.
\end{cases}, \\
&~\text { and } ~\Gamma(t,x,s,d,\kappa^{+},\kappa^{-},\mu,\xi)=(t,x-\xi, s- \nu Q(\xi),d-(1-\nu)Q(\xi) ,\kappa^{+},\kappa^{-},\mu) .
 \end{aligned}
\end{equation}
We call $\mathcal{C}$ the continuation region 
$$\mathcal{C} := \left\{(t,y)\in\mathcal{D}\times[0,T]:~\mathcal{M}V<V\right\},$$
and, $\mathcal{T}$ the trade region
$$\mathcal{T} := \left\{(t,y)\in\mathcal{D}\times[0,T]:~\mathcal{M}V=V\right\}.$$
The system above is stated only as a formal motivation and is not used in the sequel. The rigorous characterization of $V$ proceeds through the iterative scheme introduced below. In a non-degenerate multidimensional setting like ours, obtaining explicit solutions to the system of HJBQVIs described above is rare, as it typically occurs only in trivial cases. This arises from the necessity to solve the related HJBQVI. Moreover, the existence and uniqueness of this system often require consideration of viscosity solutions. The lack of regularity of the value function could represent an additional challenge. To circumvent these difficulties, we opt to employ a methodology similar to that utilized in \textcite{Costa1989} and \textcite{oksendal2007}, where the authors demonstrated that the value function of their control problem resolves a related optimal stopping problem, enabling them to directly characterize an optimal strategy. 

Here, we define the value function as the limit of auxiliary functions. These functions will be used to study the original value function and to develop an implementable numerical algorithm, detailed in the next section. Define the subsets $\mathcal{A}^{N}_t(x)$ within $\mathcal{A}_t(x)$, for all $N\in \mathbb{N}$, as
\begin{equation*}
\mathcal{A}^{N}_t(x):=\left\{\left(\tau_k, \xi_k\right)_{k \geq 1} \in \mathcal{A}_t(x): \tau_k=+\infty \text { a.s. for all } k \geq N+1\right\},\quad \forall(t,x)\in[0,T]\times \mathbb{R}_+.
\end{equation*}
The associated value function $V_N$, which represents the value function when the investor can make a maximum of $N$ interventions, is defined as
\begin{equation}
\label{val_fun_approx}
V_N\left(t,x,s,d,\kappa^{+},\kappa^{-},\mu\right):=\sup _{\alpha \in \mathcal{A}^{N}_t(x)} J\left(t,x,s,d,\kappa^{+},\kappa^{-},\mu,\alpha\right),
\end{equation}
with $\left(t,x,s,d,\kappa^{+},\kappa^{-},\mu,N\right)\in[0,T]\times\mathcal{D} \times\mathbb{N}$. We also introduce the set $\Theta$ of $\mathcal{F}^{N}$-stopping times in $[0,T]$, i.e., 
$$\Theta_t:=\left\{t\leq\tau\leq T:\tau\text{ is an }\mathcal{F}^{N}-\text{stopping time}\right\}.$$
\begin{prop}
\label{prop_approx_eq_val_fun}
Let $N\in \mathbb{N}^*$. For all $t\in[0,T]$ and $y = \left(x,s,d,\kappa^{+},\kappa^{-},\mu\right)\in\mathcal{D}$, we have that
\begin{equation*}
\begin{split}
V_N\left(t,y\right)=\sup _{\tau \in \Theta_t} \mathbb{E}_{t,y}&\bigg[\mathcal{M}V_{N-1}\left(\tau,x,{S}_{\tau},D_{\tau},\Bar{\lambda}^+_{\tau}, \Bar{\lambda}^-_{\tau},\pi_{\tau}\right) \bigg],
\end{split}
\end{equation*}
where $V_0\left(t,y\right) = \mathbb{E}_{t,y}\bigg[C({S}_{\tau_S}+{D}_{\tau_S},x)\bigg]$.
\end{prop}
\begin{proof}
    Refer to the Appendix \ref{app:dpp_n}.
\end{proof}
\begin{prop}[Compact convergence]
\label{convergence}
    Let $K \subset \mathcal{D}$ be a compact set. For every $\varepsilon > 0$, there exists $N_0 \in \mathbb{N}$ such that for all $N \geq N_0$,
$$\sup_{(t, y) \in [0,T]\times K} |V_N(t, y) - V(t, y)| < \varepsilon.$$
\end{prop}
\begin{proof}
   Let $N\in \mathbb{N}$. For all $t\in[0,T]$ and $y = \left(x,s,d,\kappa^{+},\kappa^{-},\mu\right)\in K$. Since $\mathcal{A}^{N}\subseteq\mathcal{A}^{N+1}_t(x)\subseteq \mathcal{A}_t(x)$ for all $N\in\mathbb{N}$, then $V_N\leq V_{N+1}\leq V$. Hence, using the monotone convergence theorem, $\lim_{N\rightarrow +\infty}V_N$ exists and $\lim_{N\rightarrow +\infty}V_N\leq V<+\infty$. Consider $\alpha = (\tau_k,\xi_k)_{k\geq 1} \in \mathcal{A}_t(x)$, where, without loss of generality, $\xi_k>0$ for every $k$ with $\tau_k<\tau_S$, interventions of size zero changing neither the state nor the reward. Let $n_{\alpha}$ be the number of interventions of $\alpha$ in $[t,\tau_S)$ and set $\bar{x}_K:=\sup\{x\in\mathbb{R}_+:y\in K\}$. We introduce the following constants
$$B_K := \bar{x}_K\sup_{(t,y)\in[0,T]\times K}\mathbb{E}_{t,y}\Big[\sup_{u\in[t,T]}\big({S}_{u}+{D}_{u}\big)^{+}\Big],$$and $$B'_K := \bar{x}_K\sup_{(t,y)\in[0,T]\times K}\mathbb{E}_{t,y}\Big[\big({S}_{\tau_S}+{D}_{\tau_S}\big)^{-}\Big]+\int_0^{Q(\bar{x}_K)}y f(y)\,\mathrm{d}y+c_0,$$
where the expectations involve the uncontrolled price, which dominates every controlled price pathwise since interventions only lower it. Both constants are finite and uniform on $[0,T]\times K$ by Proposition \ref{model_construction}(1) and Assumptions \ref{existence_uniqueness} and \ref{Q_moments}, the price going below zero at $\tau_S$ only by the overshoot of the jump arriving at that time. Since $C(p,\xi)\leq p^{+}\xi-c_0$ for $\xi>0$ and the traded quantities sum to $x\leq\bar{x}_K$,
$$J(t,y,\alpha)\ \leq\ B_K-c_0\,\mathbb{E}_{t,y}[n_{\alpha}],$$
while $C(p,x)\geq -p^{-}\bar{x}_K-\int_0^{Q(\bar{x}_K)}yf(y)\,\mathrm{d}y-c_0$ gives $V\geq V_0\geq -B'_K$ on $[0,T]\times K$. Hence any $\alpha$ with $\mathbb{E}_{t,y}[n_{\alpha}]>K_0:=(B_K+B'_K)/c_0$ satisfies $J(t,y,\alpha)<-B'_K\leq V_0(t,y)$, so that restricting the supremum defining $V$ to $\{\mathbb{E}_{t,y}[n_{\alpha}]\leq K_0\}$ leaves $V$ unchanged on $[0,T]\times K$, and we assume this bound from now on. Set $G_N:=\{n_{\alpha}\leq N-1\}$, so that $$\mathbb{P}_{t,y}(G_N^c)\leq K_0/(N-1)$$ by Markov's inequality, uniformly in $(t,y)\in[0,T]\times K$ and in $\alpha$. Recall that the bankruptcy time $\tau_S$ is defined as $\tau_S = \inf\left\{u \in [t, T] : P_u < 0\right\}\wedge T$. The price is influenced by the same exogenous order flow $(n^+, n^-)$, in addition to the impulse controls, which always drive the price down. The impulse controls in the strategy $\alpha$ are more frequent than those in the strategy $\alpha_N$. Hence, $\tau_S \leq \tau^{N}_S$. Let $\alpha_N=(\widetilde{\tau}_k,\widetilde{\xi}_k)_{1\leq k\leq N} \in \mathcal{A}^{N}_t(x)$ define the strategy
\begin{equation*}
        (\widetilde{\tau}_k,\widetilde{\xi}_k) = \left\{\begin{array}{ll}
\left(\tau_k,\xi_k\right) \quad,\quad \text { if }1\leq k \leq N- 1\text { and }\tau_k<\tau_S, \\ 
(\tau_S,x-\sum_{1\leq k \leq N-1}\xi_k\mathbbm{1}_{\{\tau_k<\tau_S\}})\quad,\quad \text { else}, 
\end{array}\right.
    \end{equation*}
where $\tau_S=\tau^{\alpha}_S$. In accordance with Definition \ref{admissible_strat}, the last pair represents the liquidation of the residual inventory in a single trade, executed no later than the bankruptcy time $\tau^{\alpha_N}_S\geq\tau^{\alpha}_S$ of the truncated strategy, so that $\alpha_N\in\mathcal{A}^{N}_t(x)$. On $G_N$, the strategies $\alpha$ and $\alpha_N$ coincide on $[t,\tau_S]$ and all the differences estimated below vanish. Using the definition \eqref{prob_formulation} of the reward function $J$, we obtain that
\begin{equation}\label{ineg_conv1}\resizebox{\textwidth}{!}{$
    \begin{aligned} 
 J\left(t,y,\alpha\right)&-J\left(t,y,\alpha_N\right) \\\leq \mathbb{E}_{t,y}\bigg[&\sum_{\tau_{N}\leq \tau_k < \tau_S} C({S}^{\alpha}_{\tau_k} + {D}^{\alpha}_{\tau_k} ,\xi_{\tau_k})+|C({S}^{\alpha}_{\tau_S} + {D}^{\alpha}_{\tau_S},X^{\alpha}_{\tau_S})-C({S}^{\alpha_N}_{\tau_S} + {D}^{\alpha_N}_{\tau_S},x-\sum_{1\leq k \leq N-1}\xi_k\mathbbm{1}_{\{\tau_k<\tau_S\}})|\bigg]. 
 \end{aligned}$}
\end{equation}
The tail satisfies $\sum_{\tau_{N}\leq \tau_k < \tau_S}\xi_k\leq \bar{x}_K\,\mathbbm{1}_{G_N^c}$, so that $\mathbb{E}_{t,y}\big[(\sum_{\tau_{N}\leq \tau_k < \tau_S}\xi_k)^2\big]\leq \bar{x}_K^2\,K_0/(N-1)$, uniformly in $(t,y)$ and $\alpha$. 
Based on the proof of Proposition \ref{finite_val_fun}, we know that $C(p,x)\leq px$, for all $(p,x)\in \mathbb{R}\times\mathbb{R}_+$. Hence, 
\begin{equation*}
    \begin{split}
        \mathbb{E}_{t,y}\bigg[\sum_{\tau_{N}\leq \tau_k < \tau_S} C({S}^{\alpha}_{\tau_k} + {D}^{\alpha}_{\tau_k} ,\xi_{\tau_k})\bigg]&\leq \mathbb{E}_{t,y}\bigg[\sup_{u\in[t,T]}\left({S}^{\alpha}_{u} + {D}^{\alpha}_{u}\right)\sum_{\tau_{N}\leq \tau_k < \tau_S} \xi_{\tau_k}\bigg]\\&\leq \sqrt{\mathbb{E}_{t,y}\bigg[\sup_{u\in[t,T]}\left({S}_{u} + {D}_{u}\right)^2\bigg]\mathbb{E}_{t,y}\bigg[(\sum_{\tau_{N}\leq \tau_k < \tau_S} \xi_{\tau_k})^2\bigg]},
    \end{split}
\end{equation*}
where the second inequality is obtained using the Cauchy-Schwarz inequality. This proves the uniform convergence of the first term to zero since $\mathbb{E}_{t,y}\bigg[\sup\limits_{u\in[t,T]}\left({S}_{u} + {D}_{u}\right)^2\bigg]$ is uniformly bounded on the compact set on $[0,T]\times K$ based on Assumptions \ref{existence_uniqueness} and \ref{Q_moments}, the price $P^{\alpha}$ being linearly dependent on the variables $(t,y)$. Therefore, for any $\varepsilon>0$ and $N$ large enough, uniformly in $(t,y)$ and $\alpha$,
\begin{equation}
\label{first_ineq_unif_conv}
    \mathbb{E}_{t,y}\bigg[\sum_{\tau_{N}\leq \tau_k < \tau_S} C({S}^{\alpha}_{\tau_k} + {D}^{\alpha}_{\tau_k} ,\xi_{\tau_k})\bigg]\leq \frac{\varepsilon}{3}.
\end{equation}
Next, we will prove the uniform convergence of the second term in inequality \eqref{ineg_conv1} to zero using the expressions of ${S}^{\alpha_N}_{\tau_S}$ and ${D}^{\alpha_N}_{\tau_S}$. Based on \eqref{price_dynamics}, 
\begin{equation*}
    \begin{split}
        {D}^{\alpha}_{\tau_S} &= e^{-\rho(\tau_S - \tau_{N-1})}{D}^{\alpha}_{\tau_{N-1}} \\
        &\quad + (1 - \nu)\sum_{\tau_N \leq \tau_k < \tau_S} e^{-\rho({\tau_S} - \tau_k)} Q(\xi_k) \\
        &\quad + (1 - \nu)\int_{{\tau_S} \wedge \tau_N}^{\tau_S} \int_{\mathbb{R}_+} e^{-\rho({\tau_S} - t)} Q(v) n(\mathrm{d}t, \mathrm{d}v).
    \end{split}
\end{equation*}
Knowing that ${D}^{\alpha}_{u \wedge \tau_{N-1}} = {D}^{\alpha_N}_{u \wedge \tau_{N-1}}$, for every $u \in [t, \tau_S]$, we obtain that $${D}^{\alpha}_{\tau_S} = {D}^{\alpha_N}_{\tau_S} + (1 - \nu)\sum_{\tau_N \leq \tau_k < \tau_S} e^{-\rho({\tau_S} - \tau_k)} Q(\xi_k).$$
 We proceed in the same way to prove that, ${S}^{\alpha}_{\tau_S} = {S}^{\alpha_N}_{\tau_S} + \nu\sum_{\tau_N \leq \tau_k < \tau_S} Q(\xi_k)$. Therefore, 
\begin{equation*}
    \begin{split}
    \mathbb{E}_{t,y}\bigg[|C({S}^{\alpha}_{\tau_S} + {D}^{\alpha}_{\tau_S},X^{\alpha}_{\tau_S})-C({S}^{\alpha_N}_{\tau_S} + {D}^{\alpha_N}_{\tau_S},X^{\alpha}_{\tau_S})|\bigg]\leq \bar{x}_K\,\mathbb{E}_{t,y}\bigg[\mathbbm{1}_{G_N^c}\big({P}^{\alpha_N}_{\tau_S}-{P}^{\alpha}_{\tau_S}\big)\bigg].
    \end{split}
\end{equation*}
Pathwise, $0\leq {P}^{\alpha_N}_{\tau_S}-{P}^{\alpha}_{\tau_S}\leq \sup_{u\in[t,T]}\left({S}_{u}+{D}_{u}\right)^{+}+Q(v_{\tau_S})+Q(\bar{x}_K)$: the strategy $\alpha_N$ only sells, so ${P}^{\alpha_N}$ lies below the uncontrolled price, and the overshoot of ${P}^{\alpha}$ below zero at $\tau_S$ is at most the size of the jump occurring at that time. This bound is free of $N$ and of $\alpha$, and its second moment is uniformly bounded on $[0,T]\times K$ by Assumptions \ref{existence_uniqueness} and \ref{Q_moments}. The Cauchy-Schwarz inequality and $\mathbb{P}_{t,y}(G_N^c)\leq K_0/(N-1)$ then give, for any $\varepsilon>0$ and $N$ large enough, uniformly in $(t,y)$ and $\alpha$,
\begin{equation}
\label{second_ineq_unif_conv}
    \begin{split}   \mathbb{E}_{t,y}\bigg[|C({S}^{\alpha}_{\tau_S} + {D}^{\alpha}_{\tau_S},X^{\alpha}_{\tau_S})-C({S}^{\alpha_N}_{\tau_S} + {D}^{\alpha_N}_{\tau_S},X^{\alpha}_{\tau_S})|\bigg]\leq \frac{\varepsilon}{3}.
    \end{split}
\end{equation}
Finally, knowing that the cost function satisfies, \begin{equation*}\resizebox{\textwidth}{!}{$
    \begin{aligned} C\Big({S}^{\alpha_N}_{\tau_S} + {D}^{\alpha_N}_{\tau_S},X^{\alpha}_{\tau_S}\Big)-C\Big({S}^{\alpha_N}_{\tau_S} + {D}^{\alpha_N}_{\tau_S},x-\sum_{1\leq k \leq N-1}\xi_k\mathbbm{1}_{\{\tau_k<\tau_S\}}\Big)= \int_{Q(x-\underset{1\leq k \leq N-1}{\sum}\xi_k\mathbbm{1}_{\{\tau_k<\tau_S\}})}^{Q(X^{\alpha}_{\tau_S})} \left({S}^{\alpha_N}_{\tau_S} + {D}^{\alpha_N}_{\tau_S} - y\right)f(y)\mathrm{d}y, \end{aligned}$}
\end{equation*}
we can apply the Cauchy-Schwarz inequality again to bound it,
\begin{equation*}
    \begin{split}   \mathbb{E}_{t,y}&\bigg[|C({S}^{\alpha_N}_{\tau_S} + {D}^{\alpha_N}_{\tau_S},X^{\alpha}_{\tau_S})-C({S}^{\alpha_N}_{\tau_S} + {D}^{\alpha_N}_{\tau_S},x-\sum_{1\leq k \leq N-1}\xi_k\mathbbm{1}_{\{\tau_k<\tau_S\}})|\bigg]\\&\quad\leq\sqrt{\mathbb{E}_{t,y}\bigg[\sup_{u\in[t,T]}\left(|{S}_{u} + {D}_{u}|+Q(\bar{x}_K)\right)^2\bigg]\mathbb{E}_{t,y}\bigg[(Q(X^{\alpha}_{\tau_S}) - Q(x-\sum_{1\leq k \leq N-1}\xi_k\mathbbm{1}_{\{\tau_k<\tau_S\}}))^2\bigg]}.
    \end{split}
\end{equation*}
On $G_N$, $x-\sum_{1\leq k \leq N-1}\xi_k\mathbbm{1}_{\{\tau_k<\tau_S\}}=X^{\alpha}_{\tau_S}$ and the difference vanishes, while $0\leq Q\big(x-\sum_{1\leq k \leq N-1}\xi_k\mathbbm{1}_{\{\tau_k<\tau_S\}}\big)-Q\big(X^{\alpha}_{\tau_S}\big)\leq Q(\bar{x}_K)$, so that the last expectation is at most $Q(\bar{x}_K)^2\,K_0/(N-1)$. Hence, for any $\varepsilon>0$ and $N$ large enough, uniformly in $(t,y)$ and $\alpha$,
\begin{equation}
    \label{third_ineq_unif_conv}
    \mathbb{E}_{t,y}\bigg[|C({S}^{\alpha_N}_{\tau_S} + {D}^{\alpha_N}_{\tau_S},X^{\alpha}_{\tau_S})-C({S}^{\alpha_N}_{\tau_S} + {D}^{\alpha_N}_{\tau_S},x-\sum_{1\leq k \leq N-1}\xi_k\mathbbm{1}_{\{\tau_k<\tau_S\}})|\bigg]\leq \frac{\varepsilon}{3}.
\end{equation}
As a result of inequalities \eqref{first_ineq_unif_conv}, \eqref{second_ineq_unif_conv}, and \eqref{third_ineq_unif_conv}, 
\begin{equation*}
    \begin{split}
        J\left(t,y,\alpha\right)-J\left(t,y,\alpha_N\right) \leq \varepsilon .
    \end{split}
\end{equation*}
Therefore, 
$$
V_N(t,y)\geq J(t,y, \alpha_N) \geq J(t,y, \alpha)-\varepsilon,
$$
  Since $\alpha$ and $\varepsilon$ are arbitrary, $V_N$ converges uniformly to $V$ on $[0,T]\times K$ which concludes the proof.
\end{proof}
Next, we present two formulations of the dynamic programming principle. The proof of each result directly follows from Proposition $3.3$ and Proposition $3.4$ in \textcite{BAYRAKTAR20091792}. We briefly indicate why the arguments of the latter reference carry over. Their proofs rely only on the strong Markov property of the underlying PDMP, on the characterization of its stopping times between jumps, and on the integrability of the reward, and use neither the boundedness of the intensities nor the compactness of the state space. In our setting, the PDMP property is provided by Theorem \ref{ks_equation_final} together with the controlled dynamics \eqref{price_dynamics}, the integrability of the reward by the estimates of Propositions \ref{model_construction} and \ref{finite_val_fun}, and the state-dependent horizon $\tau_S$ enters only through the terminal payoff, which is integrable by the same estimates.
\begin{prop}[Dynamic Programming Principle \textRoman{1}]
    Let $t\in [0,T]$ and $y = \left(x,s,d,\kappa^{+},\kappa^{-},\mu\right)\in\mathcal{D}$. The value function $V$ is the smallest solution of the dynamic programming equation
\begin{equation}
\label{prog_dyn1}
V\left(t,y\right)=\sup _{\tau \in \Theta_t} \mathbb{E}_{t,y}\bigg[\mathcal{M}V\left(\tau,x,S_{\tau},D_{\tau},\Bar{\lambda}^{+}_{\tau}, \Bar{\lambda}^{-}_{\tau},\pi_{\tau}\right) \bigg],
\end{equation}
such that, $V \geq V_0$.
\end{prop}
\begin{prop}[Dynamic Programming Principle \textRoman{2}]
\label{prog_dyn2}
    Let $N \in \mathbb{N}$, $y = (x,s,d,\kappa^{+},\kappa^{-},\mu) \in \mathcal{D}$, and $t \in [0,T]$. Let $\tau_1^+ := \inf\big\{s\geq t: N^+_t<N^+_s\big\}$, $\tau_1^-:= \inf\big\{s\geq t: N^-_t<N^-_s\big\}$, $\sigma_1 := \tau^+_1 \wedge \tau^-_1 $, and $W_0 = V_0$. Then, the value function $V$ is the pointwise limit of the sequence
    \begin{equation*}
    \begin{aligned}
        W_{N}(t,y) := \sup_{s \in [t,T]} \mathbb{E}_{t,y} \bigg[ &\mathbbm{1}_{\{s \geq \sigma_1\}}W_{N-1}(\sigma_1, x, S_{\sigma_1}, D_{\sigma_1},  \Bar{\lambda}^{+}_{\sigma_1}, \Bar{\lambda}^{-}_{\sigma_1}, \pi_{\sigma_1}) \\
        &\quad+ \mathbbm{1}_{\{s < \sigma_1\}}\mathcal{M}W_{N-1}(s, x, S_{s}, D_{s}, \Bar{\lambda}^{+}_{s}, \Bar{\lambda}^{-}_{s}, \pi_{s}) \bigg].
    \end{aligned}
    \end{equation*}
    Furthermore, $V$ is the smallest solution of the dynamic programming equation
    \begin{equation*}
    \begin{aligned}
        V(t,y) = \sup_{s \in [t,T]} \mathbb{E}_{t,y} \bigg[ &\mathbbm{1}_{\{s \geq \sigma_1\}} V(\sigma_1, x, S_{\sigma_1}, D_{\sigma_1}, \Bar{\lambda}^{+}_{\sigma_1}, \Bar{\lambda}^{-}_{\sigma_1}, \pi_{\sigma_1}) + \mathbbm{1}_{\{s < \sigma_1\}} \mathcal{M}V(s, x, S_{s}, D_{s}, \Bar{\lambda}^{+}_{s}, \Bar{\lambda}^{-}_{s}, \pi_{s}) \bigg],
    \end{aligned}
    \end{equation*}
    such that, $V \geq V_0$. 
\end{prop}
\begin{rque}
    It is important to note that in Proposition \ref{prog_dyn2}, the supremum is taken over deterministic times in $[t,T]$ rather than over the stopping times in $\Theta_t$. This distinction arises from the characterization of the stopping times for piecewise deterministic Markov processes (refer to Theorem $33$ in \textcite{Bremaud}).
\end{rque}
\section{Characterization of the Value Function}
\label{Characterization of the Value Function}
In this section, we will investigate the regularity of the value function $V$, which will pave the way for constructing an optimal strategy.
\begin{prop}
\label{increasing_vf}
    Let $t\in [0,T]$, $i\in E$ and $\left(x,s,d,\kappa^{+},\kappa^{-},\mu\right)\in\mathcal{D}$. The following results hold:
    \begin{enumerate}
        \item $s_0\mapsto V\left(t,x,s_0,d,\kappa^{+},\kappa^{-},\mu\right)$ and $d_0\mapsto V\left(t,x,s,d_0,\kappa^{+},\kappa^{-},\mu\right)$ are non-decreasing functions on $\mathbb{R}$.
         \item $\frac{\partial V}{\partial s}(t,y) = x$ on any open subset of $[0,T]\times\mathcal{D}$ on which the bankruptcy probability vanishes.
    \end{enumerate}
\end{prop}
\begin{proof}
    Refer to the Appendix \ref{app:carac_vf}.
\end{proof}
\begin{lemma}
\label{cont_expectation_bankruptcy}
Let $t_0 \in [0,T]$ and $y_0 = \left(x, s, d, \kappa^{+}, \kappa^{-}, \mu\right) \in \mathcal{D}$ with $s+d\neq0$. Then, for every $\varepsilon>0$,
$$\lim_{(t_1,y_1)\rightarrow (t_0,y_0)}\mathbb{P}\left(\left|\tau^1_S -\tau^0_S\right|>\varepsilon\right)=0,$$
and $(t,y)\mapsto\mathbb{E}_{t,y}[e^{-\rho{\tau_S}}]$ is continuous at every such point.
\end{lemma}

\begin{proof}
Let $(t_0,t_1) \in [0,T]^2$, $y_0 = \left(x_0, s_0, d_0, \kappa^{+}_0, \kappa^{-}_0, \mu_0\right) \in \mathcal{D}$, and $y_1 = \left(x_1, s_1, d_1, \kappa^{+}_1, \kappa^{-}_1, \mu_1\right) \in \mathcal{D}$. We suppose, without loss of generality, that $t_1\geq t_0$. Using the dynamics \eqref{uncontrolled_sde} of the uncontrolled price process $P$, we have that, for all $u\in[t_0,T]$,
\begin{equation*}
\begin{split}
P^1_u &- P^0_u \\&= s_1 -s_0+ d_1 e^{-\rho(u-t_1)} - d_0e^{-\rho(u-t_0)} \\&\quad+\nu \int_{t_1}^u\int_{\mathbb{R}_+}Q(v) n^1(\mathrm{d}s,\mathrm{d}v)-\nu \int_{t_0}^u\int_{\mathbb{R}_+}Q(v) n^0(\mathrm{d}s,\mathrm{d}v) \\&\quad+(1-\nu)\int_{t_1}^u\int_{\mathbb{R}_+}e^{-\rho(u-s)}Q(v) n^1(\mathrm{d}s,\mathrm{d}v) -(1-\nu)\int_{t_0}^u\int_{\mathbb{R}_+}e^{-\rho(u-s)}Q(v) n^0(\mathrm{d}s,\mathrm{d}v)\\&=\mathcal{R}(u,t_0,t_1,y_0,y_1),\quad \mathbb{P}-a.s,\end{split}
\end{equation*}
with $P^1_t = S^1_t + D^1_t = s_1+d_1$, and $P^0_t = S^0_t + D^0_t = s_0+d_0$. Standard calculations show that $\sup_{u\in[t_0,T]}\mid\mathcal{R}(u,t_0,t_1,y_0,y_1)\mid$ converges in probability to zero when $(t_1,y_1)\rightarrow (t_0,y_0)$. Fix $\varepsilon>0$. On $\{\tau^0_S<T\}$, by definition of $\tau^0_S$, there is a random time $u^*\in[\tau^0_S,(\tau^0_S+\varepsilon)\wedge T]$ with $\zeta:=-P^0_{u^*}>0$, and
$$\sup_{u\in[t_0,T]}|\mathcal{R}|<\zeta\ \Longrightarrow\ P^1_{u^*}=P^0_{u^*}+\mathcal{R}(u^*,t_0,t_1,y_0,y_1)<0\ \Longrightarrow\ \tau^1_S\leq u^*\leq\tau^0_S+\varepsilon,$$
while $\tau^1_S\leq T=\tau^0_S$ on $\{\tau^0_S=T\}$. Hence
$$\{\tau^1_S>\tau^0_S+\varepsilon\}\ \subseteq\ \{\tau^0_S<T\}\cap\Big\{\sup_{u\in[t_0,T]}|\mathcal{R}|\geq\zeta\Big\}.$$
On $\{\tau^0_S>t_0+\varepsilon\}$ set $m_\varepsilon:=\inf_{u\in[t_0,\tau^0_S-\varepsilon]}P^0_u\geq0$. Between two arrivals $P^0$ follows a monotone flow, so the infimum is attained at $t_0$, at an arrival time or its left limit, or at $\tau^0_S-\varepsilon$. Each value is strictly positive almost surely: $P^0_{t_0}=s_0+d_0>0$ on the event considered, a value $0$ along a flow segment gives $\tau^0_S\leq\tau^0_S-\varepsilon$, and a value $0$ at an arrival requires $Q(v)$ to equal the pre-jump price, an event of probability zero, the law of $Q(v)$ being diffuse by Assumptions \ref{ass_density}. Hence $m_\varepsilon>0$ almost surely,
$$\sup_{u\in[t_0,T]}|\mathcal{R}|<m_\varepsilon\ \Longrightarrow\ \inf_{u\in[t_0,\tau^0_S-\varepsilon]}P^1_u\ \geq\ m_\varepsilon-\sup_{u\in[t_0,T]}|\mathcal{R}|\ >\ 0\ \Longrightarrow\ \tau^1_S>\tau^0_S-\varepsilon,$$
while $\tau^1_S\geq t_1\geq t_0\geq\tau^0_S-\varepsilon$ on $\{\tau^0_S\leq t_0+\varepsilon\}$. Hence
$$\{\tau^1_S<\tau^0_S-\varepsilon\}\ \subseteq\ \{\tau^0_S>t_0+\varepsilon\}\cap\Big\{\sup_{u\in[t_0,T]}|\mathcal{R}|\geq m_\varepsilon\Big\}.$$
Combining the two inclusions, for every $\delta>0$,
$$\mathbb{P}\left(\left|\tau^1_S -\tau^0_S\right|>\varepsilon\right)\ \leq\ \mathbb{P}\Big(\sup_{u\in[t_0,T]}|\mathcal{R}|\geq\delta\Big)+\mathbb{P}\big(\{\tau^0_S<T\}\cap\{\zeta<\delta\}\big)+\mathbb{P}\big(\{\tau^0_S>t_0+\varepsilon\}\cap\{m_\varepsilon<\delta\}\big).$$
Letting $(t_1,y_1)\rightarrow(t_0,y_0)$ and then $\delta\rightarrow0$, the first term vanishes by convergence in probability and the last two by $\zeta>0$ and $m_\varepsilon>0$ almost surely, so
$$\lim_{(t_1,y_1)\rightarrow (t_0,y_0)}\mathbb{P}\left(\left|\tau^1_S -\tau^0_S\right|>\varepsilon\right)=0.$$
Since $u\mapsto e^{-\rho u}$ is bounded and continuous, it follows that
\begin{equation*}
\begin{split}
    \lim_{(t_1,y_1)\rightarrow (t_0,y_0)}\mathbb{E}_{t_1,y_1}[e^{-\rho \tau^1_S}] &=\mathbb{E}_{t_0,y_0}[e^{-\rho \tau^0_S}].
    \end{split}
\end{equation*}
In conclusion, $(t,y)\mapsto\mathbb{E}_{t,y}[e^{-\rho{\tau_S}}]$ is continuous at every $(t_0,y_0)$ with $s_0+d_0\neq0$.
\end{proof}
\begin{lemma}
\label{continue_u_0}
Let $t \in [0,T]$ and $y = \left(x, s, d, \kappa^{+}, \kappa^{-}, \mu\right) \in \mathcal{D}$. The value function without interventions, $V_0(t, y) = \mathbb{E}_{t,y}\bigg[C(S_{\tau_S} + D_{\tau_S}, x)\bigg]$, is continuous on $[0, T] \times \mathcal{D}^*$, where $\mathcal{D}^*:=\{y\in\mathcal{D}:s+d\neq0\}$, as well as on the closed region $\{y\in\mathcal{D}:s+d\leq 0\}$.
\end{lemma}
\begin{proof}
Let $t\in [0,T]$, and $y = \left(x,s,d,\kappa^{+},\kappa^{-},\mu\right)\in\mathbb{R}_+\times\mathbb{R}^2\times(\mathbb{R}_+^d)^2\times\mathcal{S}$.\\We know that
\begin{equation*}
    \begin{split}
        V_0\left(t,y\right)  &=  x\mathbb{E}_{t,y}[{S}_{\tau_S}+{D}_{\tau_S}] - \int_0^{Q(x)}vf(v)\mathrm{d}v -c_0 \\& = x(s+d\mathbb{E}_{t,y}[e^{-\rho(\tau_S-t)}])+x\nu\mathbb{E}_{t,y}\bigg[\int_{t}^{{\tau_S}}\int_{\mathbb{R}_+} Q(v) n(\mathrm{d}u,\mathrm{d}v)\bigg] \\&\quad+x(1-\nu)\mathbb{E}_{t,y}\bigg[\int_{t}^{{\tau_S}}\int_{\mathbb{R}_+} e^{-\rho({\tau_S}-u)}Q(v) n(\mathrm{d}u,\mathrm{d}v)\bigg] - \int_0^{Q(x)}vf(v)\mathrm{d}v -c_0 .             
    \end{split}
\end{equation*}
Hence, given that $Q$ is continuous on $\mathbb{R}_+$ and that $(t,y)\mapsto\mathbb{E}_{t,y}[e^{-\rho{\tau_S}}]$ is continuous on $[0,T]\times \mathcal{D}$, the continuity of $V_0$ on $[0, T] \times (\mathbb{R}_+^d)^2 \times \mathcal{S}$ will depend on the continuity of the function
$$(t, \kappa^+, \kappa^-, \mu) \mapsto k(t, \kappa^+, \kappa^-, \mu) = \mathbb{E}_{t,y}\bigg[\int_{t}^{\tau_S}\int_{\mathbb{R}_+} e^{-\rho(\tau_S - u)} Q(v) \, n(\mathrm{d}u, \mathrm{d}v)\bigg].$$
This is because the proof of the continuity of the mapping 
$$(t, \kappa^+, \kappa^-, \mu) \mapsto \mathbb{E}_{t,y}\bigg[\int_{t}^{\tau_S}\int_{\mathbb{R}_+} Q(v) \, n(\mathrm{d}u, \mathrm{d}v)\bigg]$$
is analogous. We define a functional operator \( I_0 \) by its action on a test function \( w \) as 
\begin{equation*}
    \begin{split}
        \label{I0_operator}
I_0 w(t,\kappa^+,\kappa^-,\mu) & =\sum_{i=1}^d\mathbb{E}_{t,y}\bigg[\int_{t}^{\sigma_1\wedge{\tau_S}}\int_{\mathbb{R}_+} \pi_u(i) e^{-\rho({\tau_S}-u)}Q(v) \left(\lambda^{i,+}_u-\lambda^{i,-}_u\right)\nu_i(\mathrm{d}v)\mathrm{d}u\bigg] \\&\quad\quad\quad\quad+ \mathbb{E}_{t,y}\bigg[\mathbbm{1}_{\{\sigma_1\leq \tau_S\}}w\left(\sigma_1,\lambda^{+}_{\sigma_1}, \lambda^{-}_{\sigma_1},\pi_{\sigma_1}\right)\bigg], 
\end{split}
\end{equation*}
with $u\in [t,T]$. Let us define $\sigma_n$ as the $n$-th jump time of $N=N^++N^-$ after time $t$. We then introduce $k_{n+1}(t, \kappa^+, \kappa^-, \mu)$ recursively as $k_{n+1}(t, \kappa^+, \kappa^-, \mu) = I_0 k_n(t, \kappa^+, \kappa^-, \mu)$, starting with $k_0(t, \kappa^+, \kappa^-, \mu) = 0$, for $n \in \mathbb{N}$. As a result, 
\begin{align*}
    k_n(t, \kappa^+, \kappa^-, \mu) &= \sum_{i=1}^d\mathbb{E}_{t,y}\bigg[\int_{t}^{\sigma_n\wedge{\tau_S}}\int_{\mathbb{R}_+} \pi_u(i) e^{-\rho({\tau_S}-u)}Q(v) \left(\lambda^{i,+}_u-\lambda^{i,-}_u\right)\nu_i(\mathrm{d}v)\mathrm{d}u\bigg]\\&= \sum_{i=1}^d\mathbb{E}_{t,y}\bigg[\int_{t}^{\sigma_n\wedge{\tau_S}}\int_{\mathbb{R}_+} e^{-\rho({\tau_S}-u)}Q(v) n(\mathrm{d}v,\mathrm{d}u)\bigg].\end{align*} Additionally, we have that, for all $n\in\mathbb{N}$,
\begin{equation*}
    \begin{split}
        |k(t,\kappa^+,\kappa^-,\mu) - k_n(t,\kappa^+,\kappa^-,\mu)|  &=\sum_{i=1}^d \mathbb{E}_{t,y}\bigg[\int_{\sigma_n \wedge \tau_S}^{\tau_S}\int_{\mathbb{R}_+} \pi_u(i) e^{-\rho(\tau_S - u)}Q(v) |\lambda^{i,+}_u - \lambda^{i,-}_u|\nu_i(\mathrm{d}v)\mathrm{d}u\bigg] \\
        &\quad \leq 2d\sup_{i \in E} \mathbb{E}_i(Q) \mathbb{E}_{t,y}\bigg[(\tau_S - \sigma_n \wedge \tau_S)\sup_{s \in [0, T]}\lambda^{i,\pm}_s\bigg] \\
        &\quad \leq 2d\sup_{i \in E} \mathbb{E}_i(Q) \sqrt{\mathbb{E}_{t,y}\bigg[\sup_{s \in [0, T]}(\lambda^{i,\pm}_s)^2\bigg]\mathbb{E}_{t,y}\bigg[(\tau_S - \sigma_n \wedge \tau_S)^2\bigg]}.
    \end{split}
\end{equation*}
with $\mathbb{E}_i(Q):=\int_{\mathbb{R}_+}Q(v)\nu_i(\mathrm{d}v)$, finite by Assumptions \ref{Q_moments}. An implication of Doob's inequality, along with the stability conditions outlined in Assumptions \ref{existence_uniqueness}, is that $\mathbb{E}[\sup_{u\in[0,T]}(\lambda^{i,\pm}_u)^2]$ is bounded. Moreover, $\tau_S-\sigma_n\wedge\tau_S\leq T\,\mathbbm{1}_{\{\sigma_n\leq T\}}$ and, by Markov's inequality together with the nonexplosion bound of Proposition \ref{model_construction}(1), $\mathbb{P}_{t,y}(\sigma_n\leq T)\leq \mathbb{E}_{t,y}[N_T]/n$, which vanishes as $n\to+\infty$, uniformly on compact sets of initial conditions. Consequently, $k_n$ converges uniformly to $k$. It remains to demonstrate that the operator $I_0$ preserves continuity, thereby establishing the continuity of $k$ and, consequently, $V_0$.\\
\textbf{If $s + d \leq 0$:}\\
The price instantaneously falls to zero, thus $\tau_S = t$ since $P$ is càdlàg in the scenario without interventions and no jumps occur at time \(t^+\). Therefore,
$$V_0(t, y) = x(s + d) - \int_0^{Q(x)} v f(v) \, \mathrm{d}v - c_0.$$
Hence, $V_0$ is continuous on $\left\{y\in\mathcal{D}:s+d\leq 0\right\}$. For the rest of the proof, we will examine the continuity of $V_0$ on $\left\{y\in\mathcal{D}:s+d> 0\right\}$. \\
\textbf{If $s \geq 0$ and $s + d > 0$}:\\
Since, before $\sigma_1$, the price follows the flow $u\mapsto s+d\,e^{-\rho(u-t)}$, which remains strictly positive when $s\geq0$ and $s+d>0$, we have $$\inf\left\{u \in [t, T] : P_u \leq 0\right\} \geq \sigma_1, \quad\mathbb{P}-a.s.$$ Hence, $\sigma_1\wedge\tau_S = \sigma_1\wedge T$, $\mathbb{P}$-a.s. Now let $\mathcal{H}^{0,k}_i$ denote the functional operator on a test function $w$, such that, for all $i\in E$ and $k\in\{+,-\}$,
\begin{equation*}
    \begin{split}
        \mathcal{H}^{0,k}_i w(u,\kappa^+,\kappa^-,\mu) &= \int_{\mathbb{R}_+}w\Big(u,\Big(\kappa^{l,+}+\mathbbm{1}_{\{k=+\}}\varphi^l_s(\tfrac{v}{m_{1,l}})+\mathbbm{1}_{\{k=-\}}\varphi^l_c(\tfrac{v}{m_{1,l}})\Big)_{l\in E}, \\&\qquad\qquad\Big(\kappa^{l,-}+\mathbbm{1}_{\{k=-\}}\varphi^l_s(\tfrac{v}{m_{1,l}})+\mathbbm{1}_{\{k=+\}}\varphi^l_c(\tfrac{v}{m_{1,l}})\Big)_{l\in E},\left(\frac{\mu_l\kappa^{l,k}\nu_l(v)}{\sum_{j=1}^d\mu_j\kappa^{j,k}\nu_j(v)}\right)_{l\in E}\Big)\,\nu_i(\mathrm{d}v),
    \end{split}
\end{equation*}
in accordance with the post-jump map of the PDMP. Utilizing the piecewise-deterministic Markov property of the processes $(\lambda^{i,+})_{i \in E}$,  $(\lambda^{i,-})_{i \in E}$, and $(\pi(i))_{i \in E}$, we can express $I_0$ as
\begin{equation*}
    \begin{split}
I_0 w(t,\kappa^+,\kappa^-,\mu) &=\sum_{i=1}^d\mathbb{E}_{t,y}\bigg[\int_{t}^{\sigma_1\wedge{T}}\int_{\mathbb{R}_+} \pi_u(i) e^{-\rho({\tau_S}-u)}Q(v) \left(\lambda^{i,+}_u-\lambda^{i,-}_u\right)\nu_i(\mathrm{d}v)\mathrm{d}u+\mathbbm{1}_{\{\sigma_1\leq T\}}w\left(\sigma_1,\lambda^{+}_{\sigma_1}, \lambda^{-}_{\sigma_1},\pi_{\sigma_1}\right)\bigg] \\&=\sum_{i=1}^d\int_{t}^{T} \pi^0_u(i) e^{\rho u} \left(\lambda^{0,i,+}_u-\lambda^{0,i,-}_u\right)\times\mathbb{E}_{t,y}\bigg[\mathbbm{1}_{\{u<\sigma_1\}}e^{-\rho{\tau_S}}\bigg]\mathrm{d}u\times\mathbb{E}_i[Q(V)] \\&\quad+ \sum_{k\{+,-\}}\sum_{i=1}^d\int_t^T\mathcal{H}_i^{0,k}w\left(u,\lambda^{0,+}_{u}, \lambda^{0,-}_{u},\pi^0_{u}\right)\mathbb{P}_{t,y}\left(I_u = i,\sigma_1 = \tau^k_1, \sigma_1\in\mathrm{d}u\right) ,
\end{split}
\end{equation*}
with $\lambda_u^{0,i,+}=(\kappa^{i,+}-\lambda^{i}_{\infty}) e^{-\beta_i(u-t)} + \lambda^{i}_{\infty}$, $\lambda_u^{0,i,-}=(\kappa^{i,-}-\lambda^{i}_{\infty}) e^{-\beta_i(u-t)} + \lambda^{i}_{\infty}$ and $\pi^0_u$ is the unique solution to the ordinary differential equation \eqref{pdmp_ode}.\\ Let us introduce the functions $m^{\pm}_i:[0,T]^2\times (\mathbb{R}_+^d)^2\times \mathcal{S}$, and $u\in[t,T]$ such that, for all $i\in E$,
\begin{equation}
\label{probas_m}
    \begin{split}
    &m^+_i(t,u,\kappa^+,\kappa^-,\mu) = \mathbb{P}_{t,y}\left(u<\sigma_1, I_u = i, \sigma_1 = \tau_1^+\right),
    \\&m^-_i(t,u,\kappa^+,\kappa^-,\mu) = \mathbb{P}_{t,y}\left(u<\sigma_1, I_u = i, \sigma_1 = \tau_1^-\right) .
    \end{split}
\end{equation}
Next, we will determine the forms of $m^{\pm}_i$ in order to demonstrate their continuity. We know that \begin{equation}
    \label{calculation_m_plus}
\begin{split}
    \mathbb{P}_{t,y}\left(u<\sigma_1, I_u = i, \tau_1^+<\tau_1^-\right) &=\mathbb{E}_{t,y}\bigg[\mathbbm{1}_{\{I_u = i\}} \cdot\mathbb{P}\left(u<\tau^+_1<\tau^-_1\mid (I_s)_{s\in[t,u]}\right)\bigg]\end{split}\end{equation}
Utilizing the definition of a non-homogeneous Poisson process (see \textcite{Snyder1991}) and recognizing that the intensities of $(\lambda^{i,+})_{i\in E}$ and $(\lambda^{i,-})_{i\in E}$ are deterministic prior to the first jump time $\sigma_1$ due to their nature as PDMPs (see Theorem \ref{ks_equation_final}), we get that
\begin{equation*}\resizebox{\textwidth}{!}{$
    \begin{aligned} \mathbb{P}\left(u<\tau^+_1<\tau^-_1\mid (I_s)_{s\in[t,u]}\right)&=\mathbb{E}\bigg[\int_u^{+\infty} \sum_{j=1}^d \mathbbm{1}_{\{I_s=j\}} \lambda_s^{0,j,+} e^{-\int_t^s \sum_{i=1}^d \mathbbm{1}_{\{I_r=i\}} (\lambda_r^{0,i,-} + \lambda_r^{0,i,+}) \mathrm{d}r} \mathrm{d}s\mid (I_s)_{s\in[t,u]}\bigg]\\&=  e^{-\int_t^u \sum_{i=1}^d \mathbbm{1}_{\{I_r=i\}} (\lambda_r^{0,i,-} + \lambda_r^{0,i,+})\mathrm{d}r}\\&~~\times\int_u^{+\infty} \sum_{j=1}^d  \lambda_s^{0,j,+} \mathbb{E}\bigg[\mathbbm{1}_{\{I_s=j\}}e^{-\int_u^s \sum_{i=1}^d \mathbbm{1}_{\{I_r=i\}} (\lambda_r^{0,i,-} + \lambda_r^{0,i,+}) \mathrm{d}r} \mathrm{d}s\mid (I_s)_{s\in[t,u]}\bigg].
     \end{aligned}$}
\end{equation*}
The transition from the first to the second line is justified by the Fubini-Tonelli theorem. As described in \textcite{Bremaud}, the dynamics of $I$ can also be understood through the Poisson measure $n^I_{ij}$, which counts the transitions from state $i$ to state $j$. This process has an $\mathcal{F}^I$-intensity given by $t\mapsto\mathbbm{1}_{\{I_{t^-} = i\}}\psi_{ij}(t)$, where $i$ and $j$ are elements of the set $E$. This means that 
$$\mathrm{d}\left(\mathbbm{1}_{\{I_u = i\}}\right) = \sum_{j\in E} \left(\mathbbm{1}_{\{I_{u^-} = j\}}-\mathbbm{1}_{\{I_{u^-} = i\}}\right) \times n^I_{ji}(\mathrm{d}u),\quad \forall u\in [0,T].$$By applying Itô's formula, we obtain that
\begin{equation*}
    \begin{split}
        \mathrm{d}L^i_u &:= \mathrm{d}\left(\mathbbm{1}_{\{I_u = i\}} e^{-\sum_{j=1}^d\int_t^u\mathbbm{1}_{\{I_s = j\}}(\lambda^{0,j,-}_s + \lambda^{0,j,+}_s)\mathrm{d}s}\right)\\&= \bigg[-\mathbbm{1}_{\{I_{u^-} = i\}}\left(\lambda^{0,i,-}_{u^-}+\lambda^{0,i,+}_{u^-}\right) \mathrm{d}u + \mathrm{d}\left(\mathbbm{1}_{\{I_u = i\}}\right)\bigg]e^{-\sum_{j=1}^d\int_t^u\mathbbm{1}_{\{I_s = j\}}(\lambda^{0,j,-}_s + \lambda^{0,j,+}_s)\mathrm{d}s}.
    \end{split}
\end{equation*}
This results in
\begin{equation*}
    \begin{split}
    \mathrm{d}\mathbb{E}_{t,y}[L^i_u] &=
    \mathbb{E}_{t,y}\bigg(\sum_{j \neq i} L^j_{u^-} \psi_{ji}(u)  + L^i_{u^-} \psi_{ii}(u) \bigg)\mathrm{d}u-\mathbb{E}_{t,y}[L^i_u]   
\left(\lambda^{0,i,-}_{u^-}+\lambda^{0,i,+}_{u^-}\right)\mathrm{d}u\\&=\sum_{j=1}^d \mathbb{E}_{t,y}[L^j_{u^-}] \psi_{ji}(u) \mathrm{d}u-\mathbb{E}_{t,y}[L^i_u]\left(\lambda^{0,i,-}_{u^-}+\lambda^{0,i,+}_{u^-}\right)\mathrm{d}u.
    \end{split}
\end{equation*}
Denote by $\Phi(a,b)$, $a\leq b$, the time-ordered exponential associated with the matrix family $A(u):=\psi(u)-\Lambda^{0,-}_{u} - \Lambda^{0,+}_{u}$, i.e.\ the unique solution of the linear matrix ODE $\partial_b\Phi(a,b)=\Phi(a,b)A(b)$ with $\Phi(a,a)=\mathrm{I}_d$. Therefore,
\begin{equation}
\label{prob_cond_convex}
    \begin{split}
        \mathbb{E}_{t,y}[L^i_u] = \left(\mu\cdot \Phi(t,u)\right)_i.
    \end{split}
\end{equation}
where $\Lambda^{0,+}$ is a $d\times d$ diagonal matrix with $(\Lambda^{0,+})_{ii} = \lambda^{0,i,+}$. Hence, we get that
\begin{equation*}
    \begin{split}
        \mathbb{P}_{t,y}\left(u<\tau^+_1<\tau^-_1\mid (I_s)_{s\in[t,u]}\right)&=  e^{-\int_t^u \sum_{i=1}^d \mathbbm{1}_{\{I_r=i\}} (\lambda_r^{0,i,-} + \lambda_r^{0,i,+})\mathrm{d}r}\int_u^{+\infty} \sum_{j=1}^d  \lambda_s^{0,j,+} \sum_{k = 1}^d\mathbbm{1}_{\{I_u=k\}}\cdot \left(\Phi(u,s)\right)_{kj}\mathrm{d}s,
        \end{split}\end{equation*}and, based on \eqref{calculation_m_plus}, that
\begin{equation*}
    \begin{split}
    \mathbb{P}_{t,y}\left(u<\sigma_1, I_u = i, \tau_1^+<\tau_1^-\right) &=\mathbb{E}_{t,y}\left(\mathbbm{1}_{\{I_u = i\}} \cdot e^{-\int_t^u \sum_{i=1}^d \mathbbm{1}_{\{I_r=i\}} (\lambda_r^{0,i,-} + \lambda_r^{0,i,+})\mathrm{d}r}\right)\int_u^{+\infty} \sum_{j=1}^d  \lambda_s^{0,j,+} \left(\Phi(u,s)\right)_{ij}\mathrm{d}s\end{split}\end{equation*}
Therefore, for all $i\in E$ and $u\in[t,T]$,
\begin{equation*}
    \begin{split}
        m^+_i(t,u,\kappa^+,\kappa^-,\mu) &= \left(\mu\cdot \Phi(t,u) \right)_i\int_u^{+\infty} \sum_{j=1}^d  \lambda_s^{0,j,+} \left(\Phi(u,s)\right)_{ij}\mathrm{d}s.
    \end{split}
\end{equation*}
Similarly, we get that
\begin{equation*}
        m^-_i(t,u,\kappa^+,\kappa^-,\mu) = \left(\mu\cdot \Phi(t,u) \right)_i \int_u^{+\infty} \sum_{j=1}^d  \lambda_s^{0,j,-} \left(\Phi(u,s)\right)_{ij}\mathrm{d}s.
\end{equation*}
Since $\lambda_u^{0,i,+}=(\kappa^{i,+}-\lambda^{i}_{\infty}) e^{-\beta_i(u-t)} + \lambda^{i}_{\infty}$ and $\lambda_u^{0,i,-}=(\kappa^{i,-}-\lambda^{i}_{\infty}) e^{-\beta_i(u-t)} + \lambda^{i}_{\infty}$, we get a continuous dependence of $\lambda^{0,i,+}$ and $\lambda^{0,i,-}$ with respect to their initial parameters, and hence of $u\mapsto\Phi(t,u)$ as well, by continuous dependence of the solutions of linear ordinary differential equations on their coefficients. Hence, $\left(t,\kappa^+,\kappa^-,\mu\right)\rightarrow m_i^+\left(t,u,\kappa^+,\kappa^-,\mu\right)$ and $\left(t,\kappa^+,\kappa^-,\mu\right)\rightarrow m_i^-\left(t,u,\kappa^+,\kappa^-,\mu\right)$ are continuous on $[0,T]\times (\mathbb{R}_+^d)^2\times \mathcal{S}$, for all $i\in E$ and $u\in [0,T]$. Additionally, we know that $(t,y)\mapsto\mathbb{E}_{t,y}\bigg[\mathbbm{1}_{\{u<\sigma_1\}}e^{-\rho{\tau_S}}\bigg]$ is continuous on $[0,T]\times \mathcal{D}$, based on the results of Lemma \ref{cont_expectation_bankruptcy} and the fact that $(t,y)\mapsto \sigma_1$ is continuous in probability on $[0,T]\times \mathcal{D}$. Hence, conditioning on the path of $I$ as in \eqref{calculation_m_plus}, the first order arrives, given $(I_s)_{s\geq t}$, at the first point of a Poisson point process with rate $\sum_{j=1}^d\mathbbm{1}_{\{I_s=j\}}(\lambda^{0,j,+}_s+\lambda^{0,j,-}_s)$, so that, by \eqref{prob_cond_convex},
\begin{equation}
\label{deriv_m}
    \begin{split}
      \mathbb{P}_{t,y}\left(I_u = i,\sigma_1 = \tau^k_1, \sigma_1\in\mathrm{d}u\right)&= \lambda^{0,i,k}_u\,\mathbb{E}_{t,y}[L^i_u]\,\mathrm{d}u\ =:\ q^k_i(t,u)\,\mathrm{d}u,
    \end{split}
\end{equation}
and $q^k_i$ is continuous by \eqref{prob_cond_convex}, we conclude that $I_0$ preserves the continuity. This concludes this part of the proof.\\
\textbf{If $s < 0$ and $s + d > 0$}:\\
The bankruptcy time in case of no prior jumps $\tau^0_S$ satisfies the equation 
$$de^{-\rho(\tau^0_S - t)} + s = 0$$
before time $T$, which implies that
$$\tau^0_S = \left(t + \frac{1}{\rho} \ln\left(\frac{-d}{s}\right)\right)\wedge T. $$
Hence, $\sigma_1\wedge \tau_S=\sigma_1\wedge \tau^0_S$. The rest of the proof is based on similar arguments as those presented in the previous case. One can just replace $T$ by $\tau_S$.
\end{proof}
\begin{prop}
\label{cont_w_n}
The approximating sequence of value functions $\left(W_N\right)_{N\in\mathbb{N}}$ is continuous on $[0,T]\times\mathcal{D}^*$, with $\mathcal{D}^*=\{y\in\mathcal{D}:s+d\neq0\}$. 
\end{prop}
\begin{proof}
Let $t\in [0,T]$, $N\in\mathbb{N}$, and $y = \left(x,s,d,\kappa^{+},\kappa^{-},\mu\right)\in\mathcal{D}$. Define a functional operator $I_1$ by its action on a test function $w$ as
\begin{equation}
    \begin{split}
        \label{I_operator}
I_1 w(t,u,y)  :=\mathbb{E}_{t,y}\bigg[&\mathbbm{1}_{\{u \geq \sigma_1\}}w\left(\sigma_1,x,S_{\sigma_1},D_{\sigma_1},\Bar{\lambda}^{+}_{\sigma_1}, \Bar{\lambda}^{-}_{\sigma_1},\pi_{\sigma_1}\right) \\&+\mathbbm{1}_{\{u<\sigma_1\}}\mathcal{M}w\left(u,x,s,e^{-\rho(u-t)}d,\lambda^{0,+}_{u}, \lambda^{0,-}_{u},\pi^0_{u}\right) \bigg],     
\end{split}
\end{equation}
with $u\in [t,T]$, $\lambda_u^{0,i,+}=(\kappa^{i,+}-\lambda^{i}_{\infty}) e^{-\beta_i(u-t)} + \lambda^{i}_{\infty}$, $\lambda_u^{0,i,-}=(\kappa^{i,-}-\lambda^{i}_{\infty}) e^{-\beta_i(u-t)} + \lambda^{i}_{\infty}$, and $\pi^0_u = \mathbb{P}_{t,y}\left(I_u = i\mid \sigma_1>u\right)$ is the unique solution to the ordinary differential equation \eqref{pdmp_ode}.\\ Now let $\mathcal{H}^{1,k}_i$ denote the functional operator on a test function $w:[0,T]\times\mathcal{D}\mapsto \mathbb{R}$, for all $i\in E$ and $k\in\{+,-\}$, such that,
\begin{equation*}
    \begin{aligned}
        \mathcal{H}^{1,k}_i w(u,y):= \int_{\mathbb{R}_+}&w\bigg(u,x,s+ \epsilon_k\nu Q(v),d + \epsilon_k (1-\nu)Q(v),  \Big(\kappa^{l,+}+\mathbbm{1}_{\{k = +\}}\varphi^l_s(\tfrac{v}{m_{1,l}})+\mathbbm{1}_{\{k = -\}}\varphi^l_c(\tfrac{v}{m_{1,l}})\Big)_{l\in E}, \\&\quad\quad\quad\quad\Big(\kappa^{l,-}+\mathbbm{1}_{\{k = -\}}\varphi^l_s(\tfrac{v}{m_{1,l}})+\mathbbm{1}_{\{k = +\}}\varphi^l_c(\tfrac{v}{m_{1,l}})\Big)_{l\in E},\left(\frac{\mu_l\kappa^{l,k}\nu_l(v)}{\sum_{j=1}^d\mu_j\kappa^{j,k}\nu_j(v)}\right)_{l\in E}\bigg)\nu_i(\mathrm{d}v),
     \end{aligned}
\end{equation*}
where $\epsilon_+:=1$ and $\epsilon_-:=-1$. By means of the strong Markov property, 
\begin{equation*}
    \begin{aligned}
I_1 w(t,u,y) & =\sum_{i=1}^d\mathbb{P}_{t,y}\left(I_u =i,\sigma_1>u\right)\mathcal{M}w\left(u,x,s,e^{-\rho(u-t)}d,\lambda^{0,+}_{u}, \lambda^{0,-}_{u},\pi^0_{u}\right)\\&\quad +\sum_{k\in\{+,-\}}\sum_{i=1}^d\int_t^u  \mathcal{H}^{1,k}_i w\left(r,x,s,e^{-\rho(r-t)}d,\lambda^{0,+}_{r}, \lambda^{0,-}_{r},\pi^0_{r}\right)\mathbb{P}_{t,y}\left(I_r = i,\sigma_1 = \tau^k_1, \sigma_1\in\mathrm{d}r\right),\end{aligned}
\end{equation*}
Using \eqref{probas_m} and \eqref{deriv_m}, we can express $I_1$ as \begin{equation*}
    \begin{aligned} I_1 w(t,u,y) &=\sum_{i=1}^d m_i(t,u,\kappa^+,\kappa^-,\mu)\mathcal{M}w\left(u, x, s, e^{-\rho(u-t)}d,\lambda^{0,+}_{u}, \lambda^{0,-}_{u},\pi^0_{u}\right)\\&\quad +\sum_{k\in\{+,-\}}\sum_{i=1}^d\int_t^u  \mathcal{H}^{1,k}_i w\left(r, x, s, e^{-\rho(r-t)}d, \lambda^{0,+}_{r}, \lambda^{0,-}_{r},\pi^0_{r}\right) \frac{\partial}{\partial u} m^k_i(t,r,\kappa^+,\kappa^-,\mu)\mathrm{d}r,
 \end{aligned}
\end{equation*}
with $m_i(t,u,\kappa^+,\kappa^-,\mu) := m_i^+(t,u,\kappa^+,\kappa^-,\mu)+m_i^-(t,u,\kappa^+,\kappa^-,\mu) = \left(\mu\cdot \Phi(t,u)\right)_i$, by \eqref{prob_cond_convex}. The lifting function $\Gamma$, defined in \eqref{intervention}, maintains continuity, $Q$ being continuous. Although $C(p,\cdot)$ is discontinuous at $0$, replacing it by its continuous extension $\widetilde{C}(p,\xi):=p\xi-\int_0^{Q(\xi)}yf(y)\mathrm{d}y-c_0$ on $[0,x]$ leaves the supremum defining $\mathcal{M}$ unchanged, $\widetilde{C}(p,0)=-c_0$ being the limit of values attained on $(0,x]$. Since $x \mapsto a(x) = [0,x]$ defines a hemicontinuous, compact-valued correspondence with $a(x) \neq \emptyset$, applied to this extension, a consequence of the Maximum theorem (see Section $3$ in \textcite{berge1963topological}) is that the operator $\mathcal{M}$ preserves continuity. 
We apply the arguments presented in Lemma \ref{continue_u_0} to prove that the first jump operator $I_1$ preserves continuity. 
Using the Maximum Theorem again, since $t \mapsto [t,T]$ also maps to a non-empty set and defines a hemicontinuous compact-valued correspondence, 
we find that the operator $\sup_{u \in [t,T]} I_1 w(t,u,y)$ associated to the test function $w$ preserves continuity. 
Therefore, given that the dynamic programming principle in Proposition \ref{prog_dyn2} states $W_{N+1}(t,y) = \sup_{u \in [t,T]} I_1 W_N(t,u,y)$
and $W_0 = V_0$ is continuous (see Lemma \ref{continue_u_0}), we conclude that $W_N$ is continuous for all $N \in \mathbb{N}$, which completes the proof.
\end{proof}
\begin{corro}
\label{cont_v}
    The value function $V$ is continuous on $[0,T]\times\mathcal{D}^*$.
\end{corro}
\begin{proof}
Based on Proposition $3.2$ in \textcite{sezer_ludo}, we have that $W_N = V_N$, for all $N\in \mathbb{N}$. Moreover, $[0,T]\times \mathcal{D}^*$, endowed with the Euclidean metric, is a locally compact space. Since $W_N$ has been proven to be continuous on $[0,T]\times \mathcal{D}^*$ in Proposition \ref{cont_w_n} and that $V_N$ converges locally uniformly to $V$ based on Proposition \ref{convergence}, we conclude that $V$ is continuous on $[0,T]\times \mathcal{D}$. On the region $\{y\in\mathcal{D}:s+d\leq0\}$, $V$ is given explicitly by the boundary condition $V(t,y)=C(s+d,x)$ and is therefore continuous there as well.
\end{proof}
We conclude this section by presenting a verification theorem, which follows from the continuity of $V$ established in Corollary \ref{cont_v}, along with Proposition $4.1$ in \textcite{BAYRAKTAR20091792}. A measurable maximizing selector exists in \eqref{intervention}. For fixed $x'>0$, the map $\xi\mapsto C(\cdot,\xi)+V(\Gamma(\cdot,\xi))$ is continuous on $(0,x']$ with limit $V-c_0<\mathcal{M}V$ as $\xi\downarrow 0$, so the supremum is attained, and the measurability of the selection follows from the Maximum theorem. Moreover, the proof of Proposition \ref{convergence} shows that any optimal strategy satisfies $\mathbb{E}_{t,y}[n_{\alpha^*}]\leq K_0<+\infty$, so that $\alpha^*$ makes almost surely finitely many interventions before $\tau_S$.
\begin{theo}[Optimal Strategy] 
\label{verfication}
    Suppose that $f$ is non-decreasing. Let $t \in [0, T]$, and $y = (x, d, s, \kappa^{+}, \kappa^{-}, \mu) \in \mathcal{D}$. A strategy $\alpha^* = (\tau_k, \xi_k)_{k \geq 1}$ is defined recursively as
    \begin{itemize}     
        \item Set $\xi_1 = 0$ and $\tau_1 = t$.
        \item For $k \geq 1$,
        $$\left\{
        \begin{array}{l}
            \tau_{k+1} = \inf \left\{ s \in [\tau_k, T] : V(s, Y^*_s) = \mathcal{M} V(s, Y^*_s) \right\},  \\
            \xi_{k+1} = \arg\max\limits_{\xi} \left\{ \mathcal{M} V(\tau_{k+1}, Y^*_{\tau_{k+1}}) = C(D^{\alpha^*}_{\tau_{k+1}} + S^{\alpha^*}_{\tau_{k+1}}, \xi) + V(\Gamma(\tau_{k+1}, Y^*_{\tau_{k+1}}, \xi)) \right\},
        \end{array}
        \right.$$
        where $Y^*_s := (X^{\alpha^*}_s, S^{\alpha^*}_s,D^{\alpha^*}_s,  \Bar{\lambda}^+_s, \Bar{\lambda}^-_s, \pi_s)$, for all $s \in [0, T]$.
    \end{itemize}
    With the convention that \( \inf \{\varnothing\} = +\infty \), the strategy $\alpha^*$ is optimal for problem \eqref{val_fun}, i.e.,
    \[
    V(t, y) = \mathbb{E}_{t, y} \bigg[\sum_{\tau_k \in [t, \tau_S)} C(S^{\alpha^*}_{\tau_k} + D^{\alpha^*}_{\tau_k}, |\Delta X^{\alpha^*}_{\tau_k}|) + C(S^{\alpha^*}_{\tau_S} + D^{\alpha^*}_{\tau_S}, X^{\alpha^*}_{\tau_S})\bigg].
    \]
\end{theo}
\section{Numerical Results}
\label{Numerical Results}
Below, we present numerical examples to demonstrate the shape of the optimal exercise and continuation regions. We will also evaluate the impact of the stochastic filtering procedure on the optimal policy and exercise regions. Now, we will specify the forms of the limit order book density and intensities used in our illustrations.
\begin{itemize}
    \item \textbf{Market impact:} Drawing from the research of \textcite{BOUCHAUD200957} and \textcite{gatheral}, we propose a price impact model in which the trading size influences the price $Q$ in a concave manner, meaning that $v\mapsto V^{-1}(v)$ forms a continuous concave function from $\mathbb{R}^+$ to $\mathbb{R}^+$. This is particularly relevant when the trade size is small, which is often the case in high-frequency trading. As empirical observations suggest that the price impact function often approximates a square-root function, we will employ a power-shaped density function $x \mapsto f(x) =  c\times x^{-1+e}$, where $c$ is a positive constant and $e\geq1$. This results in a concave market impact function $v \mapsto Q(v) = \left(\frac{ev}{c}\right)^{1/e}$ and a transaction price equal to $C(p,v) = pv - \frac{e^{\frac{e+1}{e}}}{e+1} c\left(\frac{v}{c}\right)^{\frac{e+1}{e}} -c_0 = pv - c^{\prime}v^{\frac{e+1}{e}} -c_0$, with $c^{\prime} = \frac{c}{e+1} \left(\frac{e}{c}\right)^{\frac{e+1}{e}}$.

\item \textbf{Hidden liquidity:} To ensure clarity in our results, we will focus on two possible regimes $i \in \{ 1, 2 \}$ throughout the remainder of our study. Based on the findings of \textcite{chevalier2023uncovering} and \textcite{Pomponio2011}, we select an exponential distribution for the order volumes, given by $\nu_i(v) = \zeta e^{-\zeta v}$. When the volume distribution $\nu_i$ does not depend on $i$, the marks are
independent of the hidden state. In this special case, the filtering equations with marked point–process observations reduce to those with unmarked point–process observations. We aim to illustrate the behavior of the liquidating agent when market trends reverse. This is achieved through the following dynamics of the intensities $\lambda^{i,+}_{t}$ and $\lambda^{i,-}_{t}$, defined as
$$\varphi^i_s\left(\frac{v}{m_{1,i}}\right) = \varphi^i_c\left(\frac{v}{m_{1,i}}\right) = \frac{\zeta^{\eta}}{\Gamma(1+\eta)} v^{\eta},$$
where $\zeta$ and $\eta$ are positive parameters. Additionally, we define $\kappa^{2,+} = \kappa^{1,-}$ and $\kappa^{2,-} = \kappa^{1,+}$, leading to $\lambda^{2,+} = \lambda^{1,-}$ and $\lambda^{2,-} = \lambda^{1,+}$. This allows us to reduce the dimensional complexity impacting the numerical scheme. 
\end{itemize}

We propose this set of parameters for the upcoming numerical results to ensure consistency and clarity in our analysis. Whenever a variable is not explicitly mentioned as changing within a graph, or if it is not varying across different scenarios, it takes the default values specified in this set. These default values will be maintained throughout the computations unless stated otherwise.

\vspace{0.3cm}
\begin{table}[H]
    \centering
    \begin{tabular}{|c|c|c|c|c|c|c|c|c|c|}
        \hline
        \textbf{Parameters} &\(\rho\) & \(c\) & \(e\)  & \(c_0\) & \(\nu\) & \(\lambda_{\infty}^1\) & \(\beta\) & \(\eta\) \\
        \hline
        \textbf{Values} &0.1 & 1.0 & 3.0 & 0.1 & 0.8 & 1.0 & 0.5 & 0.05 \\
        \hline
    \end{tabular}
    \caption{Default parameter values used for numerical results.}
    \label{tab:parameters}
\end{table}
\begin{rque}
     The parameters governing the Markov-modulated Hawkes processes could be estimated jointly with the transition rate matrix of the underlying hidden Markov chain as described in \textcite{wu2022}. Note that for a fixed market regime, model calibration could follow the methodology proposed in \textcite{alfonsi_calibration}.
\end{rque}
\subsection{Filter Sample Paths}
\begin{figure}[H]
    \centering
    \begin{subfigure}[b]{0.45\textwidth}
        \centering
        \includegraphics[width=\textwidth]{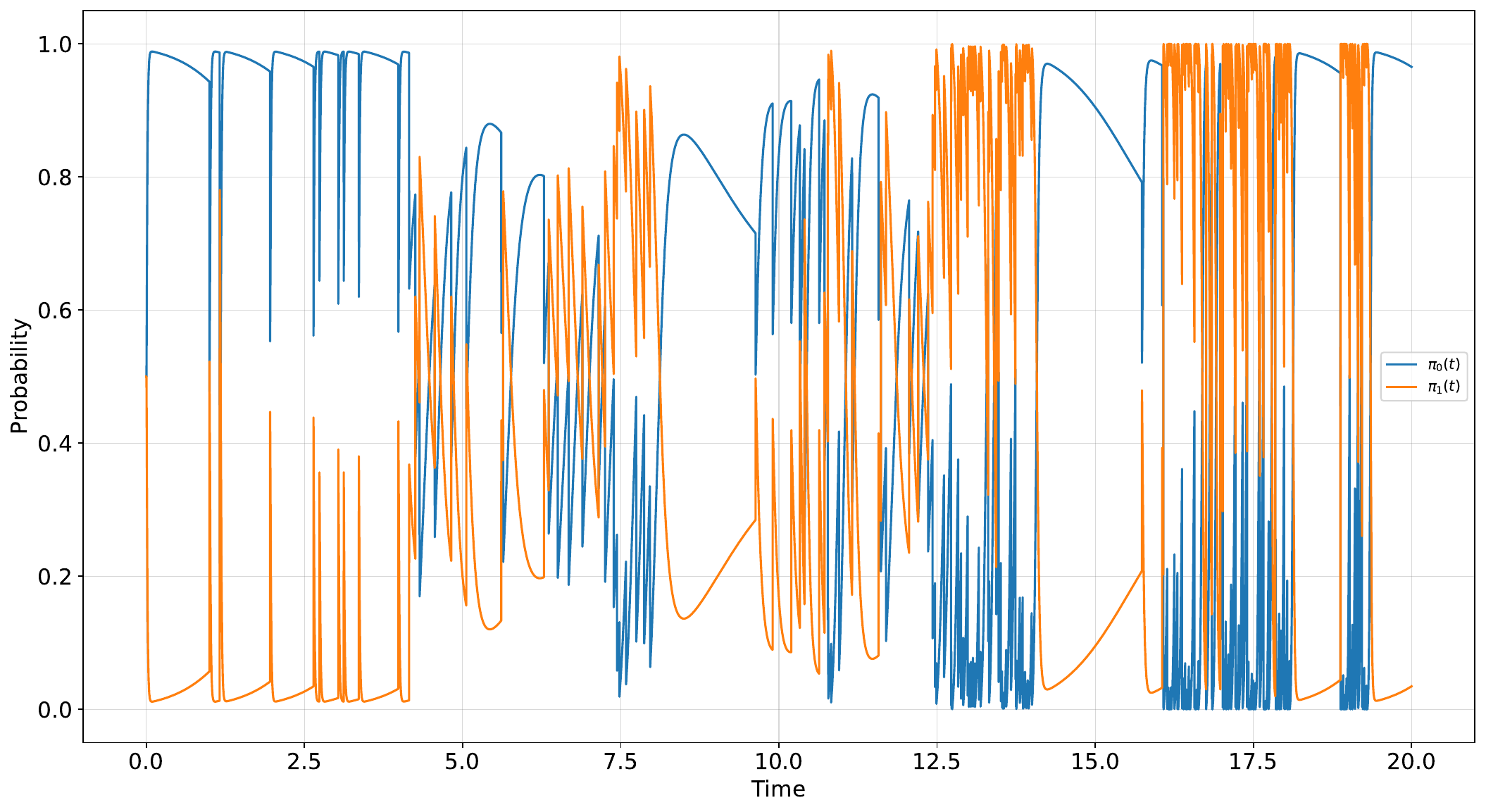}
    \end{subfigure}
    \begin{subfigure}[b]{0.46\textwidth}
        \centering
        \includegraphics[width=\textwidth]{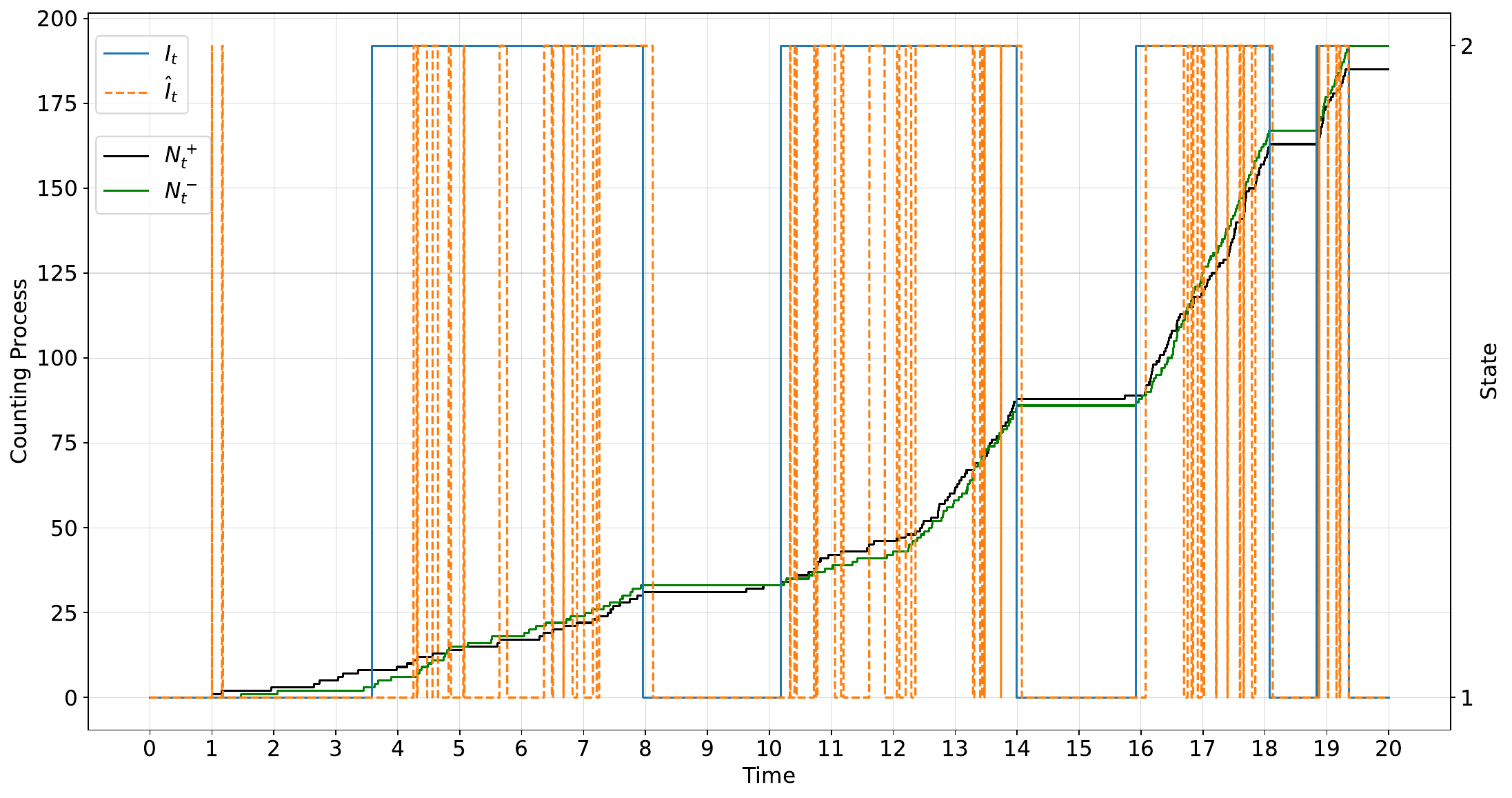}
    \end{subfigure}
       \caption{Illustration of the filter's PDMP system trajectory.}
       \label{fig:pdmp_filter}
\end{figure}

We analyze a two-state homogeneous Markov chain through sample path analysis. When fully observed, the chain exhibits a clear sequence of transitions between states, providing a detailed view of its temporal evolution. Each transition is directly observable, facilitating straightforward interpretation of the chain's dynamics. We choose the scenario where the transition rate matrix $\Psi(t)$ at time $t\in[0,T]$ is
 $$\Psi(t) = \begin{pmatrix}
-0.2 & 0.2 \\
0.2 & -0.2
\end{pmatrix}.$$ 
In the context of our Markov chain with discrete states, the accuracy of our state estimation is illustrated using the Maximum A Posteriori (MAP) estimator, which selects the state with the highest posterior probability $\hat{I}_t = \arg \max_{i\in \{0,1\}} \pi_t(i)$, where $\pi_t(i)$ refers to the filter \eqref{filter}. This approach ensures that the estimate remains within the discrete state space, providing practical and computational advantages for illustration purposes. However, going forward, we will rely on the Least Squares Estimator (LSE) for its theoretical optimality in minimizing the expected squared error, despite it yielding a weighted average that may not correspond to an actual discrete state. This ensures a more precise estimation in our subsequent analyses. These estimators serve only to visualize the latent regime, as in Figure \ref{fig:pdmp_filter}.
\subsection{Optimal Liquidation Strategy}
The numerical implementation involves discretizing the domain $\mathcal{D}$ and the time horizon $[t, T]$. Subsequently, we compute the deterministic supremum in $$W_{N+1}(t,y) = \sup_{u \in [t,T]} I_1 W_N(t,u,y),$$ for all $(t, y) \in [0, T] \times \mathcal{D}$. Note that the results presented here use $N = 20$. However, we observed that the value function typically converges after approximately $13$ iterations, in the sense that the maximal absolute difference between successive iterates over the computational grid falls below a fixed tolerance. The one-dimensional integrals in $u$ and $v$ appearing in $I_1$, through the kernels $q^k_i$ of \eqref{deriv_m} and the operators $\mathcal{H}^{1,k}_i$, are evaluated by numerical quadrature, the supremum over $u\in[t,T]$ is taken over the time grid, and off-grid values of $W_N$ are obtained by interpolation. In the following section, we will primarily discuss the exercise and continuation regions of the approximated optimal liquidation strategy. 

\begin{figure}[H]
    \centering
        \includegraphics[width=0.9\linewidth]{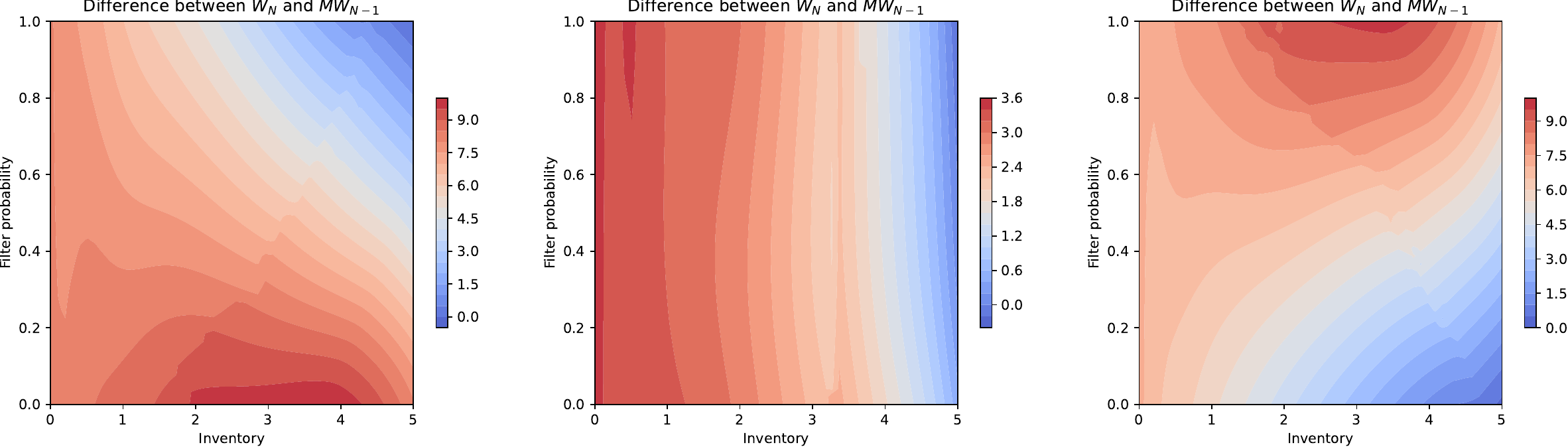}
       \caption{Contour plot of the exercise/continuation region of $W_N$ as a function of the filter's probability to be in regime $1$ and the inventory. The buy and sell intensities $(\kappa^{1,+}, \kappa^{1,-})$ are set to $(5,1)$ in the left figure, $(3, 3)$ in the central figure, and $(1,5)$ in the right figure.}
       \label{fig:exercice_filter_x_2}
\end{figure}

\paragraph{Market Risk.} Figure \ref{fig:exercice_lambdas_2} below shows the behavior of the optimal strategy with respect to the buy and sell intensities. Under the configuration introduced in the beginning of this section, the buy and sell intensities in Regime $2$ are reversed compared to Regime $1$. Specifically, $\left(\lambda^{2,+}, \lambda^{2,-}\right) = \left(\lambda^{1,-}, \lambda^{1,+}\right)$, meaning the roles of the buy and sell intensities are swapped between the two regimes.
\begin{figure}[H]
    \centering
        \includegraphics[width=0.9\linewidth]{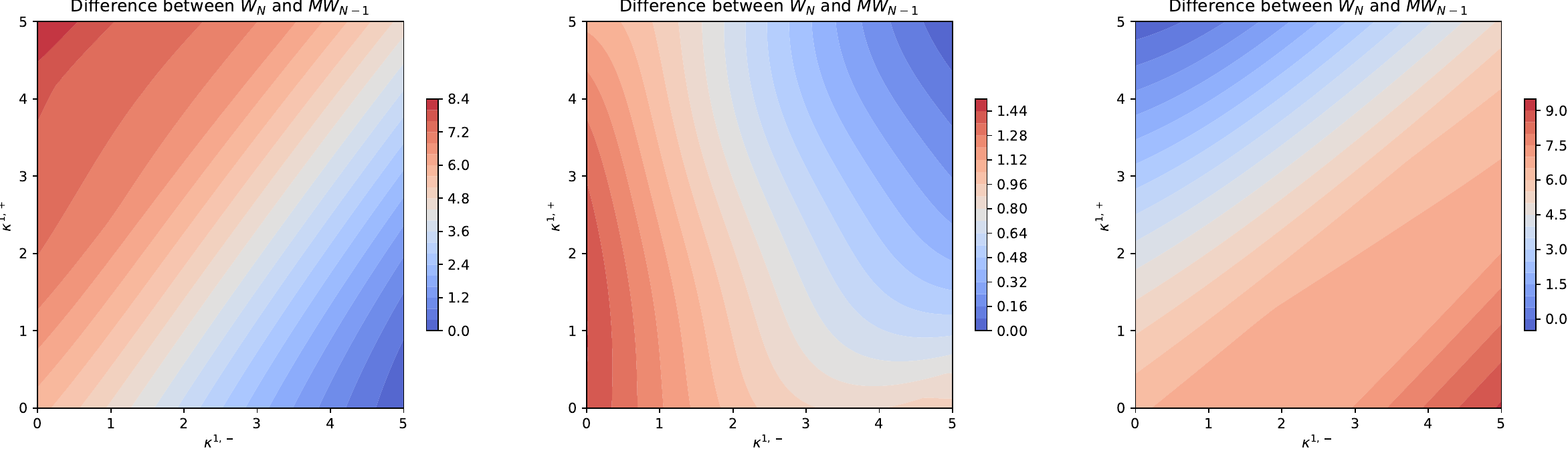}
       \caption{Contour plot of the exercise/continuation region of $W_N$ as a function of buy intensity $\kappa^{1,+}$ and sell intensity $\kappa^{1,-}$. The prior probability $\pi_t(0) = \mu_0$ is set to $0.9$ in the left figure, $0.5$ in the central figure, and $0.1$ in the right figure.}
       \label{fig:exercice_lambdas_2}
\end{figure} 
The continuation and exercise regions, as depicted in Figure \ref{fig:exercice_lambdas_2}, are influenced by the agent’s beliefs $\pi = \left(\pi(0), \pi(1)\right)$ regarding the prevailing market regime.
\begin{itemize}
    
    \item \textbf{Left-hand side figure:} Here, the probability filter is fixed at $0.9$, signaling a high likelihood of being in regime $1$. A higher $\kappa^{1,+}$ (buy intensity in regime $1$) corresponds to an upward price trend. In this case, no sell orders take place. Conversely, the higher $\kappa^{1,-}$ (sell intensity in regime $1$), the more likely the agent is to place a sell order (blue region).
    \item \textbf{Right-side figure:} On the right-hand side plot, the probability filter is fixed at $0.1$, signaling a high likelihood of being in regime $2$, where buy and sell intensities have likely switched from regime $1$ to $2$. In this case, a higher $\kappa^{1,+}$ (or lower $\kappa^{2,-}$) corresponds to a downward price trend, leading the agent to post sell orders (blue region). We observe a clear symmetry with the right-hand side plot. The optimal strategy involves liquidating when the price is going downward in regime $2$ and upward in regime $1$. This reflects the agent’s adaptive strategy to hedge against market risks and to exit positions when a downward price trend is expected.
    \item \textbf{Middle figure:} This is further confirmed by the middle plot, where the agent has no prior belief about the market state. In this case, the agent liquidates when both buy and sell intensities in regime $1$ (and consequently in regime $2$) increase. This behavior is justified by the fact that the likelihood of an imbalance in intensities is higher when both are elevated, compared to when both are low, allowing the agent to hedge against market risk.
    
\end{itemize}
We continue our analysis by examining the dependence of the approximated optimal strategy on the remaining inventory and the agent’s prior beliefs regarding the market liquidity state, as depicted in Figure \ref{fig:exercice_filter_x_2}.
Note that as the prior probability of a downward price movement increases (low $\pi_t(0)$ in the left-hand side figure), the agent’s incentive to liquidate intensifies, driven by the need to mitigate potential losses. Moreover, the agent exhibits sensitivity to inventory levels, with larger inventories prompting liquidation regardless of market liquidity conditions. This is reflected in the expansion of the selling region as the remaining inventory $x$ increases, highlighting the agent’s tendency to execute trades more aggressively when managing larger positions.
\paragraph{Liquidity Risk.} We now examine the market impact of the trading agent in relation to the trade sizes. 

\vspace{0.2cm}
\begin{figure}[H]
    \centering
        \includegraphics[width=0.37\textwidth]{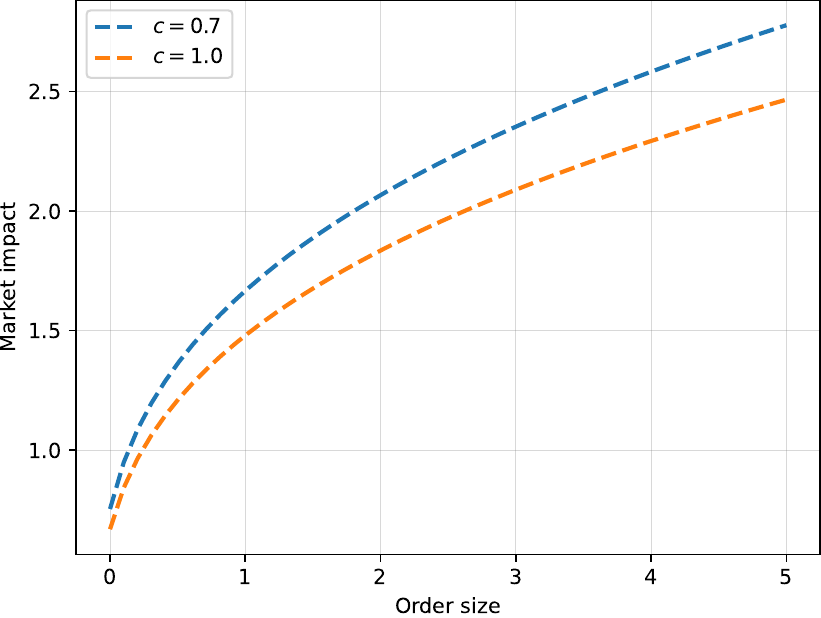}
        \caption{Representation of the market impact $Q$ as a function of the order size for two price impact parameters: $c = 0.7$ and $c = 1.0$.}
        \label{market_impact}
\end{figure}
\begin{figure}[H]
    \centering
    \begin{subfigure}[b]{0.9\textwidth}
        \centering
        \includegraphics[width=\textwidth]{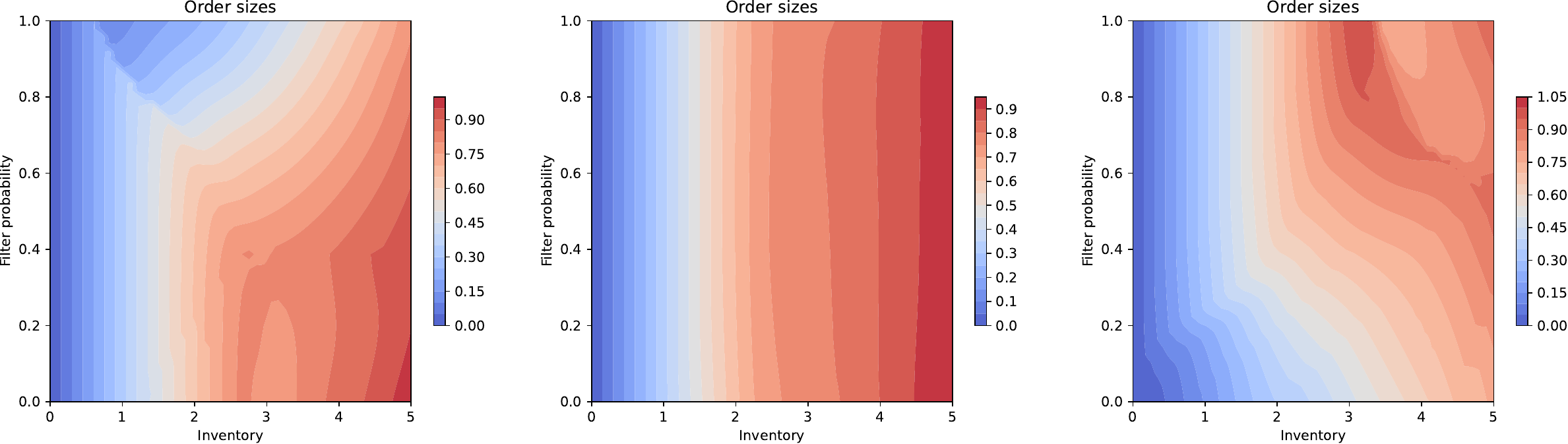}
    \end{subfigure}
    \begin{subfigure}[b]{0.9\textwidth}
        \centering
        \includegraphics[width=\textwidth]{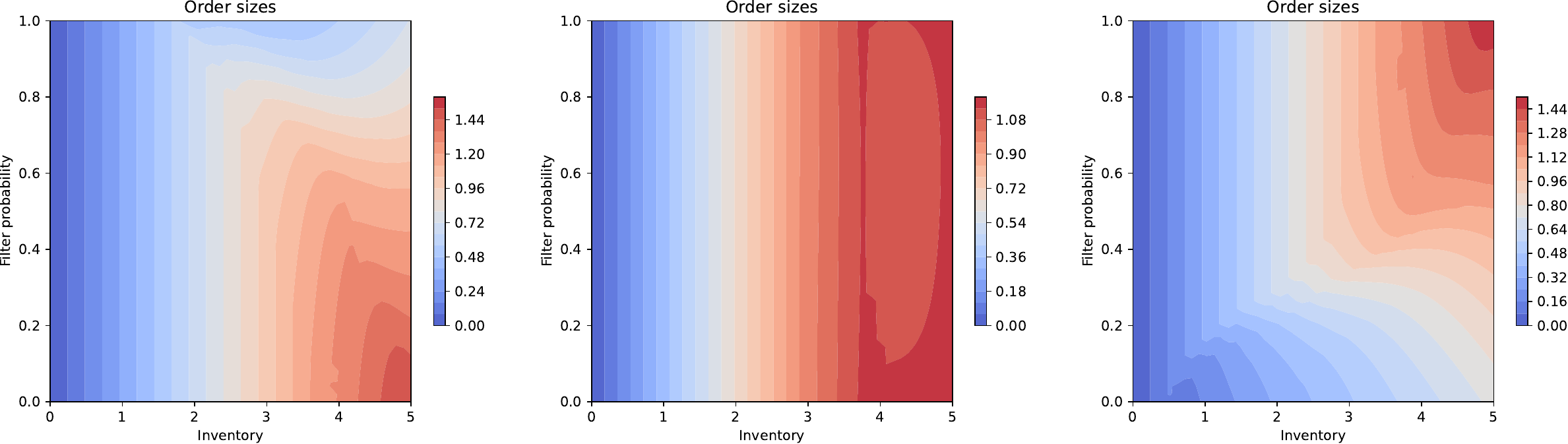}
    \end{subfigure}
       \caption{Contour plot of the order sizes as a function of the filter's probability of being in regime $1$ and the inventory. The buy and sell intensities $(\kappa^{1,+},\kappa^{1,-})$ are set to $(5,1)$ in the left figure, $(3,3)$ in the central figure, and $(1,5)$ in the right figure. The upper plots correspond to a price impact parameter $c = 0.7$, while the lower plots correspond to $c = 1$.}
       \label{fig:order_size}
\end{figure}

 
Figure \ref{market_impact} illustrates the power-law shape of the price impact function $Q$, highlighting its concave nature. In this analysis, we vary the parameter $c$ to investigate its influence on the optimal order sizes. Figure \ref{fig:order_size} illustrates the order sizes that maximize the intervention operator $\mathcal{M}$ and reveals that an increased price impact leads to a decrease in the average order size.
More interestingly, its analysis shows that the order sizes are initially proportional to the inventory, after which they stabilize. This behavior confirms the effect of the concavity of the impact function $Q$, indicating that as price sensitivity increases, the relationship between order sizes and inventory adjusts in a manner consistent with the underlying market dynamics. This effect is more pronounced when the price impact is higher, as depicted in the upper plots for $c = 0.7$.

\begin{rque}
    As shown in Proposition \ref{increasing_vf}, the dimensions of our problem reduce by one since the fundamental price variable $S$ can be separated from the other variables. Thus, our problem is equivalent to finding the exercise and continuation regions of the function $g$ introduced in the proof of the aforementioned Proposition as:
    \begin{equation*}
    \begin{split}
        V(t,y) = xs+\sup_{\alpha\in\mathcal{A}_t(x)}g(t,x,d,\kappa^+,\kappa^-,\mu,\alpha), \quad \forall (t,y)\in[0,T]\times\mathcal{D}.
    \end{split}
\end{equation*}
We recall that $g$ retains a residual dependence on $s$ through the bankruptcy time $\tau_S$ (see Proposition \ref{increasing_vf}). The dimension reduction below is therefore exact on the region where the bankruptcy probability vanishes, and the residual dependence is neglected in the numerical scheme. The corresponding intervention operator $\widetilde{\mathcal{M}}$ is defined as 
\begin{equation*}\resizebox{\textwidth}{!}{$
    \begin{aligned} 
    &\widetilde{\mathcal{M}}\varphi\left(t, x, d, \kappa^{+}, \kappa^{-}, \mu\right) :=  
    \begin{cases}
        \sup\limits_{\xi \in a(x)} C(d, \xi) -\nu Q(\xi)(x-\xi) + \varphi\left(\widetilde{\Gamma}\left(t, x, d, \kappa^{+}, \kappa^{-}, \mu, \xi\right)\right), &\text{if } a(x) \neq \{\emptyset\}, \\
        +\infty, &\text{otherwise},
    \end{cases} \\ &\text { with }  ~\widetilde{\Gamma}(t,x,d,\kappa^{+},\kappa^{-},\mu,\xi)=(t,x-\xi,d-(1-\nu)Q(\xi) ,\kappa^{+},\kappa^{-},\mu) .
\end{aligned}$}
\end{equation*}
Thus, the continuation region $\mathcal{C}$ and the trade region $\mathcal{T}$ are equivalent to
$$\mathcal{C} = \left\{(t,y)\in\mathcal{D}\times[0,T]:~\widetilde{\mathcal{M}}g<g\right\},~~\text{and}~~\mathcal{T} = \left\{(t,y)\in\mathcal{D}\times[0,T]:~\widetilde{\mathcal{M}}g=g\right\}.$$
As a result, the fixed point operator $I_1$, defined in \eqref{I_operator}, is influenced accordingly and can be expressed as
\begin{equation*}
    \begin{split}
    I_1 V(t,u,y) &= I_1 g(t,u,y) + \mathbb{E}_{t,y}\bigg[\mathbbm{1}_{\{u<\sigma_1\}}xs + \mathbbm{1}_{\{u\geq\sigma_1\}}xS_{\sigma_1}\bigg]= xs + \widetilde{I}_1 g(t,u,y),
    \end{split}
\end{equation*}
where, for $\epsilon_+:=1$ and $\epsilon_-:=-1$,\begin{equation*}
    \begin{aligned} 
\widetilde{I}_1 g(t,u,y) &:= I_1g(t,u,y) + x\mathbb{E}_{t,y}\bigg[\mathbbm{1}_{\{u\geq\sigma_1\}}(S_{\sigma_1}-s)\bigg]\\&=\sum_{i=1}^d m_i(t,u,\kappa^+,\kappa^-,\mu)\widetilde{\mathcal{M}}V\left(u, x, e^{-\rho(u-t)}d,\lambda^{0,+}_{u}, \lambda^{0,-}_{u},\pi^0_{u}\right)\\  &\qquad+\sum_{k\in\{+,-\}}\sum_{i=1}^d\int_t^u  \bigg[\mathcal{H}^{1,k}_i V\left(r, x, e^{-\rho(r-t)}d, \lambda^{0,+}_{r}, \lambda^{0,-}_{r},\pi^0_{r}\right) + {\color{blue}\epsilon_k} \nu \mathbb{E}_i(Q)x\bigg] q^k_i(t,r)\,\mathrm{d}r.
\end{aligned}
\end{equation*}
Taking all this into account, the numerical implementation boils down to  computing the optimal strategy that maximizes $g$.
\end{rque}

{\raggedright
\printbibliography}

\appendix
\section{Proofs of Section \ref{model_setup}}
\label{prop: exuni}
\begin{proof}[Proof of Proposition \ref{existence_intensities}]
Write $\lambda^{k}_{t}(v):=\sum_{i\in E}\mathbbm{1}_{\{I_{t}=i\}}\lambda^{i,k}_{t}\,\nu_i(v)$ for the kernel \eqref{kernel_def}, and let $i\in E$ and $k\in\{+,-\}$. The candidate intensities $\lambda^{i,k}$ are the solutions of \eqref{intensity_dynamics}, all driven by the same observed pair $(n^{+},n^{-})$. Write $N:=N^{+}+N^{-}$, let $T_0:=0<T_1<T_2<\cdots$ be the jump times of $N$ and set $T_\infty:=\lim_{m}T_m$.

\smallskip
\noindent (1) We first prove the moment bounds and nonexplosion, valid for any $\mathcal{F}$-adapted solution. Between two jump times no order arrives and \eqref{intensity_dynamics} reduces to $y'=-\beta_i(y-\lambda^{i}_{\infty})$. On $[T_m,T_{m+1})$,
$$\lambda^{i,k}_t=\lambda^{i}_{\infty}+\big(\lambda^{i,k}_{T_m}-\lambda^{i}_{\infty}\big)\,e^{-\beta_i(t-T_m)},$$
so each candidate stays between $\lambda^{i,k}_{T_m}$ and $\lambda^{i}_{\infty}$. At the $\ell$-th order, of side $k_\ell$ and volume $v_\ell$, the candidate $\lambda^{i,k}$ jumps by $\varphi^{i}_{\mathrm{s}}(v_\ell/m_{1,i})$ if $k_\ell=k$ and by $\varphi^{i}_{\mathrm{c}}(v_\ell/m_{1,i})$ otherwise. Hence, on $[0,T_m]$,
$$\lambda^{i,k}\ \leq\ \kappa^{i,k}+\lambda^{i}_{\infty}+\sum_{\ell=1}^{m}\big(\varphi^{i}_{\mathrm{s}}+\varphi^{i}_{\mathrm{c}}\big)(v_\ell/m_{1,i}),$$
a sum with finite first and second moments by (A1)--(A2). Set $a:=\max_{i,j\in E,\,\ell\in\{\mathrm{s},\mathrm{c}\}}\int_{\mathbb{R}_+}\varphi^{i}_{\ell}(v/m_{1,i})\,\nu_j(v)\,\mathrm{d}v<+\infty$ by (A2). Integrating \eqref{intensity_dynamics} on $[0,t\wedge T_m]$, taking expectations (compensation holds for nonnegative predictable integrands) and using $\lambda^{k}_{s^-}(v)\leq\sum_{j\in E}\lambda^{j,k}_{s^-}\,\nu_j(v)$, we obtain
$$\mathbb{E}\big[\lambda^{i,\pm}_{t\wedge T_m}\big]\ \leq\ \kappa^{i,\pm}+\beta_i\lambda^{i}_{\infty}\,t+a\int_0^{t}\sum_{j\in E}\mathbb{E}\big[\lambda^{j,+}_{s\wedge T_m}+\lambda^{j,-}_{s\wedge T_m}\big]\,\mathrm{d}s .$$
The left-hand side is finite by the kick bound, so Gr\"onwall's lemma gives $\sum_{i\in E}\mathbb{E}\big[\lambda^{i,+}_{t\wedge T_m}+\lambda^{i,-}_{t\wedge T_m}\big]\leq C_t$ with $C_t$ independent of $m$. Hence, by compensation,
$$m\,\mathbb{P}(T_m\leq t)\ \leq\ \mathbb{E}\big[N_{t\wedge T_m}\big]\ =\ \mathbb{E}\int_0^{t\wedge T_m}\!\big(\lambda^{+}_s+\lambda^{-}_s\big)\,\mathrm{d}s\ \leq\ C_t\,t,$$
so $\mathbb{P}(T_\infty\leq t)=0$ for every $t\geq0$ and $N^{\pm}$ do not explode. Fatou's lemma and monotone convergence then give $\sup_{s\leq t}\mathbb{E}[\lambda^{i,\pm}_{s}]\leq C_t$ and $\mathbb{E}[\int_0^{t}\lambda^{\pm}_s\,\mathrm{d}s]\leq C_t\,t$. For the second moments, fix $t\geq0$ and set, for $u\leq t$,
$$G_m(u)\ :=\ \max_{i\in E,\,k\in\{+,-\}}\mathbb{E}\Big[\sup_{s\leq u}\big(\lambda^{i,k}_{s\wedge T_m}\big)^{2}\Big],$$
finite by the kick bound and Cauchy--Schwarz. Bounding $e^{-\beta_i(u-s)}\leq1$ in the closed-form expression of $\lambda^{i,k}$ (displayed below \eqref{thinning_def}) and splitting each jump integral into a compensated part plus its compensator, we get, for every $s\leq u$,
$$\lambda^{i,k}_{s\wedge T_m}\ \leq\ \lambda^{i}_{\infty}+\kappa^{i,k}+\big|\widetilde{A}^{i,k}_{s\wedge T_m}\big|+C^{i,k}_{u\wedge T_m},$$
where $\widetilde{A}^{i,k}_{r}:=\int_0^{r}\!\int_{\mathbb{R}_+}\varphi^{i}_{\mathrm{s}}\big(\tfrac{v}{m_{1,i}}\big)\big(n^{k}(\mathrm{d}s,\mathrm{d}v)-\lambda^{k}_{s^-}(v)\,\mathrm{d}v\,\mathrm{d}s\big)+\int_0^{r}\!\int_{\mathbb{R}_+}\varphi^{i}_{\mathrm{c}}\big(\tfrac{v}{m_{1,i}}\big)\big(n^{k'}(\mathrm{d}s,\mathrm{d}v)-\lambda^{k'}_{s^-}(v)\,\mathrm{d}v\,\mathrm{d}s\big)$ and $C^{i,k}_{r}:=\int_0^{r}\!\int_{\mathbb{R}_+}\big[\varphi^{i}_{\mathrm{s}}\big(\tfrac{v}{m_{1,i}}\big)\lambda^{k}_{s^-}(v)+\varphi^{i}_{\mathrm{c}}\big(\tfrac{v}{m_{1,i}}\big)\lambda^{k'}_{s^-}(v)\big]\,\mathrm{d}v\,\mathrm{d}s$ is nondecreasing. By $(x+y+z)^{2}\leq3(x^{2}+y^{2}+z^{2})$,
$$G_m(u)\ \leq\ 3\max_{i,k}\big(\lambda^{i}_{\infty}+\kappa^{i,k}\big)^{2}+3\max_{i,k}\mathbb{E}\Big[\sup_{s\leq u}\big(\widetilde{A}^{i,k}_{s\wedge T_m}\big)^{2}\Big]+3\max_{i,k}\mathbb{E}\big[\big(C^{i,k}_{u\wedge T_m}\big)^{2}\big].$$
The processes $n^{+}$ and $n^{-}$ never jump at the same time, being thinned from the independent $M^{+}$ and $M^{-}$, so the stopped martingale $\widetilde{A}^{i,k}_{\cdot\wedge T_m}$ is square-integrable with predictable quadratic variation
$$\langle\widetilde{A}^{i,k}\rangle_{u\wedge T_m}=\int_0^{u\wedge T_m}\!\!\int_{\mathbb{R}_+}\Big[\big(\varphi^{i}_{\mathrm{s}}\big)^{2}\big(\tfrac{v}{m_{1,i}}\big)\,\lambda^{k}_{s^-}(v)+\big(\varphi^{i}_{\mathrm{c}}\big)^{2}\big(\tfrac{v}{m_{1,i}}\big)\,\lambda^{k'}_{s^-}(v)\Big]\,\mathrm{d}v\,\mathrm{d}s.$$
Doob's $L^{2}$ inequality, the same kernel bound and $a_{2}:=\max_{i,j\in E,\,\ell\in\{\mathrm{s},\mathrm{c}\}}\int_{\mathbb{R}_+}\big(\varphi^{i}_{\ell}\big)^{2}\big(\tfrac{v}{m_{1,i}}\big)\,\nu_j(v)\,\mathrm{d}v<+\infty$ from (A2) give
$$\mathbb{E}\Big[\sup_{s\leq u}\big(\widetilde{A}^{i,k}_{s\wedge T_m}\big)^{2}\Big]\ \leq\ 4\,\mathbb{E}\big[\langle\widetilde{A}^{i,k}\rangle_{u\wedge T_m}\big]\ \leq\ 4\,a_{2}\,C_t\,t.$$
Cauchy--Schwarz in time and $\big(\sum_{j=1}^{2d}x_j\big)^{2}\leq 2d\sum_{j=1}^{2d}x_j^{2}$ give
$$\mathbb{E}\big[\big(C^{i,k}_{u\wedge T_m}\big)^{2}\big]\ \leq\ 4\,t\,a^{2}d^{2}\int_0^{u}G_m(s)\,\mathrm{d}s .$$
Hence $G_m(u)\leq K_1(t)+K_2(t)\int_0^{u}G_m(s)\,\mathrm{d}s$ and $G_m(u)\leq K(t)$ by Gr\"onwall's lemma. Since $\sup_{s\leq u}\lambda^{i,k}_{s\wedge T_m}$ increases to $\sup_{s\leq u}\lambda^{i,k}_{s}$ as $T_m\uparrow+\infty$, monotone convergence concludes.

We now build the solution by recursion and prove uniqueness (thinning, cf.\ \textcite{massioule_bremaud}). Set $T_0:=0$ and $\lambda^{i,k}_{0}:=\kappa^{i,k}$. Suppose the solution is built on $[0,T_m]$. For $t>T_m$, let the candidates follow the flow above from their values at $T_m$. The resulting kernel has total mass
$$\int_{\mathbb{R}_+}\lambda^{k}_{t^-}(v)\,\mathrm{d}v\ =\ \lambda^{I_{t^-},k}_{t^-}\ \leq\ \max_{i\in E}\big(\lambda^{i,k}_{T_m}\vee\lambda^{i}_{\infty}\big)\ =:\ B^{k}_m\ <\ +\infty.$$
Set $A^{k}_{m,h}:=\big\{(t,v,z):\ t\in(T_m,T_m+h],\ z\leq\lambda^{k}_{t^-}(v)\big\}$. Then
$$\mathbb{E}\big[M^{k}(A^{k}_{m,h})\,\big|\,\mathcal{F}_{T_m}\big]\ \leq\ h\,B^{k}_m\ <\ +\infty,\qquad h>0,$$
so the accepted points form an a.s.\ locally finite set without accumulation at $T_m$, and
$$T_{m+1}:=\inf\big\{t>T_m:\ (t,v,z)\ \text{point of}\ M^{k},\ k\in\{+,-\},\ z\leq\lambda^{k}_{t^-}(v)\big\}\ >\ T_m\quad\text{a.s.},$$
the infimum being attained at a point $(T_{m+1},v_{m+1},z_{m+1})$ of $M^{k_{m+1}}$. Set
$$\lambda^{i,k}_{T_{m+1}}:=\lambda^{i,k}_{T_{m+1}^-}+\varphi^{i}_{\mathrm{s}}\big(\tfrac{v_{m+1}}{m_{1,i}}\big)\mathbbm{1}_{\{k=k_{m+1}\}}+\varphi^{i}_{\mathrm{c}}\big(\tfrac{v_{m+1}}{m_{1,i}}\big)\mathbbm{1}_{\{k\neq k_{m+1}\}}.$$
The resulting process solves \eqref{intensity_dynamics}--\eqref{thinning_def} on $[0,T_\infty)$, is $\mathcal{F}$-adapted as a measurable function of $(I,M^{+},M^{-})$, and $T_\infty=+\infty$ by the first part. 

Conversely, let $(n^{\pm},\lambda)$ solve \eqref{intensity_dynamics}--\eqref{thinning_def}, with jump times $(T_m)_{m\geq0}$. On $(T_m,T_{m+1})$ the candidates follow the flow started from $\lambda_{T_m}$, which determines the barrier in \eqref{thinning_def}. By \eqref{thinning_def}, the jumps of $n^{k}$ are exactly the points of $M^{k}$ lying below this barrier. No such point lies in $(T_m,T_{m+1})$, as it would be a jump of $n^{k}$ before $T_{m+1}$, and exactly one lies at $T_{m+1}$, the time coordinates of the points of $(M^{+},M^{-})$ being pairwise distinct. The next jump time, its side, its volume and, by \eqref{intensity_dynamics}, the post-jump state are thus measurable functions of $\lambda_{T_m}$ and $(I,M^{+},M^{-})$, common to all solutions. Since $\lambda_0=\kappa$, induction on $m$ yields that all solutions coincide on $[0,T_\infty)=\mathbb{R}_+$. The kernels \eqref{kernel_def} are identified in Proposition \ref{compensator_immersion}, whose proof uses only \eqref{thinning_def}.

\smallskip
\noindent (2) Let $\mathcal{T}_I:=\{t>0:\ I_t\neq I_{t^-}\}$, a countable $\mathcal{F}^{I}_{\infty}$-measurable set, and, for countable $D\subset\mathbb{R}_+$, let $g(D):=\mathbb{P}\big(M^{+}\text{ or }M^{-}\text{ has a point with time coordinate in }D\big)$. The intensity measure of $D\times[0,n]^{2}$ is $\mathrm{Leb}(D)\,n^{2}=0$ for every $n\in\mathbb{N}$, so monotone convergence gives $g(D)=0$. The jump times of $N^{\pm}$ are time coordinates of points of $(M^{+},M^{-})$, and $\mathcal{T}_I$ is independent of $(M^{+},M^{-})$, so conditioning on $\mathcal{F}^{I}_{\infty}$ and Fubini yield
$$\mathbb{P}\big(\exists\,t\in\mathcal{T}_I:\ \Delta N^{+}_t+\Delta N^{-}_t>0\big)\ \leq\ \mathbb{E}\big[g(\mathcal{T}_I)\big]\ =\ 0 .$$
Hence $[\mathbbm{1}_{\{I=i\}},N^{k}]_t=\sum_{s\leq t}\Delta\mathbbm{1}_{\{I_s=i\}}\,\Delta N^{k}_s=0$ for every $i\in E$ and $k\in\{+,-\}$.

\smallskip
\noindent (3) On $\{I_{t^-}=i\}$, the acceptance region $\big[0,\sum_{j=1}^d\mathbbm{1}_{\{I_{t^-}=j\}}\lambda^{j,k}_{t^-}\nu_j(v)\big]$ in \eqref{thinning_def} equals $\big[0,\lambda^{i,k}_{t^-}\nu_i(v)\big]$, so, for each point of $M^{k}$, $\mathbbm{1}_{\{I_{t^-}=i\}}\,n^{k}=\mathbbm{1}_{\{I_{t^-}=i\}}\,n^{i,k}$. Summing over $i\in E$ with $\sum_{i\in E}\mathbbm{1}_{\{I_{t^-}=i\}}=1$ yields the identity.
\end{proof}
\section{Proofs of the Results in Section \ref{The filtering equations}}
\label{filtering_proofs}

  Note that the dynamics of $I$ can alternatively be represented using the Poisson processes $(n_{lm}^I)_{(l,m)\in E^2}$. The process $n_{lm}^I$ counts transitions from state $l$ to state $m$ and has an $\mathcal{F}^I$-intensity given by $t \mapsto \mathbbm{1}_{\{I_t = l\}} \psi_{lm}(t)$, where $(l, m) \in E^2$. Thus, $I$ has the following semi-martingale decomposition:
$$\mathrm{d}I_t = \sum_{m=1}^{d} (m - I_{t^-}) n_{I_{t^-} m}^I(\mathrm{d}t),$$ for all $t \geq 0.$ This will help us derive the Kushner–Stratonovich equation. Here, we denote $\mathcal{F}^{N} = \left\{\mathcal{F}^{N}_t\right\}_{t\geq 0}$ as the natural filtration associated to a general $m$-variate point measures $\left(n^1, \ldots, n^m\right)$ such that
    $$
    \mathcal{F}_t^{N} := \underset{\substack{k=1}}{\bigvee^m}\sigma\left(n^{k}((0,s]\times A); s \leq t, A \in \mathcal{B}(\mathbb{R}_+)\right). 
    $$ 
    \subsection{No common jumps case}
    Next, we determine, in the case of no common jumps between $\left(n^1, \ldots, n^m\right)$, the dynamics of $\pi_t(\varphi) = \mathbb{E}\Big[\varphi(I_t)\mid \mathcal{F}^N_t\Big]$ through the use of the innovation method as described in \textcite{Bremaud}. We apply Proposition \ref{ks_equation} to establish Theorem \ref{ks_equation_final} which corresponds to the case $\left(n^1, \ldots, n^m\right)=\left(n^+, n^-\right)$.
\begin{lemma}[Projection lemma]
\label{projection_lemma}
Let $\{\mathcal{F}_t\}_{t\geq 0}\subset\{\mathcal{G}_t\}_{t\geq 0}$ be two filtrations and let $\psi$ be a $\mathcal{G}$-progressively measurable process with $\mathbb{E}\big[\int_0^T|\psi_s|\,\mathrm{d}s\big]<+\infty$. Then the process
$$t\longmapsto \mathbb{E}\bigg[\int_0^t\psi_s\,\mathrm{d}s\ \Big|\ \mathcal{F}_t\bigg]-\int_0^t{}^{o}\psi_s\,\mathrm{d}s$$
is an $\mathcal{F}$-martingale, where ${}^{o}\psi$ denotes the $\mathcal{F}$-optional projection of $\psi$ (see, e.g., Lemma 8.4 in \textcite{bain2008fundamentals} and \textcite{Bremaud}.)
\end{lemma}
\begin{prop}
\label{ks_equation}
Let $\varphi\in \mathcal{C}^2(\mathbb{R})$ and $t\in[0,T]$.  
Let $(n^1,\ldots,n^m)$ be an $m$-variate marked point process, 
$\mathbb{P}$-nonexplosive, and let $\mathcal{F}^N$ denote its internal history.  
Let $\mathbb{G}=\{\mathcal{G}_t\}_{t\geq0}$, with $\mathcal{G}_t := \mathcal{F}^N_t\vee\mathcal{F}^I_t$. Assume that for each $1\le k\le m$, the random measure $n^k$ admits a $(\mathbb{P},\mathbb{G})$-predictable intensity kernel $(t,v)\mapsto\lambda^k_t(v)$ of the form $\lambda^k_t(v) = \ell^k_t(I_{t^-},v)$, where, for each state $i\in E$, the map $(t,v)\mapsto \ell^k_t(i,v)$ is $\mathcal{F}^N$-predictable (this is the situation of Proposition \ref{compensator_immersion}). We write, as a shorthand,
$$
\mathbb{E}\big[n^k(\mathrm{d}t,\mathrm{d}v)\mid\mathcal{F}^N_{t^-}\big]
=
\pi_{t^-}(\lambda^k(v))\,\mathrm{d}v\,\mathrm{d}t,\quad \pi_{t^-}(\lambda^k(v)) := \sum_{i=1}^d\pi_{t^-}(i)\,\ell^k_t(i,v),
$$
meaning that the $\mathcal{F}^N$-dual predictable projection of $n^k(\mathrm{d}t,\mathrm{d}v)$ is $\pi_{t^-}(\lambda^k(v))\,\mathrm{d}v\,\mathrm{d}t$; this identity follows from the $\mathbb{G}$-intensity kernel by Lemma \ref{projection_lemma}. Assume moreover that $[n^k,n^{k^{\prime}}]=0$ for all $k\neq k^{\prime}$.  
Then the filtering process $\pi_t(\varphi)$ satisfies the marked 
Kushner--Stratonovich equation
\begin{equation}
    \resizebox{\textwidth}{!}{$
    \begin{aligned}
\mathrm{d}\pi_t(\varphi)
&=
\sum_{\substack{j,l=1\\j\neq l}}^d 
\big(\varphi(j)-\varphi(l)\big)\,
\psi_{lj}(t)\,
\widehat{\mathbbm{1}_{\{I_t=l\}}}\,
\mathrm{d}t \\
&\quad+
\sum_{k=1}^m
\int_{\mathbb{R}_+}
\frac{
\pi_{t^-}(\varphi\,\lambda^k(v))
+
\pi_{t^-}(D^k(v))
-
\pi_{t^-}(\varphi)\,\pi_{t^-}(\lambda^k(v))
}{
\pi_{t^-}(\lambda^k(v))
}\times
\Big(
n^k(\mathrm{d}t,\mathrm{d}v)
-
\pi_{t^-}(\lambda^k(v))\,\mathrm{d}t\mathrm{d}v
\Big).
\end{aligned}$}
\end{equation}
Here, $\big(D^k_t(v)\big)_{(t,v)\in[0,T]\times \mathbb{R}_+}$, $1\le k\le m$, are the $\mathcal{F}^N$-predictable kernels defined, in the same dual-projection sense, by  
$$
\mathbb{E}\big[\mathrm{d} M^I_t\,
n^k(\mathrm{d}t,\mathrm{d}v)\mid\mathcal{F}^N_{t^-}\big]
=
D^k_t(v)\,\mathrm{d}v\,\mathrm{d}t,
$$
that is, the $\mathcal{F}^N$-dual predictable projection of the covariation measure $\sum_{s\leq t}\Delta M^I_s\, n^k(\{s\},\mathrm{d}v)$ is $D^k_t(v)\,\mathrm{d}v\,\mathrm{d}t$,
and the process $(M^I_t)_{t\in[0,T]}$ is the martingale component of
$\varphi(I_t)$, namely
$$
M^I_t
=
\int_0^t
\sum_{\substack{j=1\\j\neq I_s}}^d
\big(\varphi(j)-\varphi(I_s)\big)
\Big(
n^I_{I_s j}(\mathrm{d}s)
-
\psi_{I_s j}(s)\,\mathrm{d}s
\Big),
\quad\forall t\in[0,T].
$$
\end{prop}
\begin{proof}
   Let $\varphi\in \mathcal{C}^2(\mathbb{R})$. We can derive a semi-martingale representation for $\varphi(I_t)$ by applying Dynkin’s formula. We get that
    \begin{equation}
    \label{proof1_dyn2}
\varphi\left(I_t\right)=\varphi\left(I_0\right)+\int_0^t \mathcal{L}_s^I \varphi\left(I_{s}\right) \mathrm{d}s+M^I_t, \quad t\in[0,T],
\end{equation}
where $\mathcal{L}^I$ denotes the infinitesimal generator associated to $I$, such that
$$
\mathcal{L}_t^I \varphi(l)=\sum\limits_{\substack{j=1 \\ j\neq l}}^d\big(\varphi(j) - \varphi(l)\big)\psi_{lj}(t),\quad \forall (t,l)\in[0,T]\times E,
$$ and, the stochastic process $\left(M^I_t\right)_{t\in[0,T]}$ is given by
$$
M^I_t=\int_0^t\sum\limits_{\substack{j=1 \\ j\neq I_s}}^d \left(\varphi(j)-\varphi\left(I_{s}\right)\right) \bigg[n^I_{I_{s}j}(\mathrm{d}s)-\psi_{I_{s}j}\mathrm{d}s\bigg], \quad t\in[0,T] .
$$
Observing that the process $\left(\varphi(I_t)\right)_{t\in[0,T]}$ is bounded, it follows that $M^I$ is an integrable $\mathcal{F}^I$-martingale, and that
\begin{equation*}
\mathbb{E}\bigg[\int_0^t\mid\mathcal{L}^I\varphi(I_s)\mid\mathrm{d}s\bigg]<+\infty.
\end{equation*}
Applying Lemma \ref{projection_lemma} with $\mathcal{F}=\mathcal{F}^N$, $\mathcal{G}=\mathbb{G}$ and $\psi_s = \mathcal{L}^I_s\varphi(I_s)$, which is integrable by the display above, we obtain that the process $\widetilde{M}^I$ defined by the  martingale problem 
\begin{equation}
    \label{proofks_filter_dyn}
    \mathrm{d}\widetilde{M}^I_t = \mathrm{d}\widehat{\varphi(I_t)}-\widehat{\mathcal{L}^I \varphi\left(I\right)} \mathrm{d}t, \quad t\in[0,T]
\end{equation} is an optional $\mathcal{F}^{N}$-martingale. For each $1\le k\le m$, the random measure $n^k$ admits the $\mathcal{F}^N$-dual
predictable projection
$$
\mathbb{E}\big[n^k(\mathrm{d}t,\mathrm{d}v)\mid \mathcal{F}^N_{t^-}\big]
=
\pi_{t^-}(\lambda^k(v))\,\mathrm{d}v\,\mathrm{d}t.
$$
By the martingale representation theorem (see Theorem T9 of Section~{\MakeUppercase{\romannumeral 3}} and
Theorem T8 of Section~{\MakeUppercase{\romannumeral 8}} in \textcite{Bremaud}), there exist
$\mathcal{F}^N$-predictable kernels $(K^k_t(v))_{(t,v)\in[0,T]\times\mathbb{R}_+}$ such that
$$
\mathrm{d}\widetilde{M}^I_t
=
\sum_{k=1}^m
\int_{\mathbb{R}_+}
K^k_t(v)
\Big(
n^k(\mathrm{d}t,\mathrm{d}v)
-
\pi_{t^-}(\lambda^k(v))\,\mathrm{d}v\,\mathrm{d}t
\Big),
\quad \forall t\in[0,T],
$$
and for all $t\in[0,T]$ and all $V\in\mathcal{B}(\mathbb{R}_+)$,
$$
\int_0^t\int_V 
\big|K^k_s(v)\big|\,
\pi_{s^-}(\lambda^k(v))\,\mathrm{d}v\,\mathrm{d}s
<+\infty.
$$
For each $1\le k\le m$ and $n\ge1$, define the $\mathcal{F}^N$-predictable stopping times
$$
\tau^k_n
=
\begin{cases}
\displaystyle
\inf\bigg\{t\ge0:\int_0^t\int_{\mathbb{R}_+}
\big(1+\lvert K^k_s(v)\rvert\big)\,
\pi_{s^-}(\lambda^k(v))\,\mathrm{d}v\,\mathrm{d}s
\ge n\bigg\},
&\text{if this set is nonempty},\\[8pt]
+\infty, &\text{otherwise}.
\end{cases}
$$
Fix $1\le k\le m$ and $V\in\mathcal{B}(\mathbb{R}_+)$. Define the
$\mathcal{F}^N$-adapted finite–variation process
$$
A^k(t,V)
:=
\int_{(0,t]\times V} n^k(\mathrm{d}s,\mathrm{d}v),
\quad \forall t\in[0,T].
$$
Since $A^k(\cdot,V)$ jumps only when $n^k$ has a jump with mark in $V$, an
integration-by-parts argument applied to $A^k(t,V)\,\varphi(I_t)$ together
with the decomposition of $\varphi(I_t)$ yields
$$
\begin{aligned}
\mathrm{d}\big(A^k(t,V)\,\varphi(I_t)\big)
&=
A^k(t^-,V)\,\mathrm{d}\varphi(I_t)
+
\varphi(I_{t^-})\int_V n^k(\mathrm{d}t,\mathrm{d}v)
+
\mathrm{d}\big[M^I,\,A^k(\cdot,V)\big]_t.
\end{aligned}
$$
Since the quadratic covariation only accumulates at common jumps of $M^I$ and
$A^k(\cdot,V)$, there exists a predictable kernel $D^k_t(v)$ such that
$$
\mathrm{d}\big[M^I,\,A^k(\cdot,V)\big]_t
=
\int_V D^k_t(v)\,n^k(\mathrm{d}t,\mathrm{d}v),
$$
and hence
$$
\begin{aligned}
\mathrm{d}\big(A^k(t,V)\,\varphi(I_t)\big)
&=
A^k(t^-,V)\,\mathcal{L}^I_t\varphi(I_t)\,\mathrm{d}t
+
A^k(t^-,V)\,\mathrm{d}M^I_t
\\
&\quad+
\int_V\big(\varphi(I_{t^-})+D^k_t(v)\big)
\Big(
n^k(\mathrm{d}t,\mathrm{d}v)
-
\lambda^k_t(v)\,\mathrm{d}v\,\mathrm{d}t
\Big)
\\
&\quad+
\int_V\big(\varphi(I_{t^-})+D^k_t(v)\big)
\lambda^k_t(v)\,\mathrm{d}v\,\mathrm{d}t.
\end{aligned}
$$
We now perform the same integration-by-parts computation for the filtered
quantity $A^k(t,V)\,\widehat{\varphi(I_t)}$. Using
$$
\mathrm{d}\widehat{\varphi(I_t)}
=
\pi_t(\mathcal{L}^I\varphi)\,\mathrm{d}t
+
\mathrm{d}\widetilde{M}^I_t,
$$
and the martingale representation of $\widetilde{M}^I$, we obtain
$$
\begin{aligned}
\mathrm{d}\big(A^k(t,V)\,\widehat{\varphi(I_t)}\big)
&=
A^k(t^-,V)\,\mathrm{d}\widehat{\varphi(I_t)}
+
\widehat{\varphi(I_{t^-})}\int_V n^k(\mathrm{d}t,\mathrm{d}v)
+
\mathrm{d}\big[A^k(\cdot,V),\,\widehat{\varphi(I_\cdot)}\big]_t
\\[3pt]
&=
A^k(t^-,V)\,\pi_t(\mathcal{L}^I\varphi)\,\mathrm{d}t
+
A^k(t^-,V)\,\mathrm{d}\widetilde{M}^I_t
\\
&\quad+
\widehat{\varphi(I_{t^-})}\int_V
\Big(
n^k(\mathrm{d}t,\mathrm{d}v)
-
\pi_{t^-}(\lambda^k(v))\,\mathrm{d}v\,\mathrm{d}t
\Big)
\\
&\quad+
\widehat{\varphi(I_{t^-})}
\int_V \pi_{t^-}(\lambda^k(v))\,\mathrm{d}v\,\mathrm{d}t
+
\mathrm{d}\big[A^k(\cdot,V),\,\widehat{\varphi(I_\cdot)}\big]_t.
\end{aligned}
$$
Since $A^k(\cdot,V)$ jumps only when $n^k$ jumps with marks in $V$, the
quadratic covariation satisfies
$$
\mathrm{d}\big[A^k(\cdot,V),\,\widehat{\varphi(I_\cdot)}\big]_t
=
\int_V K^k_t(v)\,n^k(\mathrm{d}t,\mathrm{d}v),
$$
and hence
$$
\begin{aligned}
\mathrm{d}\big(A^k(t,V)\,\widehat{\varphi(I_t)}\big)
&=
A^k(t^-,V)\,\pi_t(\mathcal{L}^I\varphi)\,\mathrm{d}t
+
A^k(t^-,V)\,\mathrm{d}\widetilde{M}^I_t
\\
&\quad+
\int_V\big(\widehat{\varphi(I_{t^-})}+K^k_t(v)\big)
\Big(
n^k(\mathrm{d}t,\mathrm{d}v)
-
\pi_{t^-}(\lambda^k(v))\,\mathrm{d}v\,\mathrm{d}t
\Big)
\\
&\quad+
\int_V\big(\widehat{\varphi(I_{t^-})}+K^k_t(v)\big)
\pi_{t^-}(\lambda^k(v))\,\mathrm{d}v\,\mathrm{d}t.
\end{aligned}
$$
Because $(\varphi(I_t))_{t\in[0,T]}$ and
$\big(A^k(t\wedge\tau^k_n,V)\big)_{t\in[0,T]}$ are bounded, the processes
$$
t\mapsto
\int_0^{t\wedge\tau^k_n} A^k(s^-,V)\,\mathrm{d}M^I_s
~~\text{and}~~
t\mapsto
\int_{(0,t\wedge\tau^k_n]\times V}
\big(\varphi(I_{s^-})+D^k_s(v)\big)
\Big(
n^k(\mathrm{d}s,\mathrm{d}v)
-
\lambda^k_s(v)\,\mathrm{d}v\,\mathrm{d}s
\Big)
$$
are $\mathcal{F}^N$-martingales. Similarly, by the integrability condition on
$K^k$ and the definition of $\tau^k_n$, the processes
$$
t\mapsto
\int_0^{t\wedge\tau^k_n} A^k(s^-,V)\,\mathrm{d}\widetilde{M}^I_s
~~\text{and}~~
t\mapsto
\int_{(0,t\wedge\tau^k_n]\times V}
\big(\widehat{\varphi(I_{s^-})}+K^k_s(v)\big)
\Big(
n^k(\mathrm{d}s,\mathrm{d}v)
-
\pi_{s^-}(\lambda^k(v))\,\mathrm{d}v\,\mathrm{d}s
\Big)
$$
are $\mathcal{F}^N$-martingales. Since
$\mathbb{E}\big[A^k(t\wedge\tau^k_n,V)\big]<+\infty$, we get
$$
\widehat{A^k(t\wedge\tau^k_n,V)\,\varphi(I_{t\wedge\tau^k_n})}
=
A^k(t\wedge\tau^k_n,V)\,\widehat{\varphi(I_{t\wedge\tau^k_n})},
\quad \forall t\in[0,T],
$$
that is, the finite–variation parts of the two decompositions coincide after
conditioning on $\mathcal{F}^N$. By uniqueness of the Doob–Meyer decomposition
for semimartingales (see Section~III.3 in \textcite{protter2005stochastic}), we
obtain, for all $t\le \tau^k_n\wedge T$,
$$
\int_V\big(\widehat{\varphi(I_{t^-})}+K^k_t(v)\big)
\pi_{t^-}(\lambda^k(v))\,\mathrm{d}v
=
\int_V
\widehat{\big(\varphi(I_{t^-})+D^k_t(v)\big)\lambda^k_t(v)}\,
\mathrm{d}v.
$$
Since this holds for every $V\in\mathcal{B}(\mathbb{R}_+)$, we deduce (at the
level of densities in $v$) that
$$
\big(\widehat{\varphi(I_{t^-})}+K^k_t(v)\big)\pi_{t^-}(\lambda^k(v))
=
\widehat{\varphi(I_{t^-})\lambda^k_t(v)}
+
\widehat{D^k_t(v)}
-
\widehat{\varphi(I_{t^-})}\,\pi_{t^-}(\lambda^k(v)),
\quad \text{a.e. }v,
$$
and therefore
$$
K^k_t(v)
=
\frac{
\widehat{\varphi(I_{t^-})\lambda^k_t(v)}
+
\widehat{D^k_t(v)}
-
\widehat{\varphi(I_{t^-})}\,\pi_{t^-}(\lambda^k(v))
}{
\pi_{t^-}(\lambda^k(v))
}.
$$
Using the predictability of $K^k$ and the definition of the filter
$\pi_{t^-}$, we can rewrite this as
$$
K^k_t(v)
=
\frac{
\pi_{t^-}\big(\varphi\,\lambda^k(v)\big)
+
\pi_{t^-}\big(D^k(v)\big)
-
\pi_{t^-}(\varphi)\,\pi_{t^-}(\lambda^k(v))
}{
\pi_{t^-}(\lambda^k(v))
}.
$$

Finally, from the definition of $\widetilde{M}^I$ we have
$$
\mathrm{d}\pi_t(\varphi)
=
\pi_t\big(\mathcal{L}^I\varphi\big)\,\mathrm{d}t
+
\mathrm{d}\widetilde{M}^I_t,
$$
and substituting the martingale representation of $\widetilde{M}^I$ with the
expression of $K^k_t(v)$ yields the dynamics of the filter:
$$
\begin{aligned}
\mathrm{d}\pi_t(\varphi)
&=
\pi_t\big(\mathcal{L}^I\varphi\big)\,\mathrm{d}t
\\
&\quad+
\sum_{k=1}^m
\int_{\mathbb{R}_+}
\frac{
\pi_{t^-}\big(\varphi\,\lambda^k(v)\big)
+
\pi_{t^-}\big(D^k(v)\big)
-
\pi_{t^-}(\varphi)\,\pi_{t^-}(\lambda^k(v))
}{
\pi_{t^-}(\lambda^k(v))
}
\Big(
n^k(\mathrm{d}t,\mathrm{d}v)
-
\pi_{t^-}(\lambda^k(v))\,\mathrm{d}v\mathrm{d}t
\Big).
\end{aligned}
$$
\end{proof}
For completeness, we conclude this section by outlining the derivation of the Zakai and Duncan-Mortensen-Zakai equations. In the subsequent discussion, we define $\mathcal{F}^I$ as a sub-$\sigma$-field generated by the process $I$, where $\mathcal{F}^I_t$ is defined as $\sigma\left(I_s; s \in[0, t]\right)$ for all $t\in[0,T]$. We also consider the enlarged filtration $\mathbb{G} = \{\mathcal{G}_t\}_{t \ge 0}$ of $\mathcal{F}^N$, with $\mathcal{G}_t = \mathcal{F}_t^N \vee \mathcal{F}_t^I$ for all $t \ge 0$. The filtration $\mathbb{G}$ encapsulates the information available to a hypothetical observer who, in addition to the order flow, observes the trajectory of $I$ up to the present time. We employ the change of measure technique: fixing a reference probability density $\widetilde{\nu}>0$ on $\mathbb{R}_+$ (see Assumptions \ref{ass_density}). In the setting of Section \ref{model_setup} we take the uniform mixture $\widetilde{\nu}=\tfrac{1}{d}\sum_{i\in E}\nu_i$, so that $\nu_i\leq d\,\widetilde{\nu}$ for every $i\in E$. We construct a new probability measure under which the observation measures $n^1,\dots,n^m$ become independent Poisson random measures with intensity $\widetilde{\nu}(v)\,\mathrm{d}v\,\mathrm{d}t$, independent of $I$. To achieve this, we define the Doléans-Dade exponential $Z = \left(Z_t\right)_{t\in[0,T]}$ as
\begin{align*}
Z_t
:=
&\prod_{k=1}^m
\exp\Bigg\{
 -\int_0^t\int_{\mathbb{R}_+}
   \log\bigg(\frac{\lambda^{k}_{s^-}(v)}{\widetilde{\nu}(v)}\bigg)\,
   n^k (\mathrm ds,\mathrm dv)
 -\int_0^t
   \Big(1-\int_{\mathbb{R}_+}\lambda^{k}_s(v)\mathrm dv\Big)\,\mathrm ds
\Bigg\},
\quad \mathbb P\text{-a.s.}
\end{align*}
\begin{prop}
\label{prop_Z_martingale}
Assume that $\mathbb{E}^{\mathbb{P}}[Z_T]=1$; sufficient conditions, applicable to the unbounded intensities considered here, are given in \textcite{sokolAlex}. Then $Z$ is a positive 
$(\mathbb{P},\mathcal{G})$-martingale satisfying
\begin{equation*}
\begin{split}
\mathrm{d}Z_t
&=
Z_{t^-}
\sum_{k=1}^m
\int_{\mathbb{R}_+}
\Big(
\frac{\widetilde{\nu}(v)}{\lambda^{k}_{t^-}(v)}-1
\Big)
\Big[
  n^k(\mathrm{d}t,\mathrm{d}v)
  - \lambda^{k}_t(v)\,\mathrm{d}v\,\mathrm{d}t
\Big]
\end{split}
\end{equation*}
\end{prop}
\begin{proof}
Refer to \textcite{sokolAlex}.
\end{proof}
The previous proposition allows us to utilize the Girsanov theorem, incorporating the Radon-Nikodym density $Z$ to establish a new probability measure $\mathbb{Q}$ on $(\Omega, \mathcal{G})$ given by $\frac{\mathrm{d}\mathbb{Q}}{\mathrm{d}\mathbb{P}}|_{\mathcal{G}_t} = Z_t$, for all $t\in[0,T]$. Using the Kallianpur–Striebel formula (see \textcite{bain2008fundamentals}), we obtain that, for all $\varphi \in \mathcal{C}^2(\mathbb{R})$,
\begin{equation}
\label{unconditional_density}
\pi_t(\varphi)=\frac{\mathbb{E}^{\mathbb{Q}}\Big[Z^{-1}_t \varphi\left(I_t\right) \mid \mathcal{F}^{N}_t\Big]}{\mathbb{E}^{\mathbb{Q}}\Big[Z^{-1}_t \mid \mathcal{F}^{N}_t\Big]}=\frac{\sigma_t(\varphi)}{\sigma_t(1)},\quad t\geq 0.
\end{equation}
\begin{prop}[Zakai equation]
\label{prop_dynamic_filter}
     The process $\left(\sigma_t(\varphi), t \geq 0\right)$ satisfies the stochastic differential equation  \begin{equation}\label{zakaieq}
    \begin{aligned}  
\mathrm{d}\sigma_t(\varphi)
&=
\sigma_{t^-}(\mathcal{L}^I(\varphi))\mathrm{d} t\\
&\quad+
\sum_{k=1}^m
\int_{\mathbb{R}_+}
\frac{
    \sigma_{t^-}\big(\varphi\,\lambda^{k}(v)\big)
  + \sigma_{t^-}\big(D^k(v)\big)
  - \sigma_{t^-}(\varphi)\,\widetilde{\nu}(v)
}{\widetilde{\nu}(v)}
\Big[
    n^k(\mathrm{d}t,\mathrm{d}v)
  - \widetilde{\nu}(v)\,\mathrm{d}v\mathrm{d}t
\Big],
\end{aligned}
\end{equation}
with $\varphi\in \mathcal{C}^2(\mathbb{R})$ and $
\mathcal{L}_t^I \varphi(l)=\sum\limits_{\substack{j=1 \\ j\neq l}}^d(\varphi(j) - \varphi(l))\psi_{lj}(t),$ for all $(t,l)\in[0,T]\times E
$.
\end{prop}
\begin{proof}
Since 
$$\sigma_t(\varphi)
=
\pi_t(\varphi)\sigma_t(\mathbf{1})
=
\mathbb{E}^{\mathbb{Q}}\!\left[Z_t^{-1}\mid\mathcal{F}^N_t\right]\pi_t(\varphi),$$
we get through the Stieltjes--Lebesgue formula that
\begin{equation}
\label{proof_zakai1}
\begin{split}
\mathrm{d}\sigma_t(\varphi)
&=
\mathrm{d}\pi_t(\varphi)\,\sigma_{t^-}(\mathbf{1})
+\pi_{t^-}(\varphi)\,\mathrm{d}\sigma_t(\mathbf{1})
+\mathrm{d}\big[\pi_t(\varphi),\sigma_t(\mathbf{1})\big].
\end{split}
\end{equation}
Applying the Itô formula to $Z_t^{-1}$, we obtain the $(\mathbb{Q},\mathbb{G})$-martingale
\begin{equation*}
\begin{split}
\mathrm{d}Z^{-1}_t
&=
Z^{-1}_{t^-}
\sum_{k=1}^m
\int_{\mathbb{R}_+}
\bigg(\frac{\lambda^k_{t^-}(v)}{\widetilde{\nu}(v)}-1\bigg)
\Big[
   n^k(\mathrm{d}t,\mathrm{d}v)
 - \widetilde{\nu}(v)\,\mathrm{d}v\mathrm{d}t
\Big].
\end{split}
\end{equation*}
Hence the dynamics of the $\mathcal{F}^N$-optional process $\sigma_t(\mathbf{1})$
under $\mathbb{Q}$ is
\begin{equation*}
\begin{split}
\mathrm{d}\sigma_t(\mathbf{1})
&=
\sigma_{t^-}(1)
\sum_{k=1}^m
\int_{\mathbb{R}_+}
\bigg(\frac{\pi_{t^-}(\lambda^k(v))}{\widetilde{\nu}(v)}-1\bigg)
\Big[
   n^k(\mathrm{d}t,\mathrm{d}v)
 - \widetilde{\nu}(v)\,\mathrm{d}v\mathrm{d}t
\Big].
\end{split}
\end{equation*}
Following Theorem~\ref{ks_equation_final}, the covariation between
$\pi_t(\varphi)$ and $\sigma_t(\mathbf{1})$ is
\begin{equation*}
    \begin{aligned}
\mathrm{d}\big[\pi_.(\varphi),\sigma_.(\mathbf{1})\big]_t
=
\sum_{k=1}^m
\int_{\mathbb{R}_+}
&\frac{
  \pi_{t^-}\big(\varphi\,\lambda^k(v)\big)
 +\pi_{t^-}\big(D^k(v)\big)
 -\pi_{t^-}(\varphi)\pi_{t^-}(\lambda^k(v))
}{
  \pi_{t^-}(\lambda^k(v))
}
\\&\quad\times\sigma_{t^-}(\mathbf{1})
\bigg(\frac{\pi_{t^-}(\lambda^k(v))}{\widetilde{\nu}(v)}-1\bigg)
n^k(\mathrm{d}t,\mathrm{d}v)
\end{aligned}
\end{equation*}
Plugging the previous expressions into \eqref{proof_zakai1} yields
\begin{equation*}
    \begin{aligned}  
\mathrm{d}\sigma_t(\varphi)
&=
\sigma_{t^-}(\mathcal{L}^I(\varphi))\mathrm{d} t+
\sum_{k=1}^m
\int_{\mathbb{R}_+}
\frac{
    \sigma_{t^-}\big(\varphi\,\lambda^{k}(v)\big)
  + \sigma_{t^-}\big(D^k(v)\big)
  - \sigma_{t^-}(\varphi)\,\widetilde{\nu}(v)
}{\widetilde{\nu}(v)}
\Big[
    n^k(\mathrm{d}t,\mathrm{d}v)
  - \widetilde{\nu}(v)\,\mathrm{d}v\mathrm{d}t
\Big],
\end{aligned}
\end{equation*}
This concludes the proof.
\end{proof}
\color{black}
Proposition \ref{prop_dynamic_filter} enables us to describe the dynamics of the unnormalized filter $\left(\sigma_t(\varphi)\right)_{t\in[0,T]}$ by defining it as the unique strong solution to the Zakai equation \eqref{zakaieq}. This equation exhibits linearity and is driven by $n^1,\dots,n^m$ instead of the innovation processes $$\int_0^t \int_{\mathbb{R}_+} n^k(\mathrm{d}s,\mathrm{d}v)
-
\pi_{s^-}(\lambda^k(v))\,\mathrm{d}v\,\mathrm{d}s,
\quad k\in\{1,\dots,m\},$$ as seen in the Kushner--Stratonovich equation. This results from the expression of our filter under the previously constructed probability measure $\mathbb{Q}$. An alternative method to derive the Zakai equation would be through the change of probability technique. This involves analyzing the dynamics of the process $\left(\varphi(I_t)\widetilde{Z}^{\varepsilon}_t\right)_{t\in[0,T]}$ using Itô calculus to obtain the stochastic differential equation governing $t\mapsto \sigma_t(\varphi) = \mathbb{E}^{\mathbb{Q}}\bigg[Z^{-1}_t \varphi\left(I_t\right) \mid \mathcal{F}^{N}_t\bigg]$, where $\widetilde{Z}^{\varepsilon}_t = \frac{Z^{-1}_t}{1+\varepsilon Z^{-1}_t}$ for all $t\in[0,T]$ and $\varepsilon>0$. The introduction of the transformed process $\widetilde{Z}^{\varepsilon}$ is primarily motivated by its boundedness, which facilitates the derivation of martingales that are easy to manipulate. The desired outcome is obtained by letting $\varepsilon$ approach $0$ (see \textcite{bain2008fundamentals}). Nevertheless, given that we have previously established the Kushner--Stratonovich equation, it is more straightforward to derive the Zakai equation via the Kallianpur-Striebel formula, as $\sigma_t(\varphi) = \pi_t(\varphi)\sigma_t(\mathbf{1})$ holds for all $t\in[0,T]$. 
\subsection{Common jumps case}
Next, we prove the filtering results to formulate Theorem \ref{ks_equation_final} in the case of common jumps. We introduce new definitions and notations that will be helpful in formulating our results.
\begin{nota}
Let $(n^+,n^-)$ be a marked point process with $(\mathbb{P},\mathbb{G})$-predictable intensity kernels $\bar\lambda^{k}$, $k\in\{+,-\}$, with
$$\bar\lambda^{k}_t(v):= \sum_{i =1}^d \mathbbm{1}_{\left\{I_{t^-}=i\right\}}\, \lambda^{i,k}_{t^-}\, \nu_i(v),\quad \forall (t,v)\in\mathbb{R}_+^2.$$
Moreover, we define the point process $\widetilde n$ such that 
$$\widetilde n(\mathrm{d}t,\mathrm{d}v)
:=
n^+(\mathrm{d}t,\mathrm{d}v)\,n^-(\mathrm{d}t,\mathrm{d}v).$$
$\widetilde n$ admits $\mathcal{F}^N$–predictable compensator of the form
$\mathbb{E}\big[\widetilde n(\mathrm{d}t,\mathrm{d}v)\mid\mathcal{F}^N_{t^-}\big]
=
\widetilde\lambda_t(v)\mathrm{d}v\,\mathrm{d}t$. We recall that, in the model of Section \ref{model_setup}, $n^+$ and $n^-$ have almost surely no common jumps (Proposition \ref{model_construction}), so that $\widetilde n\equiv0$ there. This subsection is included only for completeness.
\end{nota}
\begin{theo}[Kushner–Stratonovich equation]
    Let $\varphi\in \mathcal{C}^2(\mathbb{R})$, and $t\in[0,T]$. The filter \eqref{filter} equation can be expressed as
    \begin{equation*}
    \begin{aligned}
\mathrm{d}\pi_t(\varphi)
&=
\sum_{\substack{j,l=1\\j\neq l}}^d 
(\varphi(j)-\varphi(l))\,\psi_{lj}(t)\,
\widehat{\mathbbm{1}_{\{I_t=l\}}}\,\mathrm{d}t
\\[3pt]
&\quad+\int_{\mathbb{R}_+}
\frac{
\pi_{t^-}(\varphi\,\widetilde{\lambda}(v))
+\pi_{t^-}(\widetilde{D}(v))
-\pi_{t^-}(\varphi)\pi_{t^-}(\widetilde{\lambda}(v))
}{
\pi_{t^-}(\widetilde{\lambda}(v))
}
\Big(
\widetilde n(\mathrm{d}t,\mathrm{d}v)
-
\pi_{t^-}(\widetilde{\lambda}(v))\,
\mathrm{d}v\,\mathrm{d}t
\Big)
\\
&\quad+
\sum_{k\in\{+,-\}}
\int_{\mathbb{R}_+}
\frac{
\pi_{t^-}\big(\varphi\,(\bar \lambda^{k}(v)-\widetilde{\lambda}(v))\big)
+
\pi_{t^-}\big(D^k(v)-\widetilde{D}(v)\big)
-
\pi_{t^-}(\varphi)\,\pi_{t^-}(\bar \lambda^{k}(v)-\widetilde{\lambda}(v))
}{
\pi_{t^-}(\bar \lambda^{k}(v)-\widetilde{\lambda}(v))
}
\\
&\hspace{3.5cm}\times
\Big(
(n^{k}-\widetilde n)(\mathrm{d}t,\mathrm{d}v)
-
\pi_{t^-}(\bar \lambda^{k}(v)-\widetilde{\lambda}(v))\mathrm{d}v
\,\mathrm{d}t
\Big),
\end{aligned}
\end{equation*}
where, for all $k\in\{+,-\}$, $D^k$ and $\widetilde{D}$ are determined by
\begin{equation*}
    \begin{split}
\mathbb{E}\left[\Delta M^I_t\,n^{k}(\mathrm{d}t,\mathrm{d}v)\mid\mathcal{F}^N_{t^-}\right]
&=
D^k_t(v)\,\mathrm{d}v\,\mathrm{d}t~~\text{and}~~
\mathbb{E}\left[\Delta M^I_t\,\widetilde n(\mathrm{d}t,\mathrm{d}v)\mid\mathcal{F}^N_{t^-}\right]=
\widetilde D_t(v)\,\mathrm{d}v\,\mathrm{d}t.
\end{split}
\end{equation*}
\end{theo}
\begin{proof}
The result follows directly from Proposition \ref{ks_equation} and the fact that all driver processes in the proposition have no mutual covariations, that is, for all $t \in [0,T]$ and all $k \neq k^{\prime} \in \{+,-\}$,
$$[n^k_t - \widetilde{n}_t, \widetilde{n}_t] = [n^{k^{\prime}}_t - \widetilde{n}_t, \widetilde{n}_t] = [n^k_t - \widetilde{n}_t, n^{k^{\prime}}_t - \widetilde{n}_t] = 0.$$
\end{proof}

\section{Proofs of the Results in Section \ref{The Separated Impulse Control Problem}}
\label{app:dpp_n}
\begin{proof}[Proof of Proposition \ref{prop_approx_eq_val_fun}]
    Let $t\in[0,T]$, $N\in \mathbb{N}^*$ and $y = (x,s,d,\kappa^{+},\kappa^{-},\mu) \in \mathcal{D} $. We define $U_0$ and $U_N$, such that\begin{equation*}
        \begin{split}
            U_0\left(t,y\right) &= \mathbb{E}_{t,y}\bigg[ C\left({S}_{\tau_S}+{D}_{\tau_S},x\right)\bigg]\\U_N\left(t,y\right)&=\sup_{\tau \in \Theta_t} \mathbb{E}_{t,y}\bigg[\mathcal{M}V_{N-1}\left(\tau,x,{S}_{\tau},{D}_{\tau},\Bar{\lambda}^+_{\tau}, \Bar{\lambda}^-_{\tau},\pi_{\tau}\right)\bigg].
        \end{split}
    \end{equation*}
For $\alpha = (\tau_k,\xi_k)_{k\geq 1} \in \mathcal{A}^{N}_t(x)$, we have that
$$
\begin{aligned}
J\left(t,y,\alpha\right)&=\mathbb{E}_{t,y}\bigg[\mathbbm{1}_{\{\tau_1<\tau_S^N\}}C\left({S}^{\alpha}_{\tau_1} + {D}^{\alpha}_{\tau_1},|\Delta X_{\tau_1}|\right) \bigg]\\&\quad+\mathbb{E}_{t,y}\bigg[\mathbb{E}\bigg[\sum_{\tau \in ]\tau_1, \tau^N_S)}  C\left({S}^{\alpha}_{\tau} + {D}^{\alpha}_{\tau},|\Delta X_{\tau}|\right)+C({S}^{\alpha}_{\tau^N_S} + {D}^{\alpha}_{\tau^N_S},X_{\tau^N_S}) \mid \mathcal{F}^{N}_{\tau^+_1}\bigg]\bigg]\end{aligned}$$ 
As a result of Definition \ref{val_fun_approx} of the value function approximation $V_{N-1}$ and Definition \ref{intervention} of the intervention operator $\mathcal{M}$, 
\begin{equation*}\resizebox{\textwidth}{!}{$
    \begin{aligned} 
J&\left(t,y,\alpha\right) \\& \leq \mathbb{E}_{t,y}\bigg[C\left({S}^{\alpha}_{\tau_1} + {D}^{\alpha}_{\tau_1},\mathbbm{1}_{\{\tau_1<\tau_S^N\}}|\Delta X_{\tau_1}|\right) + V_{N-1}(\tau^+_1,x-\mathbbm{1}_{\{\tau_1<\tau_S^N\}}|\Delta X_{\tau_1}|,{S}^{\alpha}_{\tau^+_1},{D}^{\alpha}_{\tau^+_1}, \Bar{\lambda}^+_{\tau^+_1}, \Bar{\lambda}^-_{\tau^+_1}, \pi_{\tau^+_1})\bigg] \\& \leq \mathbb{E}_{t,y}\bigg[C\left({S}^{\alpha}_{\tau_1} + {D}^{\alpha}_{\tau_1},\mathbbm{1}_{\{\tau_1<\tau_S^N\}}|\Delta X_{\tau_1}|\right) + V_{N-1}(\Gamma(\tau_1,x,{S}^{\alpha}_{\tau_1},{D}^{\alpha}_{\tau_1}, \Bar{\lambda}^+_{\tau_1}, \Bar{\lambda}^-_{\tau_1}, \pi_{\tau_1},\mathbbm{1}_{\{\tau_1<\tau_S^N\}}|\Delta X_{\tau_1}|))\bigg] \\
& \leq \mathbb{E}_{t,y}\bigg[\mathcal{M}V_{N-1}\left(\tau_1,x,{S}^{\alpha}_{\tau_1},{D}^{\alpha}_{\tau_1}, \Bar{\lambda}^+_{\tau_1}, \Bar{\lambda}^-_{\tau_1}, \pi_{\tau_1}\right) \bigg] \\
& \leq U_N\left(t,y\right).
 \end{aligned}$}
\end{equation*}
It follows that $V_N \leq U_N$.

Let $\varepsilon>0$. There exists an $\epsilon$-optimal $\tau^* \in \Theta_t$, such that
\begin{equation*}
    \begin{split}
U_N\left(t,y\right) &\leq \frac{\varepsilon}{2}+\mathbb{E}_{t,y}\bigg[\mathcal{M}V_{N-1}\left(\tau^*,x,{S}^{\alpha}_{\tau^*},{D}^{\alpha}_{\tau^*}, \Bar{\lambda}^+_{\tau^*}, \Bar{\lambda}^-_{\tau^*}, \pi_{\tau^*}\right) \bigg].
\end{split}
\end{equation*}
Additionally, there exists an $\mathcal{F}^{N}_{\tau^{*}}$-measurable $\xi^*$  taking values in $a\left(X_{\tau^*}\right)$, such that
\begin{equation*}
    \begin{split}
        \mathcal{M}V_{N-1}\left(\tau^*,x,{S}^{\alpha}_{\tau^*},{D}^{\alpha}_{\tau^*}, \Bar{\lambda}^+_{\tau^*}, \Bar{\lambda}^-_{\tau^*}, \pi_{\tau^*}\right)&\leq \frac{\varepsilon}{2} + C({S}^{\alpha}_{\tau^*}+{D}^{\alpha}_{\tau^*},\xi^*)\\&\quad + V_{N-1}(\Gamma(\tau^*,x,{S}^{\alpha}_{\tau^*},{D}^{\alpha}_{\tau^*}, \Bar{\lambda}^+_{\tau^*}, \Bar{\lambda}^-_{\tau^*}, \pi_{\tau^*},\xi^*)).
    \end{split}
\end{equation*}
Hence,
\begin{equation}
\label{ineq_tnter_vn_un}
    \begin{split}
    U_N\left(t,y\right) &\leq \varepsilon +  \mathbb{E}_{t,y}\bigg[C({S}^{\alpha}_{\tau^*}+{D}^{\alpha}_{\tau^*},\xi^*)+ V_{N-1}(\Gamma(\tau^*,x,{S}^{\alpha}_{\tau^*},{D}^{\alpha}_{\tau^*}, \Bar{\lambda}^+_{\tau^*}, \Bar{\lambda}^-_{\tau^*}, \pi_{\tau^*},\xi^*))\bigg].
    \end{split}
\end{equation}
Finally, we introduce $\bar{\alpha}$, such that
    \begin{equation*}
        \bar{\alpha} = \left\{\begin{array}{ll}
\left(\xi^*,\tau^*\right) \quad\quad, \text { if } k = 1 \\
\hat{\alpha} = 
(\tau_k,\xi_k)\quad\quad,  \text { else } 
\end{array}\right.
    \end{equation*}
where $\hat{\alpha}\in\mathcal{A}^{N}_{\tau^*}(X_{\tau^*})$.
We obtain that
\begin{equation*}
    \begin{split}
V_N\left(t,y\right) &\geq \mathbb{E}_{t,y}\bigg[C({S}_{\tau^*} + {D}_{\tau^*},\xi^*)+J^{\bar{\alpha}}\left({\tau^*}^+,x-\xi^*,{S}_{{\tau^*}^+},{D}_{{\tau^*}^+}, \Bar{\lambda}^+_{{\tau^*}^+}, \Bar{\lambda}^-_{{\tau^*}^+}, \pi_{{\tau^*}^+}\right)\bigg] .
\end{split}
\end{equation*}
Using inequality \eqref{ineq_tnter_vn_un} and the fact that $\widehat{\alpha}$ is arbitrary, we get that
\begin{equation*}
    \begin{split}
V_N\left(t,y\right) & \geq \mathbb{E}_{t,y}\bigg[V_{N-1}\left(\Gamma\left(\tau^*,x,{S}_{\tau^*},{D}_{\tau^*}, \Bar{\lambda}^+_{\tau^*}, \Bar{\lambda}^-_{\tau^*}, \pi_{\tau^*},\xi^*\right)\right)+ C({S}_{\tau^*} + {D}_{\tau^*},\xi^*)\bigg]\\
& \geq U_N\left(t,y\right) -\varepsilon.
\end{split}
\end{equation*}
As $\varepsilon$ approaches zero, we conclude the proof.
\end{proof}
\section{Proofs of the Results in Section \ref{Characterization of the Value Function}}
\label{app:carac_vf}
\begin{proof}[Proof of Proposition \ref{increasing_vf}]
\textbf{Proof of results 1:}\\
Let $t\in [0,T]$, $\epsilon>0$, $\left(x,s,\kappa^{+},\kappa^{-},\mu\right)\in\mathbb{R}^+\times\mathbb{R}\times(\mathbb{R}_+^d)^2 \times \mathcal{S}$ and $d_1\leq d_2$. We define $\alpha = (\tau_k,\xi_k)_{k\geq 1} \in \mathcal{A}_t(x)$ as an $\epsilon$-optimal strategy for $V\left(t,x,s,d_1,\kappa^{+},\kappa^{-},\mu\right) = V\left(t,y_1\right)$, i.e.,
\begin{equation*}
    \begin{split}
        V\left(t,x,s,d_1,\kappa^{+},\kappa^{-},\mu\right)&\leq J\left(t,x,s,d_1,\kappa^{+},\kappa^{-},\mu,\alpha\right) +\epsilon\\&\leq 
\mathbb{E}_{t,y_1}\bigg[\sum_{\tau \in [t, \tau_S)}C\left({S}^{\alpha}_{\tau} + {D}^{\alpha}_{\tau},|\Delta X^{\alpha}_{\tau}|\right)+C\left({S}^{\alpha}_{\tau_S} + {D}^{\alpha}_{\tau_S},X^{\alpha}_{\tau_S}\right)\bigg] + \epsilon.
    \end{split}
\end{equation*}
We know that, for all $\tau\in\Theta_t$,
        \begin{equation*}
        \begin{split}
            {D}^{\alpha}_{\tau}&= e^{-\rho(\tau-t)}D_t -(1-\nu)\sum_{t\leq \tau_k<{\tau}}e^{-\rho({\tau}-\tau_{k})}Q(\xi_k)+(1-\nu)\int_{t}^{{\tau}}\int_{\mathbb{R}_+} e^{-\rho({\tau}-u)}Q(v) n(\mathrm{d}u,\mathrm{d}v).
        \end{split}
    \end{equation*}
    Additionally, we know that the term 
\begin{equation*}\resizebox{\textwidth}{!}{$
    \begin{aligned} 
\mathbb{E}_{t,y}\bigg[\int_{t}^{\tau}\int_{\mathbb{R}_+} e^{-\rho({\tau}-u)} Q(v) \, n(\mathrm{d}u, \mathrm{d}v)\bigg] = \sum_{i=1}^d \mathbb{E}_{t,y}\bigg[\int_{t}^{\tau}\int_{\mathbb{R}_+} \pi_i(u) e^{-\rho({\tau}-u)} Q(v) (\lambda^{i,+}_u - \lambda^{i,-}_u) \nu_i(\mathrm{d}v) \, \mathrm{d}u\bigg]
 \end{aligned}$}
\end{equation*}
 is non-decreasing in $\kappa^{i,+}$ and decreasing in $\kappa^{i,-}$ on $\mathbb{R}_+$, for all $t\in[t,T]$, $\tau\in\Theta_t$ and $y\in\mathcal{D}$. Consequently, $\mathbb{E}_{t,y_2}[{D}^{\alpha}_{\tau}]\geq\mathbb{E}_{t,y_1}[{D}^{\alpha}_{\tau}]$. Since $p\mapsto C(p,a)$ is affine, hence non-decreasing, in $p$ on $\mathbb{R}$ for every fixed $a\in\mathbb{R}_+$ (see \eqref{reward_per_trade}), we get that \begin{equation*}
        \begin{split}
        \mathbb{E}_{t,y_2}\bigg[C\left({S}^{\alpha}_{\tau}+{D}^{\alpha}_{\tau},a\right)\bigg] &= \mathbb{E}_{t,y_2}\bigg[C\left(\mathbb{E}_{t,y_2}[{S}^{\alpha}_{\tau}+{D}^{\alpha}_{\tau}],a\right)\bigg]\\&\geq \mathbb{E}_{t,y_1}\bigg[C\left(\mathbb{E}_{t,y_1}[{S}^{\alpha}_{\tau}+{D}^{\alpha}_{\tau}],a\right)\bigg]\\&\geq \mathbb{E}_{t,y_1}\bigg[C\left({S}^{\alpha}_{\tau}+{D}^{\alpha}_{\tau},a\right)\bigg],
        \end{split}
    \end{equation*} 
    for all $a\in\mathbb{R}_+$. Therefore,
     \begin{equation*}
    \begin{split}
        V\left(t,x,s,d_2,\kappa^{+},\kappa^{-},\mu\right)&\geq\mathbb{E}_{t,y_2}\bigg[\sum_{\tau \in [t, \tau_S)}C\left({S}^{\alpha}_{\tau} + {D}^{\alpha}_{\tau},|\Delta X^{\alpha}_{\tau}|\right)+C\left({S}^{\alpha}_{\tau_S} + {D}^{\alpha}_{\tau_S},X^{\alpha}_{\tau_S}\right)\bigg] \\&\geq\mathbb{E}_{t,y_1}\bigg[\sum_{\tau \in [t, \tau_S)}C\left({S}^{\alpha}_{\tau} + {D}^{\alpha}_{\tau},|\Delta X^{\alpha}_{\tau}|\right)+C\left({S}^{\alpha}_{\tau_S} + {D}^{\alpha}_{\tau_S},X^{\alpha}_{\tau_S}\right)\bigg] \\&\geq V\left(t,x,s,d_1,\kappa^{+},\kappa^{-},\mu\right) - \epsilon.
    \end{split}
\end{equation*}
As $\epsilon$ approaches zero, we obtain $V\left(t,x,s,d_1,\kappa^{+},\kappa^{-},\mu\right) \leq V\left(t,x,s,d_2,\kappa^{+},\kappa^{-},\mu\right)$ for all $0 \leq d_1 \leq d_2$. Therefore, $V$ is non-decreasing in $d$ on $\mathbb{R}$. The proofs of the monotonicity in $s$ is similar to the latter proof.\\
\textbf{Proof of result 2:}\\
Let $t\in[0,T]$, $y = (x,s,d,\kappa^{+},\kappa^{-},\mu) \in \mathcal{D}$ and $\alpha\in\mathcal{A}_t(x)$. Using \eqref{reward_per_trade} and \eqref{prob_formulation}, we get that
\begin{equation*}
\begin{split}
    J\left(t,y,\alpha\right)&=\mathbb{E}_{t,y}\bigg[\sum_{\tau \in [t, \tau_S)}\bigg(|\Delta X^{\alpha}_{\tau}|({S}^{\alpha}_{\tau}+{D}^{\alpha}_{\tau}) - \int_0^{Q(|\Delta X^{\alpha}_{\tau}|)}vf(v)\mathrm{d}v-c_0\bigg)\bigg] \\&\quad + \mathbb{E}_{t,y}\bigg[\mathbbm{1}_{\{|\Delta X^{\alpha}_{\tau_S}|>0\}}\bigg(|\Delta X^{\alpha}_{\tau_S}|({S}^{\alpha}_{\tau_S}+{D}^{\alpha}_{\tau_S}) - \int_0^{Q(|\Delta X^{\alpha}_{\tau_S}|)}vf(v)\mathrm{d}v-c_0\bigg)\bigg]
\end{split}
\end{equation*}
Based on the controlled dynamics of the fundamental price ${S}^{\alpha}$ and of the price deviation ${D}^{\alpha}$ described in \eqref{price_dynamics}, the reward function $J$ can be expressed as
\begin{equation*}
\begin{split}J\left(t,y,\alpha\right)&=xs+\nu\mathbb{E}_{t,y}\bigg[\sum_{\tau \in [t, \tau_S)}|\Delta X^{\alpha}_{\tau}|\bigg(\int_{t}^{{\tau}}\int_{\mathbb{R}_+} Q(v) n(\mathrm{d}u,\mathrm{d}v)-\sum_{t\leq \tau_k<{\tau}}Q(\xi_k)\bigg)\bigg] \\&\quad +\mathbb{E}_{t,y}\bigg[\sum_{\tau \in [t, \tau_S)}|\Delta X^{\alpha}_{\tau}|{D}^{\alpha}_{\tau}- \int_0^{Q(|\Delta X^{\alpha}_{\tau}|)}vf(v)\mathrm{d}v -c_0\bigg]   
 \\&\quad+\nu\mathbb{E}_{t,y}\bigg[\mathbbm{1}_{\{|\Delta X^{\alpha}_{\tau_S}|>0\}}|\Delta X^{\alpha}_{\tau_S}|\bigg(\int_{t}^{{\tau_S}}\int_{\mathbb{R}_+} Q(v) n(\mathrm{d}u,\mathrm{d}v)-\sum_{t\leq \tau_k<{\tau_S}}Q(\xi_k)\bigg)\bigg] \\&\quad +\mathbb{E}_{t,y}\bigg[\mathbbm{1}_{\{|\Delta X^{\alpha}_{\tau_S}|>0\}}\bigg(|\Delta X^{\alpha}_{\tau_S}|{D}^{\alpha}_{\tau_S}- \int_0^{Q(|\Delta X^{\alpha}_{\tau_S}|)}vf(v)\mathrm{d}v -c_0\bigg)\bigg]\\&=:xs+g(t,x,s,d,\kappa^+,\kappa^-,\mu,\alpha).
\end{split}
\end{equation*}
Hence, the linear dependence on the initial parameter $s$ in the dynamics of $S^{\alpha}$ makes the following
separation
\begin{equation*}
    \begin{split}
        V(t,y) = xs+\sup_{\alpha\in\mathcal{A}_t(x)}g(t,x,s,d,\kappa^+,\kappa^-,\mu,\alpha),
    \end{split}
\end{equation*}
possible.
The residual dependence of $g$ on $s$ occurs only through the bankruptcy time $\tau^{\alpha}_S$, which enters the summation sets $[t,\tau_S)$, the indicators $\mathbbm{1}_{\{|\Delta X^{\alpha}_{\tau_S}|>0\}}$ and the terminal terms above. All the other terms involve only increments of $S^{\alpha}$, the process $D^{\alpha}$ and the trade sizes, none of which depends on $s$. In particular, on any open set of parameters on which the bankruptcy probability vanishes, $\tau^{\alpha}_S \equiv T$ does not vary with $s$, and $\frac{\partial V}{\partial s}(t,y) = x$ there. This completes the proof.
\end{proof}

\end{document}